\documentclass[a4paper,12pt]{article}
\pdfoutput=1
\def\parn              {  \par\noindent }

\def\parmedskipn        {  \par\medskip\noindent  }
\def\parsmallskipn      {  \par\smallskip\noindent  }
\def\al{\alpha}
\def\be{\beta}
\def\ga{\gamma} 
\def\ep{\epsilon}
\def\lam{\lambda}
\def\Lam{\Lambda}
\def\sig{\sigma}

\def\calA{{\cal A}}   
 \def\calE{{\cal E}}

  \def\calO{{\cal O}}
\def\calP{{\cal P}}  
  \def\calU{{\cal U}}

\def\ntil{\tilde{n}}
\def\vtil{\tilde{v}}

\def\Jtil{{\tilde{J}}}

\def\Omtil{\widetilde{\Omega}}

\def\phitil{\tilde{\phi}}
\def\psitil{\tilde{\psi}}

\def\adot{{\dot{a}}}
\def\bdot{{\dot{b}}}

\def\aldot{{\dot{\alpha}}}

\def\onedot{{\dot{1}}}
\def\twodot{{\dot{2}}}

\def\phat{\hat{p}}

\def\alhat{\hat{\alpha}}

\def\pbar{{\bar{p}}}

\def\wbar{{\bar{w}}}

\def\zbar{{\bar{z}}}

\def\delbar{\bar{\del}}

\def\albar{\bar{\alpha}}

\def\braket#1#2{\langle #1 | #2 \rangle}

\def\del        {  \partial }
\def\half       {  {1\over 2}  }
\def\rootof#1   {  \left( #1 \right)^{1/2}  } 
\def\trace      {  \mbox{Tr}\,  }
\def\Tr         { {\rm Tr}\,  }
\def\abs#1      {  \vert #1 \vert  }
\def\ie         {{\it i.e.}\,\,}
\def\evalat#1   {  \left\vert_{#1} \right. } 
\def\e          { {\rm e}  }

\def\comma          {\, ,}
\def\period         {\, .}
\def\lsim      {\lower .65ex \hbox{\ $\stackrel{<}{\sim}$\ } }
\def\gsim      {\lower .65ex \hbox{\ $\stackrel{>}{\sim}$\ } }
\def\det       {{\rm det}\, }

\def\braket#1#2{\langle #1 | #2 \rangle}
\def\com#1#2{{ \left[#1, #2\right] } } 

\def\matel#1#2#3  {{\langle #1 | #2 | #3 \rangle } }

\def\lrvec#1    {\hbox{$\stackrel{\leftrightarrow}{#1}$}}
\def\lvec#1     {\hbox{$\stackrel{\leftarrow}{#1}$}}

\def\vecii#1#2      {  \left(\begin{array}{c}#1\\#2\end{array}\right)  }
\def\veciii#1#2#3   {  \left(\begin{array}{c}#1\\#2\\#3\end{array}
                     \right)  }
\def\veciv#1#2#3#4  {  \left(\begin{array}{c}#1\\#2\\#3\\#4
                                 \end{array}\right)  }
\def\vecfv#1#2#3#4#5 {  \left(\begin{array}{c}#1\\#2\\#3\\#4\\#5
                                 \end{array}\right)  }

\def\matrixii#1#2#3#4            {  \left(\begin{array}{cc}#1&#2\\#3&#4
                                       \end{array}\right) }
\def\matrixiii#1#2#3#4#5#6#7#8#9 {  \left(\begin{array}{ccc}#1&#2&#3\\
                                     #4&#5&#6\\#7&#8&#9\end{array}
                               \right)  }
\def\mativ#1#2#3#4               {  \left(\begin{array}{cccc}
                                       #1\\#2\\#3\\#4\end{array}\right) }
\def\matv#1#2#3#4#5              {  \left(\begin{array}{ccccc}
                                     #1\\#2\\#3\\#4\\#5\end{array}
                              \right)  }
\def\eqabegin         {  \begin{eqnarray}  }
\def\eqaend           {  \end{eqnarray}  }
\def\nn               {  \nonumber  }
\def\bracetwo#1#2     {  \left\{ \begin{array}{l} #1 \\ #2 \end{array}
                         \right.  }
\def\bracetwocases#1#2#3#4  {   \left\{ \begin{array}{ll} #1 &
                                 \qquad #2 \\
                                 #3 & \qquad #4 \end{array} \right.  }
\def\bracebegin#1     {  \left\{ \begin{array}{#1}   }
\def\braceend         {  \end{array}\right.   }
\def\shead#1   { \parmedskipn {\bfall $\Box$\ #1}: \parmedskipn }
\def\head#1    { \parmedskipn {\bfall $\Box$\ #1}: \qquad }
\def\lhead#1   { \parmedskipn {\large\bfall $\Box$\ #1} \parmedskipn }
\def\Lhead#1   { \parmedskipn {\Large\bfall $\Box$\ #1} \parmedskipn }
\def\boxit#1#2      {  \vbox{\hrule\hbox{ \hskip -4.1pt \vrule\kern3pt 
                     \vbox
                    {  \hsize #1 \strut\kern3pt #2 \kern3pt\strut  }
                       \kern3pt  \vrule} \hrule  } }
\def\centerbox#1#2  {  \mbox{  }\par\bigskip  \hfil \boxit{#1}{#2} \hfil
                       \par\bigskip\noindent }
\def\rightbox#1#2   {  \hfill\boxit{#1}{#2}  }
\def\leftbox#1#2    {  \boxit{#1}{#2}  }
\def\fullbox#1      {  \boxit{\textwidth}{#1}  }

\newcommand{\nullify}[1]{}

\def\mpg#1#2{\begin{minipage}[t]{#1} #2  \end{minipage} }

\def\bfall{\boldmath\bf }
\def\nxt{\parsmallskipn}

\def\epsfig#1#2#3{
{\lower #3 \hbox{
 \mpg{#1}{\begin{center} \includegraphics[width=#1,clip]{#2.eps} \\
 Fig. #2\end{center} }}}}

\def\papertitlepage{\baselineskip 3.5ex \thispagestyle{empty}}
\def\Title#1{\baselineskip 1cm \vspace{1.5cm}\begin{center}
 {\Large\bf #1} \end{center} 
\vspace{0.5cm}}
\def\Authors#1{\begin{center} {\it #1} \end{center}}
\def\Abstract{\vspace{1.0cm}\begin{center} {\large\bf Abstract} 
           \end{center} \par\bigskip}
\def\Komabanumber#1#2#3{\hfill \begin{minipage}{4.2cm} UT-Komaba #1
              \parn #2 
              \parn #3 \end{minipage}}
\renewcommand{\thefootnote}{\fnsymbol{footnote}}

\renewenvironment{thebibliography}{\pagebreak[3]\par\vspace{0.6em}
\begin{flushleft}{\large \bf References}\end{flushleft}
\vspace{-1.0em}

\begin{enumerate}\if@twocolumn\baselineskip=0.6em\itemsep -0.2em
\else\itemsep -0.2em\fi\labelsep 0.1em}{\end{enumerate} }

\def\vecX{\vec{X}}

\def\slprod#1#2{\langle #1, #2 \rangle}

\def\bfall{\boldmath\bf}
\usepackage{ifpdf}
\ifpdf
\usepackage{graphicx} 
\usepackage{epsfig}
\usepackage[setpagesize=false,pagebackref=false, linktocpage, bookmarksopen=true, colorlinks=true, linkcolor=darkred,citecolor=darkred,urlcolor=darkred]{hyperref}
\usepackage{hyperref}
\newcommand{\arXiv}[2]{[\href{http://arxiv.org/abs/#1}{{\tt arXiv:#2}}]}
\newcommand{\hep}[2]{[\href{http://arxiv.org/abs/#1}{{\tt #2}}]}
\def\picture#1#2{\includegraphics[#1]{#2.pdf}}
\else
\usepackage{graphicx}
\newcommand{\arXiv}[2]{[{\tt arXiv:#2}]}
\newcommand{\hep}[2]{[{\tt #2}]}
\def\picture#1#2{\includegraphics[#1]{#2_400x400_p1.eps}}
\fi
\usepackage{latexsym}
\usepackage{amsmath}
\usepackage{amssymb}
\usepackage[usenames]{color}
\definecolor{darkgreen}{rgb}{0,0.5,0}
\definecolor{darkblue}{cmyk}{0.9,0.9,0,0}
\definecolor{darkred}{rgb}{0.6,0,0.3}
\usepackage{cite}
\definecolor{MyRed}{cmyk}{0,1,1,0.15}
\definecolor{MyBlue}{cmyk}{1,1,0,0.25}

\def\Xbar{\bar{X}}
\def\xbar{\bar{x}}
\def\act{J}
\def\barz{\bar{z}}
\def\delbar{\bar{\partial}}
\def\fn#1{\footnote{#1}}

\newcommand{\jumpp}[1]{[\![#1]\! ]_{\star}}
\def\beq#1{\begin{align}#1\end{align}}
\def\pmatrix#1#2{\left( 
\begin{array}{#1}
#2\end{array} 
\right)}

\def\braket#1{\langle #1 \rangle}

\def\ads3{Euclidean $AdS_3$}
\def\sref#1{(\ref{#1})}
\def\figref#1{figure \ref{#1}}
\def\ch{{\rm ch}}
\def\sh{{\rm sh}}
\def\T{\text{\tiny{$T$}}}
\def\iDelta{\scalebox{0.7}[1]{\text{$\mathit{\Delta}$}}}
\def\siDelta{\scalebox{0.7}[1]{\text{\scriptsize{$\mathit{\Delta}$}}}}
\def\sn{{\rm sn}}
\def\cn{{\rm cn}}
\def\dn{{\rm dn}}

\def\prone{\alpha}
\def\prtwo{\beta}
\def\ulp{{\prone _{i}^{+}}}
\def\ulm{{\prone}_{i}^{-}}
\def\dlp{\overline{\prone}{}_{i}^{+}}
\def\dlm{\overline{\prone}{}_{i}^{-}}
\def\urp{\prtwo _{i}^{+}}
\def\urm{\prtwo _{i}^{-}}
\def\drp{\overline{\prtwo}{}_{i}^{+}}
\def\drm{\overline{\prtwo}{}_{i}^{-}}
\def\ulpm{{\prone}_{i}^{\pm}}
\def\urpm{\prtwo _{i}^{\pm}}
\def\dlpm{\overline{\prone}{}_{i}^{\pm}}
\def\drpm{\overline{\prtwo}{}_{i}^{\pm}}
\def\ulpone{{\prone _{1}^{+}}}

\def\urpone{\prtwo _{1}^{+}}

\def\drpone{\overline{\prtwo}{}_{1}^{+}}

\def\ulptwo{{\prone _{2}^{+}}}
\def\ulmtwo{{\prone}_{2}^{-}}

\def\urptwo{\prtwo _{2}^{+}}
\def\urmtwo{\prtwo _{2}^{-}}

\def\drmtwo{\overline{\prtwo}{}_{2}^{-}}

\def\ulptri{{\prone _{3}^{+}}}

\def\urptri{\prtwo _{3}^{+}}

\def\sigone{\sigma_{1}}
\def\tauone{\tau_{1}}
\def\sigtwo{\sigma_{2}}
\def\tautwo{\tau_{2}}
\def\sigg{\sigma}
\def\taug{\tau}
\def\cl{{\rm cl}}
\def\bsc#1{\scalebox{0.9}{$#1$}}
\newcommand{\beqa}{\begin{eqnarray}}
\newcommand{\eeqa}{\end{eqnarray}}
\newcommand{\bea}{\begin{array}}
\newcommand{\eea}{\end{array}}
\newcommand{\beqn}{\begin{equation}}
\newcommand{\eeqn}{\end{equation}}

\def\Omegaref{\Omega^{{\rm ref}}}
\def\vechref{\vec{h}^{{\rm ref}}}
\def\vecpsiref{\vec{\psi}^{{{\rm ref}}}}
\def\bebar{\bar{\beta}}
\def\refsol{\mathbb{X}^{{\rm ref}} }

\def\kaphat{\hat{\kappa}}
\def\xsup#1{x^{(#1)}} 
\def\xbarsup#1{\bar{x}^{(#1)}} 
\def\hatL#1{\hat{#1}^L} 
\begin{document}
\papertitlepage
\vspace*{0cm}
\Komabanumber{12-4}{May, 2012}{}
\Title{Wave functions and correlation  functions  \\ for GKP  strings  from 
 integrability } 
\Authors{{\sc Yoichi Kazama\footnote[2]{{\tt kazama@hep1.c.u-tokyo.ac.jp}}
 and Shota Komatsu\footnote[3]{{\tt skomatsu@hep1.c.u-tokyo.ac.jp}}
\\ }
\vskip 3ex
 Institute of Physics, University of Tokyo, \\
 Komaba, Meguro-ku, Tokyo 153-8902 Japan \\
  }
\baselineskip .7cm

\numberwithin{equation}{section}
\numberwithin{figure}{section}
\numberwithin{table}{section}
\parskip=0.9ex

\Abstract
We develop a general  method of computing the contribution of the 
 vertex operators to the semi-classical correlation functions of heavy 
string states,  based on the state-operator correspondence and the integrable structure of the system.  Our method  requires 
only the knowledge of the local behavior of the  saddle point configuration around each vertex insertion point  and  can be applied to cases where the precise forms of the 
 vertex operators are not known.  As an important application, we compute 
 the contributions  of the vertex operators to the three-point functions
 of the large spin limit of the Gubser-Klebanov-Polyakov (GKP) strings in $AdS_3$ spacetime, left 
unevaluated  in our previous work [\href{http://arxiv.org/abs/1110.3949}{arXiv:1110.3949}] which  initiated such a study. 
Combining  with the finite part of the action already computed  previously  
and with the newly  evaluated  divergent part  of the action, we obtain 
 finite three-point functions with the expected 
 dependence of the  target space boundary coordinates on the 
dilatation charge and the spin.

\newpage
\baselineskip 3.5ex
\renewcommand{\thefootnote}{\arabic{footnote}}
\thispagestyle{empty}
\enlargethispage{2\baselineskip}
\renewcommand{\contentsname}{\hrule {\small \flushleft{Contents}}}
{\footnotesize \tableofcontents}
\nxt\hrule
\newpage
\section{Introduction }
For  the study of  conformally invariant theories, understanding of 
 two point and   three point functions constitutes the crux of the matter. 
Two point functions encode  the spectrum while three point functions determine the essential dynamics,  and together they govern the entire  theory,  at least in principle.  This of course applies to the study of AdS/CFT correspondence\cite{AdSCFT:Malda,AdSCFT:GKP,AdSCFT:Witten}, in which the conformal invariance plays the key role. 

In this article, we will be concerned with  such correlation functions in the best
studied duality of that type, namely the one between the $N=4$ super Yang-Mills (SYM) theory in four dimensions and the type IIB superstring theory 
in $AdS_5 \times S^5$ spacetime. 
During the past fifteen years since the inception
 of such a duality, the scope of the studies has been expanded from the BPS to the non-BPS sectors and from the two-point to the three-point functions. The history and the achievements 
 made through such studies are summarized, up to around 2010, in the comprehensive collection of review articles \cite{review}.  

In the course of the development, the integrability found on both sides of the correspondence played a pivotal role. On the SYM side, the problem of finding 
the perturbative spectrum of the composite operators,  equivalently their two point functions, has been mapped to the analysis of integrable spin chains and 
many precise results have been obtained. Such approach was subsequently generalized to the three point functions\cite{Okuyama:2004bd, Roiban:2004va, Alday:2005nd} and the new underlying integrable structure is now beginning to be uncovered\cite{Escobedo, Escobedo2, Caetano, Gromov:2011jh, Gromov:2012vu, Serban:2012dr, Kostov:2012jr, Foda, Foda2, Kostov2, Morphism}.  

On the other hand the existence of the classical integrability of the string sigma model in $AdS_5\times S^{5}$ was proven in the early days of the development\cite{BPR:0305} and is later used to characterize various classical solutions\cite{KZ,KMMZ,BKSZ}.
However it was rather recently that people began to understand how such classical solutions can be  utilized in the saddle point calculation  
of two point functions\cite{Tsuji, JSW, BuchTseytlin}. 
 As for the study of three point functions,  it is substantially more difficult, mainly because the relevant saddle point solutions  
are in general hard to construct. One special class for which this difficulty 
 can be  avoided is the so-called heavy-heavy-light (HHL) correlators\cite{Zarembo, Costa, RT}. In this case, 
 since the light vertex operator, carrying a small quantum number, does not 
 affect  the saddle configuration determined by the other two heavy operators, 
one can treat it  as a ``perturbation" of the two point function with 
 relative ease. In this way results for a variety of HHL correlators  have been 
obtained \cite{Hernandez:1011, Ryang:1011, Georgiou:1011, HHL_RusTsey, ParkLee:1012, BuchTsey:1012, BakChenWu:1103, Kristjansen:1103, Arnaudov:1103, Hernandez:1104, Ahn:1105}. 
Computations have also been made of  a more general yet still restricted class of three point functions, where one can approximate the AdS part of the saddle point configuration by that of the zero mode of the string \cite{KloseMcL, Buchbinder:2011jr, Ryang:2012pm}. 
This should be valid for BPS or near BPS strings with  charges in the $S^5$ 
 part.  For other studies on the three point functions from different viewpoints, see   \cite{Georgiou:2011qk, GGGP, Bissi, Bozhilov:2012be, Grignani:2012yu, Caputa:2012yj, Grignani2}.

Studies of the three point functions of  genuine  non-BPS  heavy states 
have been initiated recently in \cite{JW} for the strings rotating in $S^k$ but not 
in the AdS space,  and in our previous paper\cite{Kazama:2011cp} for the large spin limit of the GKP strings\cite{GKPstring} 
(to be called LSGKP strings) spinning in $AdS_3$ but sitting 
 at a point on a sphere.  In these works, the classical integrability 
 of the system is  fully exploited, as in the computation of the gluon scattering 
 amplitudes \cite{AM:0903, AM:0904, AGM:0911, AMSV:1002, Mald-Zhi}, and the essential information can be extracted without the explicit knowledge of the 
 saddle point configuration.  In this way, the universal contribution from the $AdS_2$ part was obtained in \cite{JW}, while the essential part of the three point coupling for the LSGKP strings, including the dependence on the  spins of the strings, has been computed
in our previous work.  However, in both of these works, the complete 
answers  for the three point functions were not yet obtained. 
For \cite{JW}, the contribution from the motion in the $S^k$  part was not 
 computed.   On the other hand, in our previous work, 
 the contributions of the divergent 
 part of the action and of the vertex  operators, which together should give 
 a finite result  including the dependence on the vertex insertion positions on the 
 boundary, were left unevaluated.  Such  computations remain  as  the most urgent 
  problems to be solved in this line of approach.

In the present work, we shall  give a solution to the problem 
left unanswered  in our previous study. 
Namely, we  develop a powerful method 
of computing the contribution of the vertex operators, which can be applied to 
 the cases where both  the precise form of the vertex operators and the exact 
 form of the the saddle point solution are not available.  Using this method, we will be
 able to complete the computation of the three point functions for  the LSGKP strings
 started in our previous work.  What we will need are only the local asymptotic behaviors of the solution near the vertex insertion points  and certain  global symmetry transformations. Then the integrability and the analyticity possessed by the system 
will allow us to unite  such informations and  compute 
 the desired contributions.

Let us be more specific and describe the essence of our method. 
The structure of the three point function in the semi-classical 
 approximation can be expressed schematically in the following way:
\begin{align}
G(x_1, x_2, x_3) &= e^{-S[X_\ast]_{{\rm finite}} }
 \left[ e^{-S[X_\ast]_\ep }  \prod_{i=1}^3 
V_i[X_\ast; x_i, Q_i ]_\ep \right] \period \label{corfundecomp}
\end{align}
It consists of the contribution of the action and that of the vertex operators, 
 evaluated on the saddle point configuration denoted  by $X_\ast$. 
The contribution of the action can be split, in an appropriate way, into 
the finite part $S[X_\ast]_{{\rm finite}}$ and the divergent  part $S[X_\ast]_\ep$. 
 $V_i[X_\ast; x_i, Q_i ]_\ep$ denotes the value of the 
$i$-th vertex  operator on $X_\ast$,   which carries  a large charge $Q_i
\sim O(\sqrt{\lam})$, where $\lam$ is the 't Hooft coupling,   and is  located at $x_i$ on the boundary of the AdS space. The subscript $\ep$ on $S[X_\ast]_\ep$ 
and the vertex operators signifies  that these quantities contain divergences which  are regularized by a small parameter $\ep$. 
As it will be reviewed in section 2,  with the use of  the Pohlmeyer 
reduction and  a newly developed  generalized Riemann bilinear identity, 
 we were able to compute in \cite{Kazama:2011cp} the quantity  $S[X_\ast]_{{\rm finite}}$, which was called the ``regularized  area".  But unfortunately the rest of the 
 contributions shown in the square bracket in (\ref{corfundecomp}), 
 in particular the one  from the vertex operators, was not obtained.

There are two major difficulties in computing  the contribution 
 of the vertex operators. One is  that even semi-classically the precise form of the conformally invariant vertex operator for the LSGKP string is not known. 
The other is the difficulty to construct  the saddle point solution  for the three point function.  We suggested in our previous work 
 that the appropriate  way to compute  the contribution of the vertex 
 operators would be to make use of the semi-classical wave functions, which are 
 related  to the vertex operators by  the state-operator correspondence. 
In particular, if one can find the action-angle variables of the system, 
 such wave functions can be constructed easily since the Hamilton-Jacobi 
equations become simple. But of course finding the action-angle 
variables is a difficult task for non-linear systems and  it is even more 
 difficult to evaluate the wave functions constructed  in terms of them 
 without knowing the saddle point solution. For these reasons 
 we could not implement  the  idea above  explicitly for the GKP strings. 

We now describe  two essential observations which allow us to overcome these difficulties. 

The first observation is that for the class of so-called finite gap solutions,  to which 
  the GKP solution belongs, one can construct the action-angle variables 
 by the method of ``separation of variables" developed by  Sklyanin\cite{Sklyanin}. For the case 
 of the string in $AdS_3$, due to the existence of the Virasoro constraints, 
 the analytic structures of the relevant quasi-momentum and the spectral 
 curve are  modified  from those   for the well-studied  case of the string in 
$AdS_3 \times S^1$\cite{Vicedo,DV1,DV2}, and one has to analyze them carefully. Nonetheless, 
we can show  that the Sklyanin's method is still applicable. 
The details of this construction will be given  in section 3.2. 

The second 
important observation is that the values of the angle variables, which 
 determine the value of the wave function corresponding to a vertex 
 operator,  can be computed from the behavior of the local solution
near the position of the  vertex operator. This is intimately related to 
the recognition that the choice of the normalization of  the so-called Baker-Akhiezer 
eigenvector, which governs the value of the angle variables in the Sklyanin's method, can be  characterized  by the global symmetry transformation that  generates the local solution  around the vertex operator from a common reference solution. 
More precise description will be given in section 3.3. 

Based on these observations, we develop a general method which leads to 
explicit procedures and formulas for  computing  the contributions of the 
vertex operators,  which generate  string configurations described by 
 finite gap solutions. To check the validity and the power of the method, we first 
 apply it to the computation of two point functions.  In this case, our method 
 is particularly simple and makes clear that essentially it constructs 
   the representation of the global conformal symmetry in terms of  the boundary coordinates  through  combination of the wave functions. 
Due to this general feature, the detailed form of the saddle point solution is not important and  we can compute the 
 two point functions for the general elliptic GKP strings,  \ie  without taking the large spin limit. 

Next  we apply the method to the main subject of our study, the three point 
functions of the LSGKP strings. The basic procedure is the same around each 
 vertex operator.  These contributions from the vertex operators 
 can be reexpressed in terms of the differences of the positions of
 the vertex operators on the boundary, 
 producing  the correct representation of the conformal symmetry on  the three point 
 function for  states carrying the spin as well as the conformal dimension. One important difference from the case of the two point function is that, 
together with such expressions made up of the combinations of position coordinates, 
there appear additional factors which carry the information 
about how the three states intertwine. 
 Remarkably, they can be expressed in terms of 
 the contour integrals around the three singularities on the 
 worldsheet,  which  played  crucial roles  in the computation of the finite part of the action performed in our previous work.  Consequently, we can evaluate  such 
 quantities in a similar fashion as before.  Moreover,  the divergent part of the action, 
which is the last remaining piece to be computed, can also be represented 
 by the same type of contour integrals through the Riemann bilinear identity 
and this makes  the mechanism of the cancellation of divergences transparent. 

The final result for the three point function for the LSGKP strings is given in (\ref{FINAL}). It exhibits the correct dependence on the boundary coordinates expected 
 of the three point functions carrying the spin and reduces, in the appropriate limit, 
 to the properly normalized two point function. Unfortunately, comparison with 
 the corresponding correlator  on the SYM side is not possible 
as it is not available at present time.

The organization of the rest of the  article is as follows.  To make this article 
more or less self-contained, we will give,  in section 2,  a brief review of the method 
 and the result of our previous work and set the notations. 
In section 3, we develop a general method for evaluating the contributions 
 of the vertex operators. Since this section contains a variety of materials, we 
 will first sketch the general strategies in subsection 3.1.  We then describe 
 in subsection 3.2 how one can construct the action-angle variables 
for a string in the (Euclidean) $AdS_3$ by applying the Sklyanin's method. 
In order to discuss the generic structure of the action-angle variables, we 
 make the analysis from the universal point of view of ``infinite gap" solutions. 
Section 3.3  explains how one can evaluate the angle variables and the wave functions  by the use of appropriate  global symmetry transformations. The general method 
 developed up to this point is then applied to the computation of the two point 
 functions in section 4.  Namely, we derive the general formula in section 4.1 and 
apply it to the case of the GKP string in section 4.2. Finally, in section 5 
we will  complete the calculation of the three point functions for the LSGKP string 
initiated in our previous paper. In section 5.1, by using the generalized Riemann bilinear
 identity,  we express the divergent part of the action in such a form 
that its cancellation with the divergent part of the wave functions is easy to understand. Then in section 5.2 the contributions of the wave functions are evaluated by applying 
our general method. These results as well as the finite part of the action 
 computed in our previous paper are put together in section 5.3 to give the final form of  the three point function. 
We conclude and indicate future directions in section 6 and 
provide a number of  appendices to give some useful details.

\section{A brief review of the previous work } 
We begin by giving a concise review of the methods and the results obtained 
 in our previous work and set the notations. 

The LSGKP strings of our interest live  in the $AdS_5$ spacetime with  
 the embedding coordinates  $(X_{-1},X_{0},X_{1},X_{2}, X_3, X_4)$ 
satisfying $-X_{-1}^2 - X_0^2 +X_1^2+ \cdots + X_4^2 = -1$. 
They  are related to the  Poincar\'e coordinates $(x^\mu, z) = (x^{0},x^{1},x^{2},x^{3},z)$ as
\beq{
 X_{-1}+X_{4}=\frac{1}{z}\comma \qquad X_{-1}-X_{4}=z+\frac{x^{\mu}x_{\mu}}{z}\comma \qquad X_{\mu}=\frac{x_{\mu}}{z}
 \label{poincare}
\comma }
where  $z=0$ corresponds to  the boundary of $AdS_5$. In what follows, we will focus on the solutions  which propagate  in the 
subspace spanned by $\vecX\equiv  (X_{-1}, X_1, X_2, X_4)$,  having the structure 
 of Euclidean $AdS_3$. The action on the Euclidean worldsheet
 parametrized by the  plane coordinate $(z, \zbar)$ is  proportional to the area $A$ and takes the form 
\begin{align}
S &= T A = 2T \int d^2z 
(\del \vecX \cdot \delbar \vecX + \Lam (\vecX \cdot \vecX +1))
\comma 
\end{align}
where $T=\sqrt{\lambda}/2\pi $  is the string tension, $\vec{A}\cdot \vec{B} \equiv 
-A_{-1} B_{-1} + A_1 B_1 + A_2 B_2 + A_4 B_4$, and $\Lam$ is 
 the Lagrange multiplier. After eliminating $\Lam$, the equation of motion 
for $\vecX$ is given by 
\begin{align}
\del \delbar \vecX &= (\del \vecX \cdot \delbar \vecX) \vecX \period
\end{align}
In addition we must impose the Virasoro constraints 
\begin{align}
\del \vecX \cdot \del \vecX &= \delbar \vecX \cdot \delbar \vecX =0 \period
\end{align}
The global isometry group for real $\vecX$ is $SO(1,3)$. However,  as we shall be  concerned with the  solutions which are  
the saddle point configurations 
for the semi-classical correlation functions, we need to allow complex solutions and the 
global symmetry group of the system should be taken  as $SO(4,C) \simeq SL(2,C)_L \times  SL(2,C)_R$. To exhibit its action it is convenient to introduce  the 
matrix with unit determinant 
\begin{align}
\mathbb{X} &\equiv \matrixii{X_+}{X}{\Xbar}{X_-} \comma \qquad 
\det \mathbb{X}=1 \comma \label{matrixg}
\end{align}
where 
\begin{align}
X_+ &= X_{-1} + X_4 \comma \qquad X_- = X_{-1} - X_4  
\comma \label{matrixg2}\\
X &= X_1 + iX_2 \comma \qquad \Xbar = X_1 -iX_2 \period\label{matrixg3}
\end{align}
Then $SL(2)_L\times SL(2)_R$ acts as 
\begin{align}
\mathbb{X}' &= V_L \mathbb{X} V_R \comma \qquad V_L \in SL(2)_L \comma \quad 
 V_R \in SL(2)_R \period \label{sltransfg}
\end{align}

A typical  LSGKP solution, which we shall call
 ``the reference solution", can be written as
\begin{align}
\refsol &= \matrixii{e^{-\kappa \tau} \cosh \rho(\sig) }{e^{\kappa \tau} \sinh \rho(\sig) }{ e^{-\kappa \tau} \sinh \rho(\sig) }{ 
 e^{\kappa \tau} \cosh \rho(\sig)  }  
\comma  \label{gB}
\end{align}
where $(\tau, \sig)$ are the Euclidean cylinder coordinates on the worldsheet
 and $\kappa$ is a positive parameter. 
 The function $\rho(\sig)$,   periodic with period $2\pi$, is  given in the interval $[ -{\pi \over 2}, {3\pi \over 2}]$ by 
\begin{align}
\rho(\sig) =\bracetwo{ \kappa \sig  \comma  \hspace{1.7cm} \left( -{\pi \over 2}  \le \sig \le {\pi \over 2}\right)}{\kappa \left( \pi  -\sig \right) 
\comma 
\quad \left({\pi\over 2} \le \sig \le {3\pi \over 2}\right)} 
 \period
\end{align}
Note that this solution  starts at $\tau=-\infty$ from the boundary and reaches  the horizon at $\tau=\infty$.  It will be useful to view the class
 of  solutions of our interest, namely 
 those which start from the boundary and end also on the boundary, 
as obtained from the  standard solution above by 
applying appropriate $SL(2)_L \times SL(2)_R$ symmetry transformations. 
There are two conserved global charges  in this system, namely the dilatation 
charge $\Delta$ and the angular momentum (spin) $S$ in the $1$-$2$ plane. They 
 are given in terms of the parameter $\kappa$ by 
\begin{align}
\Delta  &= \frac{\sqrt{\lambda}}{2\pi}\kappa \int_0^{2\pi} d\sig\,\cosh^2 \rho = \frac{\sqrt{\lambda}}{2\pi} (\kappa \pi 
 + \sinh \kappa \pi) \comma  \label{AdSE} \\
S &=  \frac{\sqrt{\lambda}}{2\pi}\kappa \int_0^{2\pi} d\sig\,\sinh^2 \rho = \frac{\sqrt{\lambda}}{2\pi} (-\kappa \pi 
 + \sinh \kappa \pi) \period \label{Spin}
\end{align}

One of the major difficulties for computing the correlation functions 
of  three or more  LSGKP strings as the external states is that the 
relevant classical saddle point configurations have not been found. 
Fortunately, this obstacle can be overcome  by making use of the classical integrability possessed by the system.  The most convenient framework turned out to be  the method of  Pohlmeyer reduction\footnote{For the evaluation of the 
contribution from the vertex operators,  which is the main focus  of the present work,   the finite gap integration method will also be indispensable .}, which proved extremely powerful in handling 
 a similar situation in the  analysis of  the minimal surface problem encountered in the computation of the gluon scattering  amplitudes  at strong coupling\cite{AM:0903, AM:0904, AGM:0911, AMSV:1002, Mald-Zhi}. 
In this method, one deals with the $SL(2)_L \times SL(2)_R$ invariant 
reduced degrees of freedom $\al, p$ and $\pbar$ defined by 
\begin{align}
e^{2\al} &= \half \del \vec{X} \cdot \delbar \vec{X}
\comma \quad p= \half \vec{N} \cdot \del^2 \vec{X} 
\comma \quad \pbar = -\half \vec{N} \cdot \delbar^2 \vec{X} 
\comma 
\end{align}
where $\vec{N}$ is a unit-normalized vector orthogonal to $\vec{X}, \del \vec{X}$ and $\delbar \vec{X}$. The information of the original equations 
 of motion as well as the Virasoro conditions can be encoded in 
 the flatness of  certain $2\times 2$ $SL(2)$ 
connection matrices  $B_z^L, B_\zbar^L, 
B_z^R, B_\zbar^R$, namely $\com{\del + B_z^L}{\delbar +B_\zbar^L} 
 =0$ and $\com{\del + B_z^R}{\delbar +B_\zbar^R} 
 =0$. These conditions lead to 
\begin{align}
&\del \delbar \al -e^{2\al} +p \pbar e^{-2\al}   =0 \comma \label{mSGeq} \\
& \del \pbar = \delbar p =0 \period \label{ppbareq}
\end{align}
Moreover, reflecting  the integrability of the system, one can construct 
 a flat connection with an arbitrary complex spectral parameter $\xi$ 
out of these connections. 
Explicitly its components are given by 
\begin{align}
B_z(\xi) &= {1\over \xi} \Phi_z +A_z \comma \qquad 
B_\zbar(\xi) = \xi \Phi_\zbar + A_\zbar \comma \label{Bxicon1}\\
A_z &\equiv   \matrixii{\half \del\al}{0}{0}{-\half \del\al}
\comma \qquad A_\zbar \equiv  \matrixii{-\half\delbar\al}{0}{0}{\half\delbar \al}
\comma \label{Bxicon2}\\
\Phi_z & \equiv  \matrixii{0}{-e^\al}{-p e^{-\al}}{0} \comma\qquad 
\Phi_\zbar  \equiv \matrixii{0}{-\pbar e^{-\al}}{-e^\al}{0} \period\label{Bxicon3} 
\end{align}
The original connections are identified as those  at the special values 
 of the spectral parameter (with a similarity transformation for $B^R$) in the manner
\begin{align}
B^L_z &= B_z(\xi=1) \comma \qquad B^L_\zbar = B_\zbar(\xi =1) 
\comma \\
B^R_z &= \calU^\dagger B_z(\xi=i) \calU 
\comma \qquad B^R_\zbar = \calU^\dagger B_\zbar(\xi =i) \calU 
\comma \\
\calU &= e^{i\pi/4}\matrixii{0}{1}{i}{0}  \period
\end{align}
Now the flatness of $B(\xi)$ is equivalent to the compatibility of the 
set of  linear equations 
\begin{align}
(\del + B_z(\xi))\psi (\xi) &= 0 \comma \qquad 
(\delbar + B_\zbar(\xi))\psi (\xi) =0 \comma   \label{eqalpxi}
\end{align}
 known as the auxiliary linear problem. Once the two independent 
 solutions $\psi(\xi)$ of this system are obtained, one can immediately get the 
solutions for the left and right auxiliary linear problems involving the 
 connections $B^L$ and $B^R$ as $
\psi^L= \psi(\xi=1) \comma \psi^R = \calU^\dagger \psi(\xi=i)$. 
More specifically, we denote the solutions as $\psi^L_{\al, a}$ and 
$\psi^R_{\aldot, \adot}$, where $\al=1,2$ and $\aldot=1,2$ refer to the 
matrix indices of  $B^L$ and $B^R$ (\ie  the $SL(2)_{L,R}$ spinor indices), while $a=1,2$ and $\adot=1,2$ label 
 the two independent solutions. 
What will be of great importance is the $SL(2)$-invariant product between 
 two spinors given by 
\begin{align}
\slprod{\psi}{\chi} &\equiv \ep^{\al\be} \psi_\al \chi_\be \comma 
\qquad (\ep^{\al\be} = -\ep^{\be\al}\comma 
\quad \ep^{12} \equiv 1) \period  \label{sl2prod}
\end{align}
One can then show that the solutions $\psi^{L,R}$ can  be normalized 
as 
\begin{align}
\slprod{\psi^L_a }{ \psi^L_b} &=  \ep_{ab} \comma \qquad 
\slprod{\psi^R_\adot}{\psi^R_\bdot} =  \ep_{\adot\bdot}
\comma 
\label{normpsisol}
\end{align}
where $\ep_{ab}$ is the  anti-symmetric tensor with $\ep_{12} \equiv 1$. 
With such normalized solutions, one can reconstruct
 the solution for the matrix 
 $\mathbb{X}$ of  the embedding coordinates as 
\begin{align}
\mathbb{X}_{a\adot} &= \psi^L_{1, a} \psi^R_{\onedot, \adot} +
\psi^L_{2, a} \psi^R_{\twodot, \adot} \period \label{reconstformula}
\end{align}

In the discussion so far, the spectral parameter $\xi$ has not played any significant role: It  merely served as a convenient device to handle  left and right problems in a unified manner. Moreover,  the Pohlmeyer reduction by itself 
does not generate the solution of the non-linear equations  of motion. 
(Afterall,  the LSGKP solution is known from the beginning.)
However, this formalism will be  extremely useful for the computation of 
 three (and higher) point  functions for which the saddle point solutions 
are not known. The reason is as follows. 
 It allows us to characterize the {\it local} behavior of the solution in the vicinity of the sources, \ie  the  vertex operators, by certain functions of $\xi$. Moreover, through 
 the analyticity property in $\xi$ these  local 
 behaviors can be interconnected  and yield  such {\it global}  information  as  the value of the area. 

The direct $SL(2)$-invariant data characterizing the LSGKP solutions  are the 
form of the functions $p(z), \pbar(\zbar)$ and the relation between $\al(z,\zbar)$ and these functions. They are given by 
\begin{align}
p(z) &= -{\kappa^2 \over 4z^2} \comma \qquad \pbar(\zbar) =
-{\kappa^2 \over 4\zbar^2}\comma  \label{ppbarforgkp}\\
e^{2\al(z,\zbar)} &= 
\sqrt{p\pbar}  \period \label{alforgkp}
\end{align}
These properties are then reflected  on  the solution of the auxiliary linear 
 problem. By making a gauge transformation 
\begin{align}
\psi &= \mathcal{A} \tilde{\psi}\comma\qquad \mathcal{A}=\pmatrix{cc}{p^{-1/4}e^{\alpha/2}&0\\0&p^{1/4}e^{-\alpha/2}} 
\comma \label{gaugetrA}
\end{align}
the equations $(\del + B_z(\xi))\psi=0$ and  $(\delbar + B_\zbar(\xi))\psi=0$ drastically simplify  and two independent solutions for $\psitil$ can easily be obtained as
\begin{align}
&\tilde{\psi}_\pm=\exp \left( \pm \frac{\kappa i}{2}\left(\xi ^{-1}\ln z-\xi \ln \zbar \right)\right)\pmatrix{c}{1\\ \pm1}\period \label{psitwoeq}
\end{align}
Now as we go around the origin once, this pair of basis functions  $(\psitil_+,\psitil_-)$ gets transformed as\footnote{In our previous work \cite{Kazama:2011cp},  the quantity $\phat(\xi)$ was denoted by $\rho(\xi)$. }
\begin{align}
\vecii{\psitil'_+}{\psitil'_-} &=M \vecii{\psitil_+}{\psitil_-}
 \comma \quad 
M = \matrixii{e^{i \phat(\xi)}}{0}{0}{
e^{-i\phat(\xi)} }  
 \comma \quad \phat(\xi) = i \kappa \pi \left( {1\over \xi} + \xi \right)
\period 
\label{monodGKP}
\end{align}
The matrix $M$,  called  the local monodromy,  is an important  characteristic of the LSGKP solution viewed as the local solution in the vicinity of the appropriate vertex operator.

As was already reviewed in the introduction, what was computed in our previous work is the 
part of the action integral called the regularized area $A_{reg}$. It was defined in the following way. First we split the total area $A$ into  the finite part and the divergent part as 
\begin{align}
A &=2 \int d^2z \del \vecX \cdot \delbar \vecX 
= 4 \int d^2z e^{2\al} 
= A_{fin} + A_{div}\comma \\
A_{fin} &= 4\int d^2 z \,\left( e^{2\alpha}-\sqrt{p\pbar}\right)
 \comma 
\qquad A_{div} = 4\int d^2z \,\sqrt{p\pbar} \period \label{decompA}
\end{align}
Then, by using the equation of motion (\ref{mSGeq}) and 
 evaluating  a certain boundary integral, 
$A_{fin}$ can be rewritten as
\beq{
A_{fin}&= 2 A_{reg} + \pi (N-2) \comma \label{area0}\\
A_{reg} &\equiv \int d^2z \,\left(e^{2\alpha}+p\pbar \,e^{-2\alpha}
-2\sqrt{p\pbar}\, \right) \period \label{area}}
For the three point function of our interest, $N$ should be set to 3.

As explained carefully  in our previous paper, since LSGKP string is completely folded,  one can consider the worldsheet  of half of the folded string which can be taken as the upper half plane 
and then smoothly extend it to the whole complex plane to account for the 
 contribution of the other half. The net result is that,   as far as the computation of the area is concerned, we may forget about the effect of folding. 

\begin{figure}
\begin{minipage}{0.32\hsize}
  \begin{center}
   \picture{clip,height=4cm}{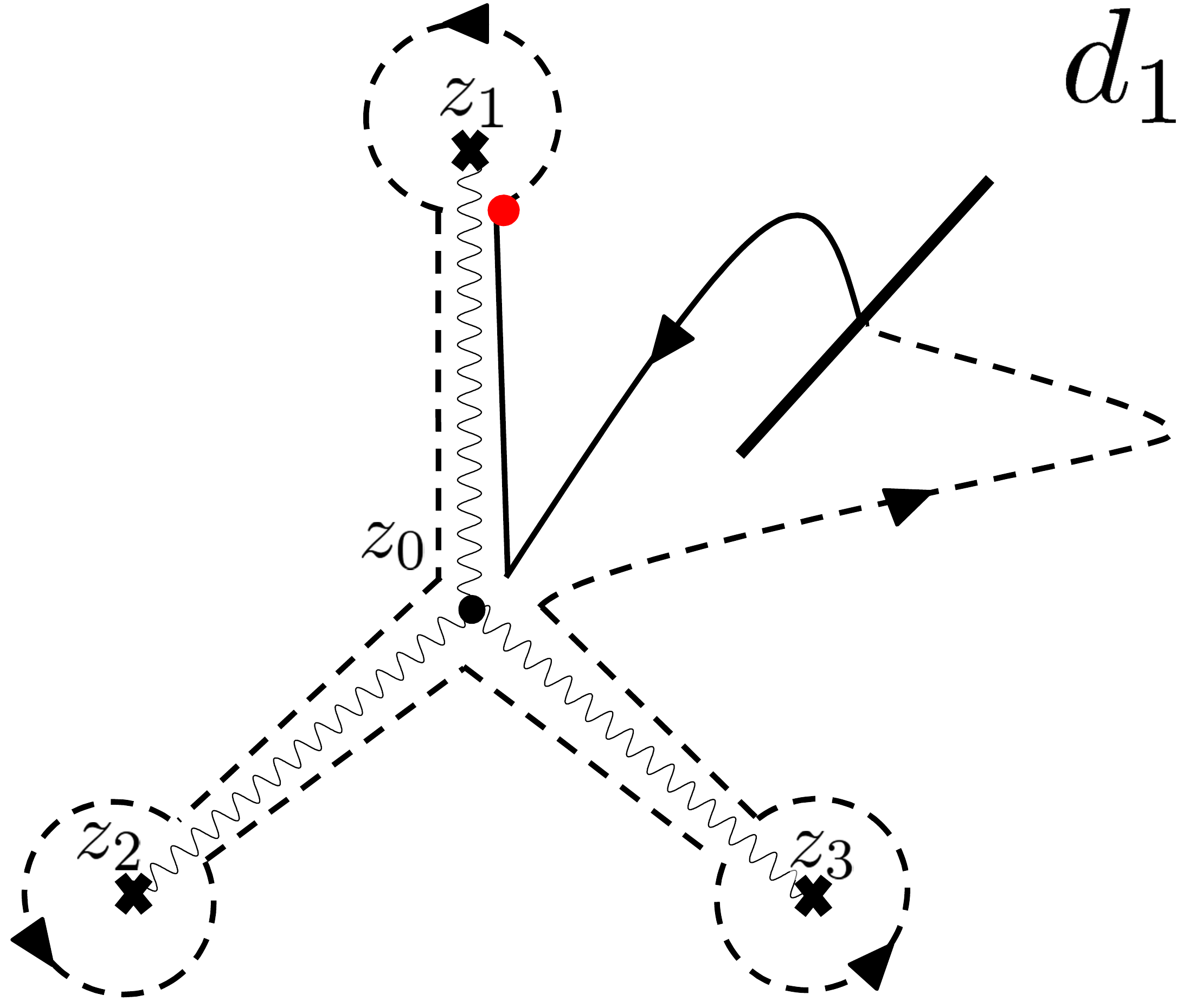}
  \end{center}
 \end{minipage}
 \begin{minipage}{0.32\hsize}
  \begin{center}
   \picture{height=4cm,clip}{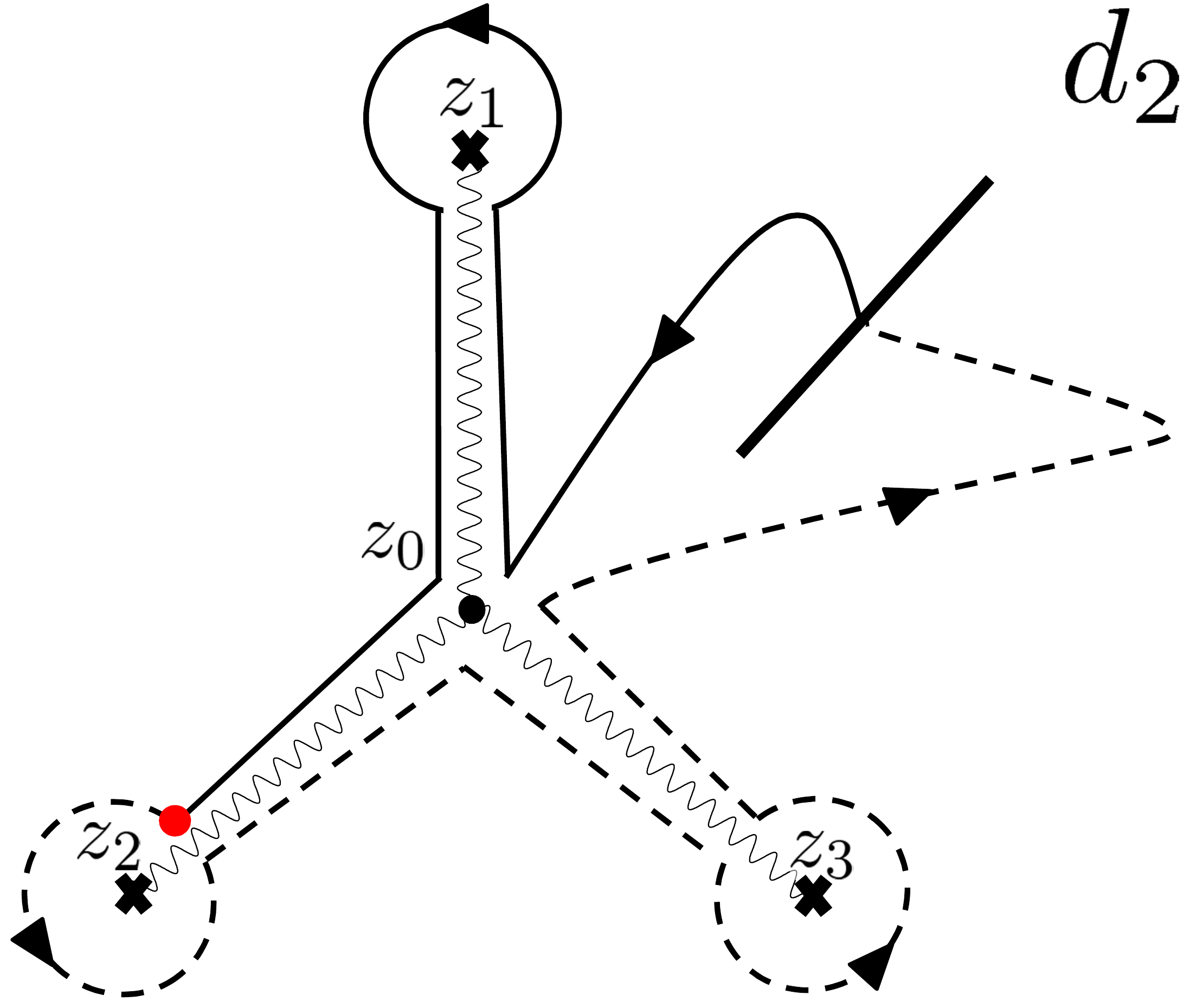}
  \end{center}
 \end{minipage}
\begin{minipage}{0.32\hsize}
  \begin{center}
   \picture{height=4cm,clip}{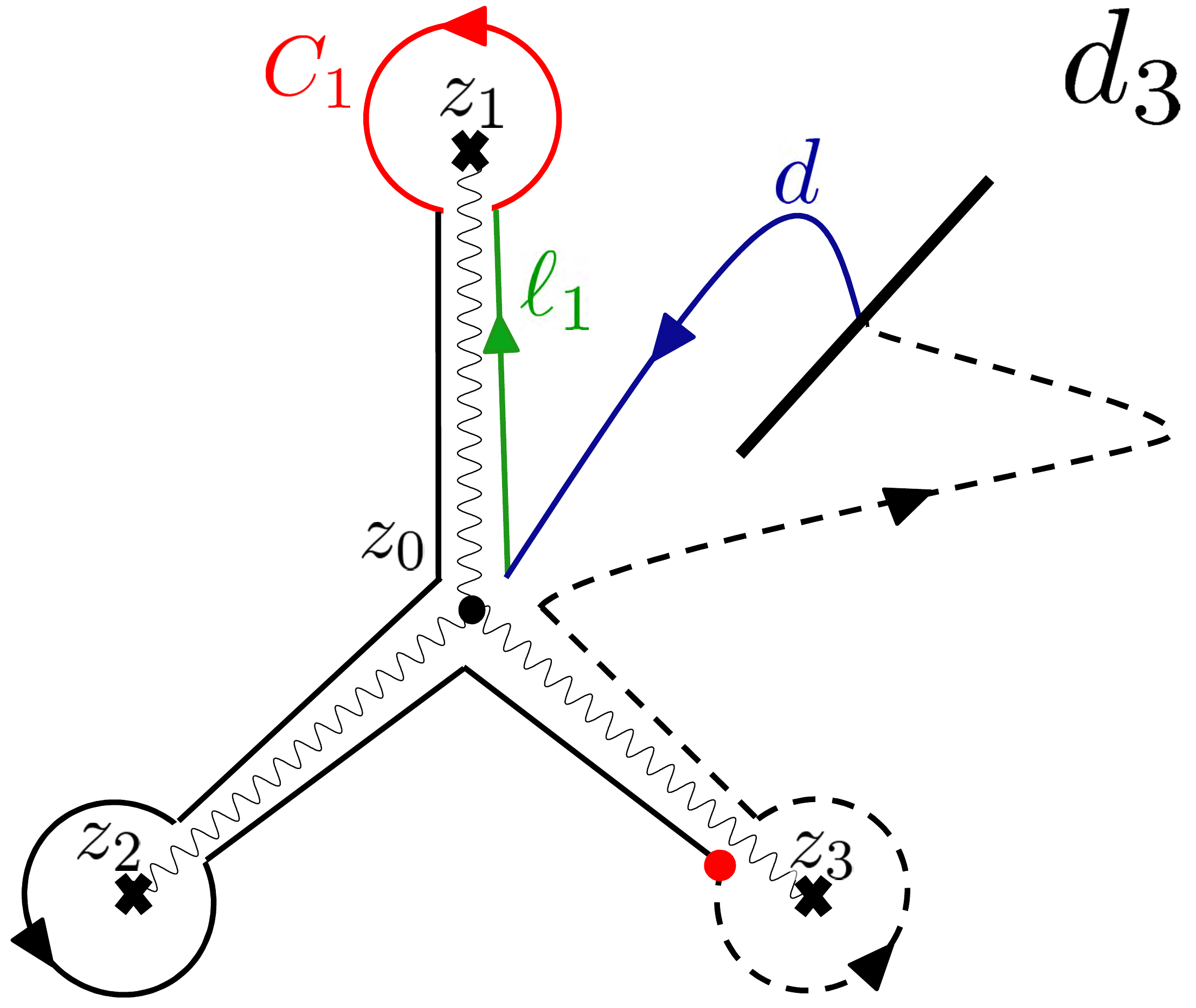}
  \end{center}
 \end{minipage}
\caption{ Definitions of the contours $d_1, d_2$ and $d_3$. 
The solid lines are on the first sheet, while the dotted lines are  on the second sheet. 
In $d_3$ we indicate  its ``components" $d$, $C_1$ and $\ell_1$. Other 
  $C_i$'s and $\ell_i$'s  are defined similarly in the vicinity of $z_i$. } 
\label{contourdj}
\end{figure}

The first step in computing $A_{reg}$ is to reexpress  this area  integral 
 in terms of certain contour integrals by developing  a  generalization of the 
Riemann bilinear identity. Let us briefly recall the basic idea. By introducing 
 a convenient variable
\begin{align}
\alhat &\equiv  \al -{1\over 4} \ln p\pbar \comma 
\end{align}
which vanishes at each vertex insertion point $z_i$, the regularized area can 
 be rewritten  as 
\begin{align}
A_{reg} &= {i \over 2} \int \lam dz \wedge \omega  \label{Areg} \\
\lam &\equiv  \sqrt{p(z)} \comma \qquad \omega \equiv u d\zbar + v dz \\
 u&=2\sqrt{\bar{p}} (\cosh 2\alhat -1) \comma \qquad 
  v = {1\over \sqrt{p}} (\del \alhat)^2 \period
\end{align}
 We have added the  $vdz$ part (which does not contribute to the area) with the property $\del u=\delbar v$ 
  to make  $\omega$ a closed 1-form.  Then introducing the integral of $\lam(z)$ as 
\begin{align}
\Lam(z) &= \int^z_{z_0} \lam(z') dz' \comma 
\end{align}
the area integral can be converted, by the Stokes theorem,  into a contour integral along a boundary  $\del D$ of a certain region $D$ as 
\begin{align}
A_{reg} &= {i \over 4} \int_D d\Lam \wedge \omega = 
{i \over 4} \int_D d(\Lam \omega) = {i \over 4} \int_{\del D}
\Lam \omega \period \label{intLamom}
\end{align}
To specify $D$, one needs to know the analytic  structure of the function $\Lam(z)$, which in turn is dictated by that of $p(z)$. As was shown in 
(\ref{ppbarforgkp}), $p(z)$  should behave around each vertex insertion point $z_i$ as 
\beq{
p(z) \overset{z\to z_i}{\sim} \frac{\delta _i^2}{(z-z_i)^2}  \comma 
\qquad (i=1,2,3)
}
for a LSGKP string with $\delta_i^2 =-\kappa_i^2/4$, and in the case of three point function it is given explicitly by 
\begin{align}
p(z) &=  \left( {\delta_1^2 z_{12} z_{13} \over z-z_1}
 +  {\delta_2^2 z_{21} z_{23} \over z-z_2} + {\delta_3^2 z_{31} z_{32} \over z-z_3}\right) {1\over (z-z_1)(z-z_2)(z-z_3)} \comma \label{pz3pt}\\
z_{ij} &\equiv z_i-z_j \period \nn
\end{align}
From this one can deduce that $\Lam(z)$ has three  logarithmic branch cuts running from the singularities at $z_i$ and  one square root 
 cut connecting two zeros of $p(z)$.   Thus, it is convenient  to take $D$ 
as the double cover  of the worldsheet with appropriate boundary $\del D$
, on which  $\Lam(z)$ is single-valued.  To evaluate the contour integral 
(\ref{intLamom}), we developed an extension of the Riemann bilinear identity, applicable  in the presence of logarithmic cuts as well as the usual square root cuts, which expresses (\ref{intLamom})  in terms of products of simple contour integrals along basic  closed paths. 
We will not display the general form of this identity here\footnote{Later in 
 section 5.1, where we evaluate the divergent part of the area, we will need 
 to recall the full form of the identity.}  but  simply 
 recall the final expression of $A_{reg}$ after the application 
 of this identity. In the case of  LSGKP strings, it gets  simplified to 
\begin{align}
A_{reg} &=
\frac{\pi}{12}+\frac{i}{4}\sum _{j=1}^3 
 \int_{C_j} \lam dz \int_{d_j}  (u d\zbar + v dz)
\period \label{riemann4} 
\end{align}
Here 
 $C_j$ is a small circle around the singularity at  $z_j$ and  the integrals $\int_{C_j}\lam dz$ can be easily evaluated. On the other hand, $d_j$ is a contour which starts from $\widehat{z}_j$, 
which is a point on the second sheet right below $z_j$, goes around the logarithmic cuts counterclockwise and reaches $z_j$ on the first sheet, as shown in \figref{contourdj}. Evaluation of the integrals  $\int_{d_j} (u d\zbar + v dz)$ 
constitutes the major non-trivial task. 

An important observation is that such an integral along $d_j$ is composed of 
 more elementary units, namely integrals along paths each of which  
 connects  a pair of singular points  $\{z_i, z_j\}$. 
 Thus, one may expect that 
the wanted information can be extracted from the properties  of the eigenfunctions of the auxiliary linear problem around the singularities  $z_i$ 
and along such paths.  

As already shown in (\ref{monodGKP}) for the LSGKP solution itself, 
 the behavior around  $z_i$ is characterized by the local monodromy 
matrix $M_i$ with a unit determinant. Each $M_i$ can be separately diagonalized as 
\begin{align}
M_i &= \matrixii{e^{i\phat_i(\xi)}}{0}{0}{e^{-i\phat_i(\xi)}}  \comma 
\end{align}
where $\phat_i(\xi)$ is of the form given in (\ref{monodGKP}).  
The local solutions of the auxiliary linear problem which form
 the eigenvectors of $M_i$ take the form 
\begin{align}
i_\pm &\sim \exp\left[ \pm \left( {1\over \xi} \int \sqrt{p(z)} dz 
 + \xi \int \sqrt{\pbar(\zbar)} d\zbar \right) \right] \comma 
\label{defiplusminus}
\end{align}
which can be normalized as $\slprod{i_+}{i_-}  =1$. 
In general $M_i$'s are not simultaneously diagonalizable, due to  the (unknown) non-trivial global behavior 
of the solutions of the auxiliary linear problem. Nevertheless, the simple 
 global relation $M_1 M_2 M_3=1$, expressing  the triviality of the monodromy  around the entire worldsheet,  is  enough to restrict the form 
 of $M_i$ sufficiently (although not completely).  This then allows one to 
 compute  the eigenvectors $i_\pm$ and further their $SL(2)$-invariant 
 products $\slprod{i_+}{j_+}, \slprod{i_-}{j_-}, \slprod{i_+}{j_-}$ 
in terms of $\phat_i(\xi)$'s. Although these expressions still contain a few 
 undetermined coefficients,  they cancel in certain products   of 
$\slprod{i}{j}$ and we obtain definite expressions.
 One such example  is  $\braket{2_-, 1_+}\braket{1_-,2_+}$, or taking the logarithm for convenience, 
\beq{
\log  \braket{2_-, 1_+}
+ \log \braket{1_-,2_+}= \log \left(\frac{ \sin{\phat_1 -\phat_2 +\phat_3
\over 2}\sin{-\phat_1 + \phat_2+\phat_3 \over 2} }{\sin \phat_1\sin \phat_2}\right) \period \label{T-like}
}
Later we will describe  how to  obtain  the information of each 
 individual invariant  from the combination of this form.

Having learned that we can express (certain combinations of) the products 
$\slprod{i}{j}$ in terms of  $\phat_i(\xi)$ characterizing local LSGKP 
behavior, we now wish to compute  $\slprod{i}{j}$ from a different 
more  dynamical point of view, namely from the actual (global) solutions of the auxiliary linear problem. Clearly, in the absence of the saddle point 
solution for the three point function,  this cannot be done directly. However, 
 it turned out to be sufficient to formally solve the auxiliary linear problem  in powers of $\xi$ and $1/\xi$, that is by using the so-called ``WKB expansions". In applying this method, the meaningful classification of the local behavior of the solutions is not by the overall signs of the exponent of the eigenfunctions,  as  in $i_\pm$ of  (\ref{defiplusminus}), but by the actual signs of the 
real part of the exponent. In other words, what is important is 
whether the solution increases or decreases 
 exponentially as it approaches $z_i$. In particular, the exponentially decreasing  ``small" solution, denoted by  $s_i$,   is  important as  it is not contaminated by  the ``big" component and  is hence unambiguous.

Skipping  all the somewhat tedious details, the end result of the WKB analysis  around $\xi=0$ 
is the formula for the $SL(2)$-invariant product of small solutions\footnote{There is  a similar formula for the expansion around $\xi=\infty$.} 
 $\slprod{s_i}{s_j}$. 
To illustrate  it with a concrete example, consider the case where 
the branch structure of  $\sqrt{p(z)}$ at $z_i$ is such
 that on the first sheet 
\begin{align}
\sqrt{p(z)} &\stackrel{z\sim z_i}{\sim} {i \kaphat_i \over z-z_i} \comma \qquad 
 \kaphat_i=\bracetwo{\kappa_i \quad \mbox{for $i=1,3$}}{-\kappa_i \quad 
 \mbox{for $i=2$}} \period 
\end{align}
Then, if ${\rm Im}\, \xi >0$,  the small solutions at $z_i$ can be  identified with 
the $i_\pm$  solutions as $s_1 \sim 1_+, s_2 \sim 2_-,
s_3 \sim 3_+$ 
and the product  $\slprod{s_i}{s_j}$ is given by 
 the formula like 
\begin{align}
\braket{s_1\comma s_2} &= 
\exp \left[ \left(\xi ^{-1}\int ^{z_2} _{z_1}\sqrt{p}dz  +\xi \int ^{z_2} _{z_1}\sqrt{\bar{p}}d\barz \right) +\frac{\xi}{2} \left(\int _{z_1} ^{z_2} ud\barz +v dz \right) +\cdots \right] \period
\end{align}
A remarkable fact is that in the exponent 
the type of contour integral of our interest $\int ( ud\barz +v dz )$ makes its appearance. Moreover, 
 as described in our previous paper,  one can make the contour of this integral to coincide 
 with the $d_j$'s  (recall \figref{contourdj})
 by making judicious ratios of $\slprod{s_i}{s_j}$'s. 
For example, the one along the contour $d_1$ can be produced in 
 \begin{align}
&\frac{\braket{s_2\comma s_3}}{\braket{s_2\comma s_1}\braket{s_1\comma s_3}}=\exp\left[\frac{1}{\xi} \int _{d_1} \lambda dz + \xi \int_{d_1}\sqrt{\bar{p}}d\barz +\frac{\xi}{2} \left(\int _{d_1} ud\barz +v dz \right) +\cdots\right] \period \label{ratioinvprod}
\end{align}
This shows that the computation of the integrals $\int_{d_j}
(ud\zbar + vdz)$,  and hence the  regularized area,  is now  reduced 
 to that of  $\slprod{s_i}{s_j}$.  

At this point, let us recall that 
 we already have some important explicit knowledge of  such $SL(2)$ invariant 
products, although not directly of $\slprod{s_i}{s_j}$. It is the 
information about  certain products of $\slprod{i}{j}$'s, such as 
$\slprod{1_+}{2_-}\slprod{1_-}{2_+}$ shown  in (\ref{T-like}). Now the problem is how to extract the information of the individual factors  $\slprod{1_+}{2_-}$ and $\slprod{1_-}{2_+}$  and
 relate them to $\slprod{s_1}{s_2}$. The key to the solution of this problem 
is to recognize that which of the $i_\pm$ is identified with the small solution $s_i$ 
depends crucially on the sign of the imaginary part\footnote{
This is because  the residue of the  pole of $\sqrt{p(z)}$ 
at the singularity is pure imaginary and hence the change of  the sign of 
${\rm Im}\, \xi$ swaps the small and the big solutions.} 
 of  $\xi$. As we already mentioned, 
if ${\rm Im}\, \xi >0$ the identification is $1_+ \sim s_1$ and $2_- \sim s_2$. 
However,  for ${\rm Im}\, \xi <0$ the identification is reversed, namely 
 $1_- \sim s_1$ and $2_+ \sim s_2$. 
Therefore, $\slprod{1_+}{2_-}$ and $\slprod{1_-}{2_+}$ 
must be identified with $\slprod{s_1}{s_2}$ in the 
domains ${\rm Im}\, \xi >0$ and ${\rm Im}\, \xi <0$ respectively. 
This analyticity property strongly characterizes  the logarithms  of  
a part of  $\slprod{1_+}{2_-}$ and $\slprod{1_-}{2_+}$ which are regular at $\xi=0,\infty$. Applying the Wiener-Hopf decomposition
\begin{align}
\frac{1}{2\pi i}\int _{-\infty}^{\infty}d\xi^{\prime}
 \frac{1}{\xi^{\prime}-\xi} \left(F(\xi ^{\prime})+G(\xi ^{\prime})\right) &= 
\bracetwo{
F(\xi ) \comma  \hspace{14pt}  ({\rm Im}\,\xi > 0)}{-G(\xi ) \comma \hspace{5pt} ({\rm Im}\,\xi < 0)} \comma 
\end{align}
with\footnote{Subscript ``n.s." denotes the part regular (nonsingular) at $\xi=0,\infty$. }  $F=\log \slprod{2_-}{1_+}_{n.s.}$ and $G=\log \slprod{1_-}{2_+}_{n.s.} $, 
 we can extract  $F$ and $G$ separately from the information of the sum $F+G$ given in (\ref{T-like}).  Using similar procedures  we can compute 
 the regular part of all the invariant products $\slprod{s_i}{s_j}$ and  the combinations  such as (\ref{ratioinvprod}). Finally, extracting the 
part proportional to $\xi$ we can express the integrals $\int_{d_j}
(ud\zbar + vdz)$ solely  in terms of the parameters 
$\kappa_i$ of the LSGKP  external strings. The final formula for 
 the regularized area, which is 
 valid for all the regimes of the parameters $\kappa_i$, is given by 
\beq{
A_{reg}&={\pi \over 12} +\pi \biggl[ -\kappa _1 K(\kappa _1)-\kappa _2 K(\kappa _2)-\kappa _3 K(\kappa _3)\nn\\
& +\frac{\kappa _1 +\kappa _2 +\kappa _3}{2}K(\frac{\kappa _1+\kappa _2+\kappa _3}{2})\nn\\
& +\frac{ |-\kappa _1 +\kappa _2 +\kappa _3|}{2} K(\frac{|-\kappa _1+\kappa _2+\kappa _3|}{2})\nn\\
& +\frac{|\kappa _1 -\kappa _2 +\kappa _3|}{2}K(\frac{|\kappa _1-\kappa _2+\kappa _3|}{2})\nn\\
& +\frac{|\kappa _1 +\kappa _2 -\kappa _3|}{2}K(\frac{|\kappa _1+\kappa _2-\kappa _3|}{2})\biggr] 
\comma  \label{regarea2}
}
where the function $K(x)$ is defined as 
\begin{align}
K(x ) &= \frac{1}{\pi} \int _{-\infty}^{\infty} d\theta \, e^{-\theta} \log \left( 1- e^{-4\pi x\cosh \theta }\right) \period 
\end{align}
%
\section{Method for evaluating the contribution of vertex operators}
The major problem left unsolved in our previous work is the 
 evaluation of the contributions of   the vertex operators which 
 generate  the  LSGKP strings  from the boundary of AdS spacetime. 
In this section, we first give a general discussion of why it is a
difficult problem and give an outline of how we are going to solve it. 
 Subsequently, we will implement the idea more concretely and develop 
 a method which can be used for  general semi-classical 
correlation functions. We will first apply it to two-point functions  in section 4
and check that it produces expected results. Then finally in section 5 
   three point  functions will be treated by the same method and 
 we will obtain  natural complete  results. 
\subsection{General strategies} 
There are two obvious obstacles  in the evaluation of the 
vertex operator contributions. 
First, as discussed in detail in our previous paper, the 
precise form of the conformally  invariant vertex operator corresponding to 
 the GKP string is not known. 
 The curvature of the AdS background, the effect of which is certainly felt 
by the macroscopic strings of our interest,  inherently 
couples the left and the right moving 
modes on the worldsheet and hence the two Virasoro symmetries  are not directly associated  with the holomorphic  and the anti-holomorphic dependence of 
 the fields.  
Furthermore, even on the boundary of the AdS space, the GKP string explicitly depends on the  $\sig$ coordinate and
 consists of  infinite number of Fourier  modes.  
This indicates that the vertex operator which creates such a string state is expected to contain infinite number of $\sig$ derivatives. Thus, although it is easy to find operators with the correct  global quantum numbers such as the dilatation charge  and the spin, it is quite difficult to construct the vertex operator  with the right conformal property  describing such an extended string state. 
The second difficulty is that we do not know any of the classical solutions which serve as the saddle configurations for the three and higher point 
correlation functions. Thus,  even if we succeed in constructing  correct vertex operators, we do not have the solutions with  which to evaluate their values. 

How can one overcome these apparently  serious difficulties?
 Below we give an outline of our strategy to solve these problems 
 by making use of the so-called finite gap method of integrable theories.

As  we have already argued  in our previous work,  
by using the state-operator 
correspondence,  we can evaluate the effect of the vertex operators 
 through  the semi-classical wave functions, which can be 
constructed in principle by solving the Hamilton-Jacobi equation. 
When the saddle point approximation is adequate, the state-operator 
 correspondence can be expressed as 
\begin{align}
V[q_{\ast}(z=0)]  e^{-S_{q_{\ast}}(\tau<\tau_0)}  &=
 \Psi[q_{\ast}(\tau_0)]  \comma \label{stateopsaddle}
\end{align}
where $q_\ast$ denotes the saddle point configuration, 
$V[q_{\ast}(z=0)]$ is the value of the vertex operator 
 inserted at the origin of the worldsheet $z=e^{\tau+i\sig} =0$,  corresponding to 
 the local cylinder time $\tau=-\infty$,  the factor $\exp[-S_{q_{\ast}}(\tau<\tau_0)]$ 
is the amplitude to develop into the state on the small circle with 
 radius $e^{\tau_0}$ and $\Psi[q_{\ast} (\tau_0)]$ is the semi-classical
 wave function describing  the   state on that circle\footnote{ 
At the end of this 
subsection, we will discuss this formula also from the point of view of the 
 renormalization of the vertex operator.}.   The variable $q$ can be the string 
 coordinates  $X_\mu(\sig)$ or any other variables which can specify the 
configuration.  In particular, if we can construct the action-angle variables 
$(\act_n, \theta_n)$ of the system and use $\{\theta_n\}$ as $q$, 
then the (Euclidean) action and the wave function can be expressed simply as 
\begin{align}
S_{\theta} &= -\int d\tau \left(  i  \sum_n \act_n \del_\tau \theta_n 
 -\calE(\{J_n\}) \right) \comma  \label{Stheta}\\
\Psi[\theta] &= \exp \left(i \sum_n \act_n \theta_n -\calE(\{J_n\}) \tau \right)  \comma \label{Psitheta}
\end{align}
where both $J_n \del_\tau \theta_n$ and $\calE(\{J_n\})$, the 
worldsheet energy,  are constant. Our main idea is to construct and evaluate the wave functions in the representation (\ref{Psitheta}) above. 

Before explaining what we need to do to achieve this, let us make an important remark  concerning   the evaluation of the action. As it will be shown in section 4, if we can 
 evaluate the action also in terms of the action-angle variables, as in (\ref{Stheta}), 
  the evaluation of the two point function becomes systematic and simple. In particular
 the cancellation of the divergences  between the wave functions and the action 
 becomes automatic. However, for the calculation of the 
action for the three point function of the LSGKP strings performed in our previous paper \cite{Kazama:2011cp} and in section 5.1, the action is expressed and computed in terms of the string coordinates  $X_\mu(\sig)$, not the angle variables.  In such a
 case, in order to combine properly with the contribution of the wave functions evaluated in terms of the angle variables,  one must make a canonical transformation and, as discussed  in Appendix D,  this in general produces an additional contribution.  Fortunately, in the case of the GKP string, we  show in Appendix E  that such an extra contribution does not affect the final result. 

Now let us describe the three problems to be solved in order to actually 
 construct and evaluate the wave functions in terms of the action-angle variables. 

\noindent (1)\quad   First problem is the  construction of  the 
action-angle variables for the string in Euclidean  $AdS_3$. 
It turns out that the direct construction through the Hamilton-Jacobi method 
is extremely difficult, if not impossible\footnote{However, for a particle 
in  Euclidean $AdS_3$, such a construction is possible and is instructive. 
See Appendix G for some detail. }.  
 However, in the case of the so-called  finite gap solutions, to which the GKP string
 belongs, we can make use of the Sklyanin's  ``separation of variable" method \cite{Sklyanin}
to construct  the action-angle variables from the positions of the poles of the suitably normalized eigenvector  (called the Baker-Akhiezer vector)  of the auxiliary linear problem.  In fact, we will argue that 
 in order to keep full grasp of all the dynamical degrees of freedom one should 
consider the problem from the perspective of   the ``infinite gap solutions" and 
understand the finite gap solutions as  suitable limits. This will be discussed 
in detail in sections 3.2.2 and  3.2.3. 
 
\noindent (2)\quad  The Sklyanin's method  referred to above  identifies 
the action-angle variables formally for finite gap solutions. 
What we need to know are the {\it values} of the action-angle variables for the specific  GKP string solutions. 
For this purpose, we first need to re-analyze the GKP solution and its large spin 
 limit using the finite gap method,  rather than in the framework of Pohlmeyer reduction we have been employing.  In contrast to the well-studied cases of  strings 
 in $AdS_3 \times S_1$ (or in $R  \times S^3$),  the finite gap method for the case of the string in (Euclidean) $AdS_3$ has not been fully developed. This is largely because 
 the structure of the associated spectral curve is more involved due to 
the different form  of the Virasoro conditions imposed on  the $AdS_3$ part.  
This point is clarified in section 3.2.1.  With this structure duly understood, 
we then develop a powerful method of computing the values of the angle 
 variables based on the use of the global symmetry transformations. 
The basic idea is to compute the angle variables of 
 a solution $\mathbb{X}$ of our interest {\it relatively} to those of  a suitable 
fixed reference solution $\refsol$. Since $\mathbb{X}$ can be obtained from $\refsol$ 
 by a global $SL(2)_L \times SL(2)_R$ symmetry transformation 
as $\mathbb{X} = V_L \refsol V_R$,  all we have to know is how the angle 
 variables shift  under such a transformation.  As will be shown  in section 3.3.1,  this can be computed explicitly in terms of the parameters of the global transformation 
  and the components of the normalization vector $\vec{n}$, which determines 
 the locations of the poles of the Baker-Akhiezer vector. 

\noindent (3) \quad   The last problem to solve is how to construct the 
suitable quantum wave functions,  corresponding to the vertex operators,  using the classical data on the action-angle variables 
 obtained by the method described above. Here again the global symmetry 
 transformations play the key role.  

Let us explain this 
 explicitly for the case of the two point function of the form $\langle V(0) V(x)\rangle$, where the vertex operator $V(0)$ corresponds to a conformal 
 primary operator on the SYM side inserted at the origin of the boundary of $AdS_3$
and $V(x)$ is  the same operator inserted at $x$ on the boundary.   
In the saddle point approximation, this is evaluated as 
$V(0)\big|_{\mathbb{X}} V(x)\big|_{\mathbb{X}}$. 
Before proceeding, let us  clarify  here how the vertex insertion points should be identified with the behavior of the saddle point solution on the boundary.  
For the LSGKP solution we are considering, 
 the string configuration is extended, not point-like,  even on the boundary and 
this appears to make such an identification ambiguous.  However, there is a natural 
 and unique choice.  As we already mentioned under item (2), a general 
 solution can be obtained from the reference solution $\refsol$ \sref{gB} 
by a global symmetry transformation.  Clearly the spin of the reference solution 
 is defined with respect to the rotation in the $x_1$-$x_2$ plane around the origin 
$(0,0)$,  which is the  $\sig=0$ point on the string.  Under the global 
 symmetry transformation, such a point $X_\mu(\sig=0)$ transforms covariantly 
and hence the two point function behaves correctly  under the 
conformal transformation of the boundary coordinates.  Therefore the vertex 
 insertion point should be identified with the point at $\sig=0$. 

Now let us first  evaluate  $V(0)\big|_{\mathbb{X}}$ using the 
corresponding wave function constructed in terms of the angle variables.  
Recall that such angle variables depend on the choice of the normalization 
vector $\vec{n}$ of the Baker-Akhiezer function. As it will be shown in section
 3.3.2, this normalization vector is uniquely determined by the requirement 
 that the wave function is unchanged under the special conformal transformation, 
namely that it is a conformal primary.  Once $\vec{n}$ is fixed, then we can 
compute the (shift of the) angle variables by the method described in the item 
(2) above. 

Next consider the evaluation of $ V(x)|_{\mathbb{X}}$.  The vertex operator 
 $V(x)$ is of course different from $V(0)$. In particular, $V(x)$ is not invariant 
 under the  special conformal transformation around the origin which leaves 
 $V(0)$ invariant. Thus  it appears that we have to re-analyze the condition 
for the normalization vector $\vec{n}$ and the proper form of the corresponding
 wave function, which is quite cumbersome.  We can actually circumvent this 
 procedure by noting that $V(x)$ is obtained from $V(0)$ by a translation, which 
 is again a global symmetry transformation. Symbolically, $V(x) = T_x V(0)$. 
Further, we can evaluate this on the solution $\mathbb{X}$  in the manner 
$V(x)\big|_\mathbb{X} = (T_x V(0))\big|_\mathbb{X}= V(0)\big|_{T^{-1}_x 
\mathbb{X}}$. In the last step, instead of evaluating the shifted vertex operator 
 on the solution $\mathbb{X}$,   equivalently we evaluate the unshifted  operator
 at the origin on the inversely translated solution $T^{-1}_x \mathbb{X}$, which is quite easy  to obtain.  This method has another important advantage that we can compute 
the shift of the angle variables relative to the same reference solution $\refsol$ which 
 emanates from the origin of the boundary. See \figref{translating}. The details of this method 
will be explained in section 3.3. 

So in this way, we can compute the contribution of the vertex operator 
for the LSGKP string inserted at any point from the local behavior of  the solution at that point. It should be clear that the method is quite general 
 and can be applied to  correlation functions for any string states
 describable by the finite gap method. 
 
Before ending this subsection, we wish to discuss the structure of the 
 divergences  that appear in the evaluation and how they cancel 
in the end result.  
The divergences that we encounter for the correlation functions are all of ultraviolet origin and hence quite local.  Therefore, their properties should be the same as in the string  theory in flat spacetime, and we need to use 
``normal-ordered" (or ``renormalized") vertex operators to 
get finite results. 
This is all standard but  in the saddle point computation the structure is somewhat unfamiliar.  The first source of divergences is the action evaluated 
on the saddle point configuration. There are charges at the vertex insertion points and the action contains  the self-energy of each local charge, 
which is  log-divergent in two 
 dimensions.  The second source of divergences is the contribution from 
 the naive (unrenormalized) vertex operators, such as $V=e^{ikX}$ for 
 a tachyon in flat space,  evaluated on the saddle.  As one can easily check 
 in the case of free string, this contribution {\it over-compensates} the one from the action and the net result is still divergent in the exponent. Finally to cancel this remainder  we need to introduce  divergent renormalization factor for each vertex. In other words,  if we prepare  the renormalized vertex operators $:V_i:$ from 
the out set, their log's  $\log :V_i[X_\ast]:$ on the saddle configuration $X_\ast$ are divergent in such a way that the correlation function  $e^{-S[X_\ast]} ( : V_1[X_\ast]: : V_2[X_\ast]:\cdots :V_n[X_\ast]:)$ becomes finite. 

Now from this point of view, the saddle point version of the state-operator 
 correspondence  (\ref{stateopsaddle})  states that 
the wave function $\Psi[q_{\ast}(\tau_0)]$ gives  nothing but the 
contribution of the  {\it renormalized}  
vertex operator evaluated on the saddle. Hence $\log \Psi$  should contain divergences  which  precisely kill those from  the action. 
We shall demonstrate this explicitly when we compute the final three point 
 function in section 5. 
\subsection{Construction of action-angle variables for a string in Euclidean
 $AdS_3$}
As explained in the previous subsection,  instead of trying to evaluate the 
contribution of the vertex operators directly,  we will 
 deal with the wave functions which are related to the vertex operators through the state-operator correspondence. 
The wave functions in the classical limit are given by the solutions of the Hamilton-Jacobi (H-J) equation. Although the solutions can be obtained rather easily  for simple systems,  such as a string moving in flat space, it is quite difficult to solve the H-J equation 
for non-linear systems, such as a string in \ads3.  However, as we shall see,  the integrability of the system  will allow us to construct the action-angle variables, in terms of which the H-J equation  simplifies  enormously.    
In this subsection  we will first discuss the classical integrability of a string in  \ads3
in the framework of the spectral curve method,  
 clarifying  the  differences from the case of 
a string in $AdS_3 \times S^1$,  which was studied in \cite{KZ}. 
Then a powerful method of  constructing the action-angle variables 
for the finite gap solutions will be explained and describe how they will be of use 
for the computation of  correlation functions. 
%
\subsubsection{Analytic structure of the quasi-momentum and the spectral curve}
The classical motion of a string in  $AdS_5 \times S^5$ is well-known to be integrable. Moreover, the  integrability persists even when the motion of a string is restricted to certain subspaces of $AdS_5\times S^5$. The Euclidean $AdS_3$, the space of 
 our interest,  is among such class of subspaces.  Although the integrability 
 of a string on a closely related space, namely $AdS_3\times S^1$, has 
 been well-studied in \cite{KZ},  the structure for case of the Euclidean $AdS_3$ 
exhibits  several important differences, which we shall clarify in the analysis 
 below. 

The classical integrability of the string in $AdS_3$ is  most  apparent when we reformulate the equation of motion as the zero-curvature condition for  a one-parameter family of connections viz.~Lax connections.  It is given by 
\beq{
&\left[ \del + J^r_z (x)\comma \delbar + J^r_{\barz} (x)\right] =0\comma\label{r-connection}\\
&J^r_z (x)\equiv \frac{1}{1-x}j_{z}\comma \hspace{33pt}J^r_{\barz}(x)\equiv \frac{1}{1+x}j_{\barz}\comma\label{lax}\\
&j_z=\mathbb{X}^{-1}\del \mathbb{X} \comma \hspace{33pt} j_{\barz} = \mathbb{X}^{-1}\delbar \mathbb{X}\comma\\
&\mathbb{X}=\pmatrix{cc}{X_{+} & X\\\bar{X}&X_{-}}\period
}
Here $j_z$ and $j_\zbar$ are the components of the left-invariant current $j$, which 
  transforms  only under $SL(2)_R$. To emphasize this transformation property, 
 we denote by $J^r_z$ and $J^r_\zbar$, with the subscript $r$,  the components of the flat Lax connection made from the ``right current" $j$.  The complex variable $x$ 
 denotes  the familiar spectral parameter.  Similarly, one can construct 
 another one-parameter family of flat connections $J^l$ from the 
``left current" $l = d \mathbb{X} \mathbb{X}^{-1}$ in the following 
 way: 
\beq{
&\left[ \del + J^l_z (x)\comma \delbar + J^l_{\barz} (x)\right] =0\comma\label{l-connection}\\
&J^l_z (x)\equiv \frac{x}{1-x}l_{z}\comma \hspace{33pt} J^l_{\barz}(x)\equiv -\frac{x}{1+x}l_{\barz}\comma\\
&l_z=\del \mathbb{X} \mathbb{X}^{-1}\comma \hspace{33pt} l_{\barz} = \delbar \mathbb{X} \mathbb{X}^{-1}\period
}
These two connections are related  to each other by the gauge transformation 
 of the form $\mathbb{X}(d + J^{r})\mathbb{X}^{-1}=d+J^{l}$.
Furthermore, as shown in Appendix A of \cite{AM:0904}, they are gauge equivalent also to the connections constructed from the Pohlmeyer reduction \sref{Bxicon1} under the identification of the spectral parameters  $x=(1-\xi^2)/(1+\xi^2) $. Although the Pohlmeyer reduction is suitable for the evaluation of the area, use of the connections $J^{r,l}$ will be  more advantageous for the construction of the action-angle variables. 

One of the manifestations of integrability is the existence of an infinite number of conserved charges. They are constructed from the path-ordered exponential of the connection $J^{r}(x)$ along a closed cycle around an insertion point of a vertex operator, 
  called the monodromy matrix:
\beq{
\Omega (x; z_0 )=\mathcal{P}\exp \left( -\oint J^{r}_z(x)dz +J^{r}_{\barz}(x)d\barz\right)\period\label{monod}
}
As indicated, the matrix $\Omega$ depends on the base point of the closed cycle $z_0$.  By virtue of the flatness of the connection,  an expansion of $\Omega(x)$ 
as a function of $x$ around some point yields an infinite number of conserved charges
as coefficients.  In particular,  expansions around $x=\infty$ and $x=0$ yield 
 global charges,  corresponding to $SL(2)_R$ and $SL(2)_L$ respectively, 
 at the leading order 
in the following way:
\beq{
&\Omega (x; z_0) = \mathbf{1} -\frac{2\pi i}{\sqrt{\lambda}x} Q_{R}+O (x^{-2}) \hspace{33pt} (x\to \infty )\comma\\
&\mathbb{X} (z_0)\Omega (x; z_0) \mathbb{X}^{-1}(z_0)= \mathbf{1} +\frac{2\pi ix}{\sqrt{\lambda}} Q_{L}+O (x^2)  \hspace{33pt} (x\to 0 )\comma
}
where the matrices  $Q_R$ and $Q_L$ are given by 
\beq{
&Q_R\equiv \frac{i\sqrt{\lambda}}{2\pi} \oint (j_z dz -j_{\barz} d\barz)\comma\hspace{33pt} Q_L\equiv \frac{i\sqrt{\lambda}}{2\pi}\oint (l_z dz -l_{\barz} d\barz)\period
}
Quantities independent of the base point $z_0$ can be extracted 
 from the eigenvalues of $\Omega$.   Since $\det \Omega=1$, these eigenvalues 
are of the form 
\beq{
u(x;z_0)\Omega (x;z_0)u(x;z_0)^{-1}\sim\pmatrix{cc}{e^{i\hat{p}(x)}&0\\0&e^{-i\hat{p}(x)}}\comma
}
where $u(x;z_0)$ is the matrix which diagonalizes $\Omega$ and 
$\hat{p}(x)$ is  the quasi-momentum\fn{The quasi-momentum is customarily 
denoted by $p(x)$ without a hat \cite{KZ}. However, to distinguish it from 
the function $p(z)$ which appears in the Pohlmeyer reduction, 
we will denote it by $\phat(x)$ .}.  One can choose the branch 
 of the logarithm appropriately so that  $\hat{p}(x)$ exhibits  the following asymptotic behavior, reflecting the asymptotics of $\Omega$ around $x=\infty$ and $x=0$:
\beq{
&\hat{p}(x)=-\frac{2\pi}{\sqrt{\lambda}x} R+O(x^{-2}) \hspace{33pt} (x\to \infty )\comma\\
&\hat{p}(x)=2\pi m +\frac{2\pi x}{\sqrt{\lambda}}  L+O(x^2)  \hspace{33pt} (x\to 0 )\period
}
Here  the conserved charges $R$ and $L$  
are the (upper) eigenvalues of $Q_{R}$ and $Q_{L}$ respectively  and $m$ is an integer. 
 Although $m$ is nonvanishing for a general string state, we will  focus on the  class of solutions with $m=0$,  to which the GKP string belongs\fn{Generalization to solutions with $m\neq 0$ is straightforward.}. 

\begin{figure}[tbp]
 \begin{center}
   \picture{clip,width=11cm}{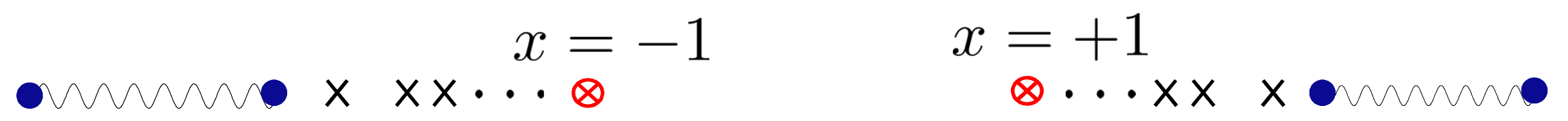}
\end{center}
\caption{ Analytic structure of the spectral curve of a string on 
$AdS_3\times S^{1}$. The wavy lines denote  square root cuts. 
There are essential singularities at $x=\pm 1$, corresponding to simple poles in $\hat{p}(x)$. The node-like points, denoted by 
crosses, accumulate to these points. 
} 
\label{analytic-0}
  \begin{center}
   \picture{width=11cm,clip}{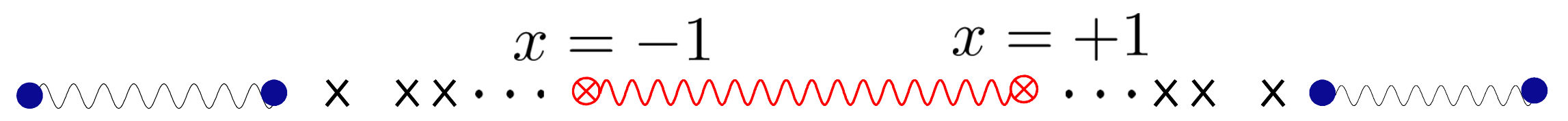}
  \end{center}
\caption{ Analytic structure of the spectral curve of a string on \ads3. 
Compared to the case of $AdS_3\times S^{1}$, the singularities 
at $x=\pm 1$ are  weakened to ``half-pole" type and a singular square root cut runs 
 between these two points.  } 
\label{analytic-structure}
\end{figure}

To discuss other conserved charges, it is important to study the analytic properties of $\hat{p}(x)$.  Such analytic structures are encoded in the spectral curve defined by
\beq{
\Gamma\,:\quad \Gamma (x,y) \equiv \det \left( y\mathbf{1}-\Omega (x;z_0)\right) =0\comma 
}
which is equivalent to $\left( y-e^{i\hat{p}(x)}\right) \left( y-e^{-i\hat{p}(x)}\right) =0$. 
As we shall show, the spectral curve $\Gamma$ has three kinds of analytic structures, namely  essential singularities, cusp-like points and node-like points (see figures \ref{analytic-0} and \ref{analytic-structure}). 
 
Let us first focus on the essential singularities.  As we shall show shortly, 
for a string on \ads3  the structure of essential singularities  turned out to be somewhat complicated. 
Therefore it is instructive to consider first a simpler case of a  string in  $AdS_3 \times S^1$, where the motion in $S^1$ provides a constant contribution to the Virasoro conditions, as we discuss below. 
 For this case,  it is known that the essential singularities arise at $x=\pm 1$, where the Lax connection \sref{lax} becomes singular \cite{KZ}. To see this, recall the definition of the monodromy matrix \sref{monod}. Near $x=\pm 1$, it behaves  as
\beq{
&\Omega (x; z_0)=\mathcal{P}\exp \left[ -\oint dz  \frac{j_z}{1-x}+O((x-1)^0)\right]  \hspace{22pt} (x\to 1)\comma\label{singmonod1}\\
&\Omega (x; z_0)=\mathcal{P}\exp \left[ -\oint dz  \frac{j_{\barz}}{1+x}+O((x+1)^0)\right]  \hspace{22pt} (x\to -1)\period\label{singmonod2}
}
The Virasoro conditions for the entire string reads 
\beq{
\del X^{\mu}\del X_{\mu}=\frac{1}{2}\Tr \left( j_{z} j_{z}\right) =\kappa ^2 \comma \hspace{22pt}\delbar X^{\mu }\delbar X_{\mu}=\frac{1}{2}\Tr \left( j_{\barz} j_{\barz}\right) =\kappa ^2 \period \label{vir}
}
where $\kappa^2$ denotes the contribution from the $S^1$ part. 
Diagonalization of \sref{singmonod1} and \sref{singmonod2} leads  to
\beq{
u(x;z_0)\Omega (x; z_0)u(x;z_0)^{-1}=\exp\left[ \frac{2 i \pi \kappa}{x\mp 1}\sigma _3+O((x\mp 1)^0)\right]\hspace{22pt}(x\to\pm 1)\period\label{limmonod}
}
\sref{limmonod} clearly shows the existence of essential singularities at $x=\pm 1$ (see \figref{analytic-0}).
 They correspond to the simple pole singularities of $\phat(x)$ of the form 
\beq{
\hat{p}(x)=\frac{2\pi \kappa }{x\mp 1} +O(1)\hspace{33pt} (x\to \pm1 )\period
}

Let us now return to the case of the \ads3. In this case, in contrast 
 to (\ref{vir}),  the Virasoro conditions read
\beq{
\del X^{\mu}\del X_{\mu}=\frac{1}{2}\Tr \left( j_z j_z\right) =0 \comma \hspace{22pt}\delbar X^{\mu}\delbar X_{\mu}=\frac{1}{2}\Tr \left( j_{\barz} j_{\barz}\right) =0 \period
}
As this corresponds to the limit of vanishing $\kappa$, we need to perform 
 a more involved analysis to extract the behavior of $\phat(x)$ around $x=\pm1$. 
The result of such an analysis,  carried out in 
Appendix A, gives  the leading singular behavior of  $\hat{p}(x)$  to be 
\beq{
\hat{p}(x)=\pm \frac{\kappa _{\pm}}{\sqrt{1\mp x}} +O((x\mp 1))\hspace{33pt} (x\to \pm1 )\comma\label{sing-euc}
}
where  the constants $\kappa_{\pm}$ are given by 
\beq{
\kappa _{+} =\frac{1}{2\pi i}\oint dz 
\left( \half \Tr \left( \del j_z \del j_z\right)\right)^{1/4} \comma\hspace{33pt}\kappa_{-} =\frac{1}{2\pi i}\oint d\barz  \left( \half \Tr \left( \delbar j_{\barz} \delbar j_{\barz}\right)\right)^{1/4}\period
}
This shows that, compared to the case of $AdS_3 \times S^1$,  the singularity is no longer of the simple pole type but  rather a weaker ``half-pole" type,  with an associated branch cut between  $x=+1$ and $x=-1$ as depicted in \figref{analytic-structure}.  For later convenience, we refer to this branch cut as the  {\it singular cut}. 

Next, let us  discuss  the remaining analytic structures, \ie  the cusp-like points and the  node-like points. Both of them are defined as the zeros of the discriminant $\Delta_{\Gamma}$ of the spectral curve given by 
\beq{
&\Delta _{\Gamma}\equiv \left( e^{i\hat{p}(x)}-e^{-i\hat{p}(x)}\right) ^2 \period
}
Note that,  although they are singular points of the spectral curve, 
the quasi-momentum $\hat{p}(x)$ is not singular at these points. 
They are classified according to the order of the zero. If the 
 order of the zero is odd, \ie $\Delta_{\Gamma}\sim (x-x^{(c)}_i)^{2r+1}$, 
then such a point is called cusp-like. If it is even, like $\Delta_{\Gamma}\sim (x-x^{(n)}_i)^{2r}$, it is called node-like. Around such a zero, the quasi-momentum 
 is approximated as 
\beq{
&e^{i\hat{p}(x)}\sim \pm\left( 1+\frac{\sqrt{\Delta_{\Gamma}}}{2}\right)
 \quad \Rightarrow \quad 
 \hat{p}(x)\sim m\pi + \frac{\sqrt{\Delta_{\Gamma}}}{2i}\hspace{22pt}m\in\mathbb{Z}\period\label{loci}
} 
This shows that at the cusp-like points\fn{By appropriately choosing the  branch of 
the  logarithm, the integer $m$ in \sref{loci} can be set to zero at the cusp-like points. We refer the reader to \cite{Vicedo} for details.} the spectral curve develops branch cuts. 
 In what follows, we call these branch cuts {\it nonsingular cuts},  to distinguish them from the singular cut  connecting the points $x=\pm 1$ defined previously. 
Another important property of a cusp-like point is that, 
as shown in Proposition 7.3 in \cite{Vicedo}, the monodromy matrix at such 
 a point  always takes the  form of a Jordan block, namely 
\beq{
\Omega (x^{(c)}_i)\sim \pmatrix{cc}{1&\ast \\ 0&1}\period \label{cusp}
}
Now consider the properties of  the node-like points. The  formula  (\ref{loci})
 shows that in this case  the spectral curve does not develop a branch cut  and such a point is characterized simply by 
\beq{
\hat{p}(x_i^{(n)})=m_i\pi\period
}
As concerns  the form of the monodromy matrix, there are two possibilities 
 at a node-like point. It either  takes the  form of a Jordan block or is proportional to the identity matrix: 
\beq{
\Omega (x^{(n)}_i)\sim \pmatrix{cc}{1&\ast \\ 0&1}\hspace{11pt} \text{or} \hspace{11pt} \pm \mathbf{1} \period
}

From the preceding discussion we see that,   for both $AdS_3\times S^1$ and $AdS_3$ cases,  the quasi-momentum $\phat(x)$ has singularities at $x=\pm 1$ and 
infinite number of node-like and/or cusp-like points accumulate  at $x=\pm 1$. 
Consequently,  the spectral curve becomes non-algebraic and is difficult to study. 
In such a case,  one often replaces the spectral curve with 
 the a simpler curve $\hat{\Sigma}$ \cite{BKSZ,DV1,DV2,Vicedo} 
defined with the use of the logarithmic derivative of $\Omega$ as 
\beq{
&\hat{\Sigma}\,:\hat{\Sigma} (x,y)\equiv \det \left( y\mathbf{1}-L(x;z_0)\right) =0\comma\\
&u(x;z_0)L(x;z_0)u^{-1}(x;z_0)\equiv -i \frac{\del}{\del x}\log \left( u(x;z_0)\Omega (x;z_0) u(x;z_0)^{-1}\right)\period
}
Following \cite{Vicedo}, we will call it the logarithmic derivative curve 
or a {\it log-curve}  for short. The log-curve preserves the 
branch cut structure of the spectral curve and has the nice property that  
the eigenvectors of $L$ is the same as those of $\Omega$. 
At the same time, the essential singularities and the node-like points 
 present on  the spectral curve are removed on the log-curve. This feature makes 
it much easier to handle compared to the spectral curve. In fact 
 for the so-called finite gap solutions, for which there are only a finite 
 number of cusp-like points, the log-curve reduces to an algebraic curve
\cite{BKSZ,DV1},  a subject well-studied  in mathematical literature. 
Of course one should remember that by going to the log-curve and neglecting 
the node-like points, some of the important information carried by the spectral 
 curve may be lost. Despite this possible drawback, the log-curve is widely used 
 in the literature because usually node-like points do not play important
 roles,  as far as  the construction of the finite gap solutions is  concerned\fn{We will discuss this point further in 3.2.3.}. 

Now for our purpose of constructing the complete set of action-angle variables 
capable of describing all the states of a string in the \ads3, it is not appropriate to 
deal with the log-curve and discard the node-like points. This is because 
the node-like points, which can be  treated as the degeneration limit of 
 the usual  branch cuts \cite{DV1}, correspond to the modes 
  of the string  which are not excited for a finite gap solution.  In other words 
 the angle variables for these modes do exist as dynamical variables. 
Therefore for us the node-like points are conceptually important and we shall 
study the spectral curve, not the log-curve, with the node-like points 
 treated as a special subset of nonsingular cuts. 

\begin{figure}[tbp]
\begin{minipage}{0.5\hsize}
  \begin{center}
   \picture{clip,height=3.3cm}{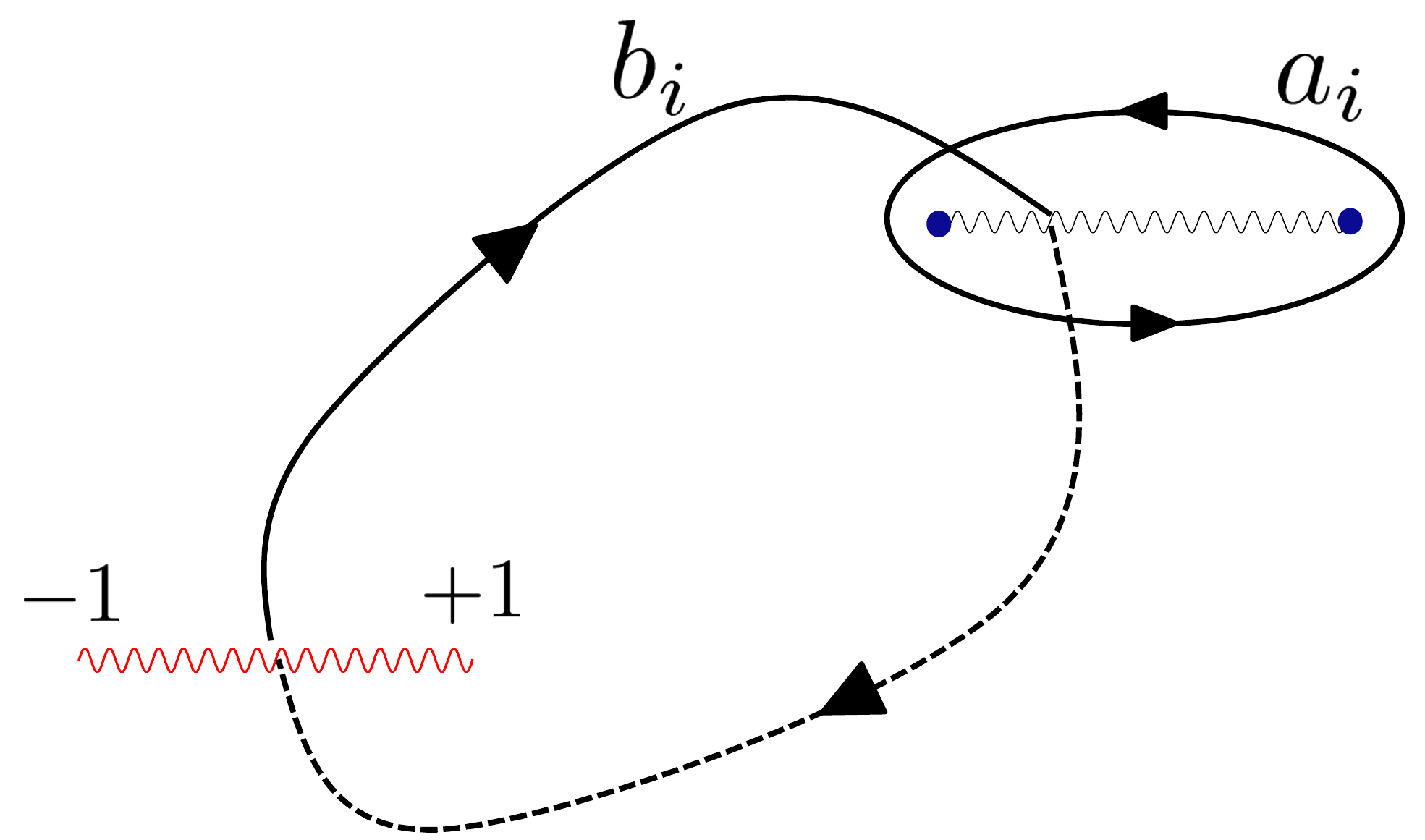}
  \end{center}
 \end{minipage}
 \begin{minipage}{0.5\hsize}
  \begin{center}
   \picture{height=3.3cm,clip}{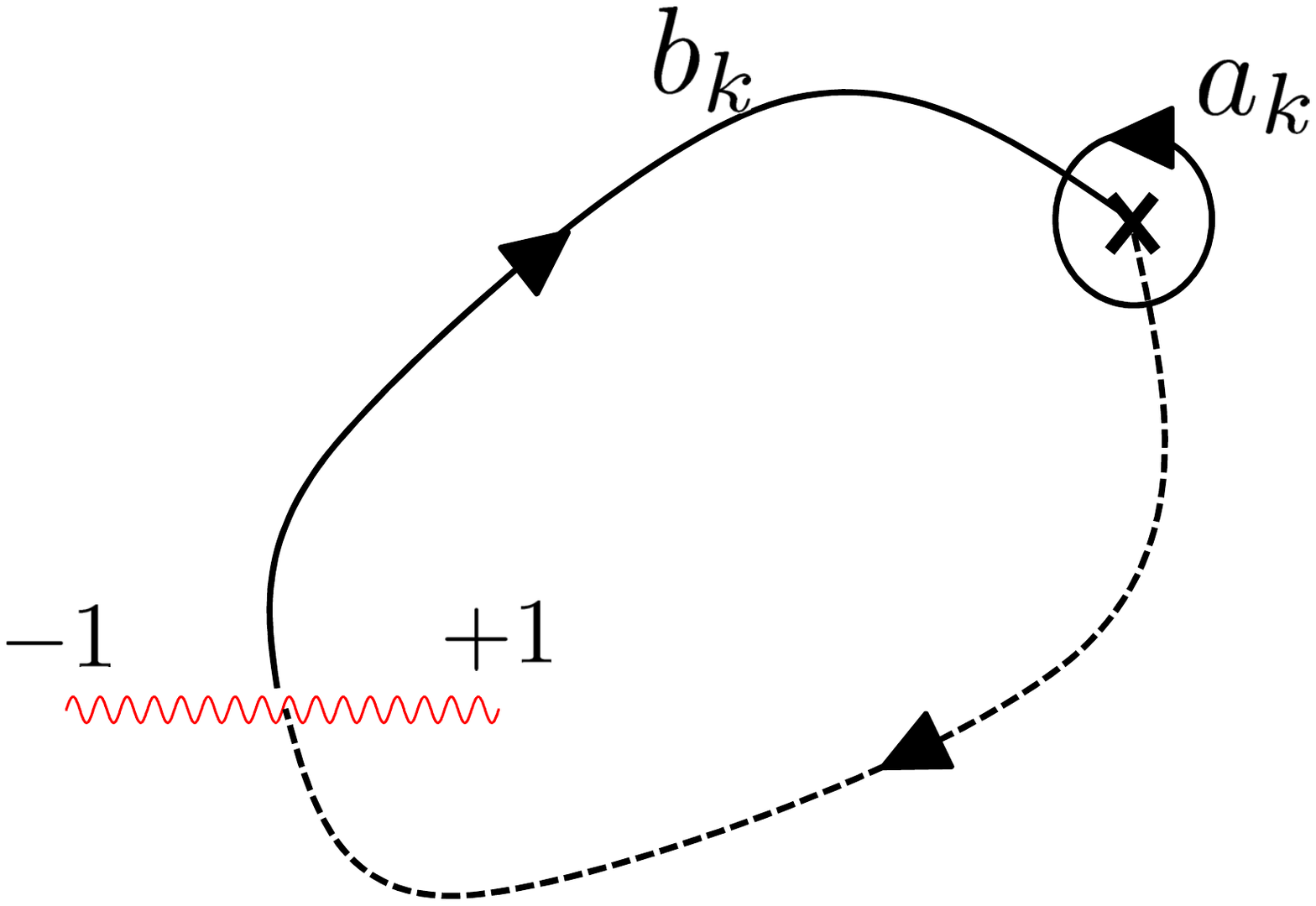}
  \end{center}
 \end{minipage}
\caption{ Definition of $a$-cycles and $b$-cycles. Shown on the right is the 
case where a non-singular cut is shrunk to a node-like point denoted by 
 a cross. } 
\label{ab-cycle}
\end{figure}
To extract nontrivial information from the spectral curve, we now introduce a basis of cycles on the spectral curve. As shown in \figref{ab-cycle}, a convenient choice is to 
define the  $a$-cycles as those  which surround nonsingular cuts counterclockwise and 
the $b$-cycles as those connecting the singular cut and the nonsingular cuts. 
Under an appropriate choice of the branch of the logarithm and the positions of the branch cuts, the integrals of the differential $d\hat{p}$ along $a$- and $b$-cycles take the following form\cite{DV1,DV2}:
\beq{
&\int_{a_i} d\hat{p} =0\comma\hspace{33pt} \int _{b_i}d\hat{p} =2\pi n_i\comma\hspace{22pt} n_i\in\mathbb{Z}\period\label{mode-number}
}
In addition to these cycles, it is convenient to introduce four more cycles $a_0, a_\infty, b_0$ and $b_\infty$. 
The cycles  $a_0$ and $a_{\infty}$  surround the points $x=0$ and $x=\infty$ counterclockwise respectively,  
while $b_0$ and $b_{\infty}$ connect the singular cut with  $x=0$ and $x=\infty$. As discussed in \cite{Vicedo}, one can treat these cycles essentially 
 on equal footing with the other cycles. Now using the $a$-type cycles,  
one can  define a set of conserved charges called {\it filling fractions} as 
\beq{
S_i\equiv \frac{i\sqrt{\lambda}}{8\pi ^2}\int_{a_i} \hat{p}(x) dz\comma \label{filling}
}
where 
\begin{align}
z=x+\frac{1}{x}  \label{Zhukovskyvar}
\end{align}
is the Zhukovsky variable. 
As it will be discussed shortly,  
when interpreted 
appropriately  as dynamical variables of a string system, $\phat(x)$ and $z$ 
 are canonically conjugate and hence the definition (\ref{filling}) is nothing but 
that of an action variable. For this reason the filling fractions are of extreme importance
 and we shall construct the angle variables as their conjugates in section 3.2.2 below. 
Among the $S_i$'s, $S_0$ and $S_\infty$ are of special  interest since 
 they correspond to the global charges $R$ and $L$ in the following way:
\beq{
&S_0 =\frac{L}{2}\comma\hspace{33pt}S_{\infty}=-\frac{R}{2}\period
}
It should be remarked that  the filing fractions for the node-like points vanish,   since $\hat{p}(x)$ is not singular at those points:
\beq{
S_{k}=\frac{i\sqrt{\lambda}}{8\pi ^2}\oint_{x_k}\hat{p}(x) dz=0\period
}
This is consistent with the interpretation of the node-like points as representing unexcited  modes
 of the string.
\subsubsection{Action-angle variables for ``infinite gap" solutions}
Now we move on to the construction of the action-angle variables. 
For a closely related system, namely the string in  $R\times S^3$, 
the action-angle variables have been  constructed in \cite{DV2,Vicedo}
 by employing the so-called Sklyanin's separation of variables \cite{Sklyanin}. 
As we shall describe below, with some modifications this method is applicable also to 
the string in \ads3. 

\begin{figure}[tbp]
 \begin{center}
   \picture{clip,height=4cm}{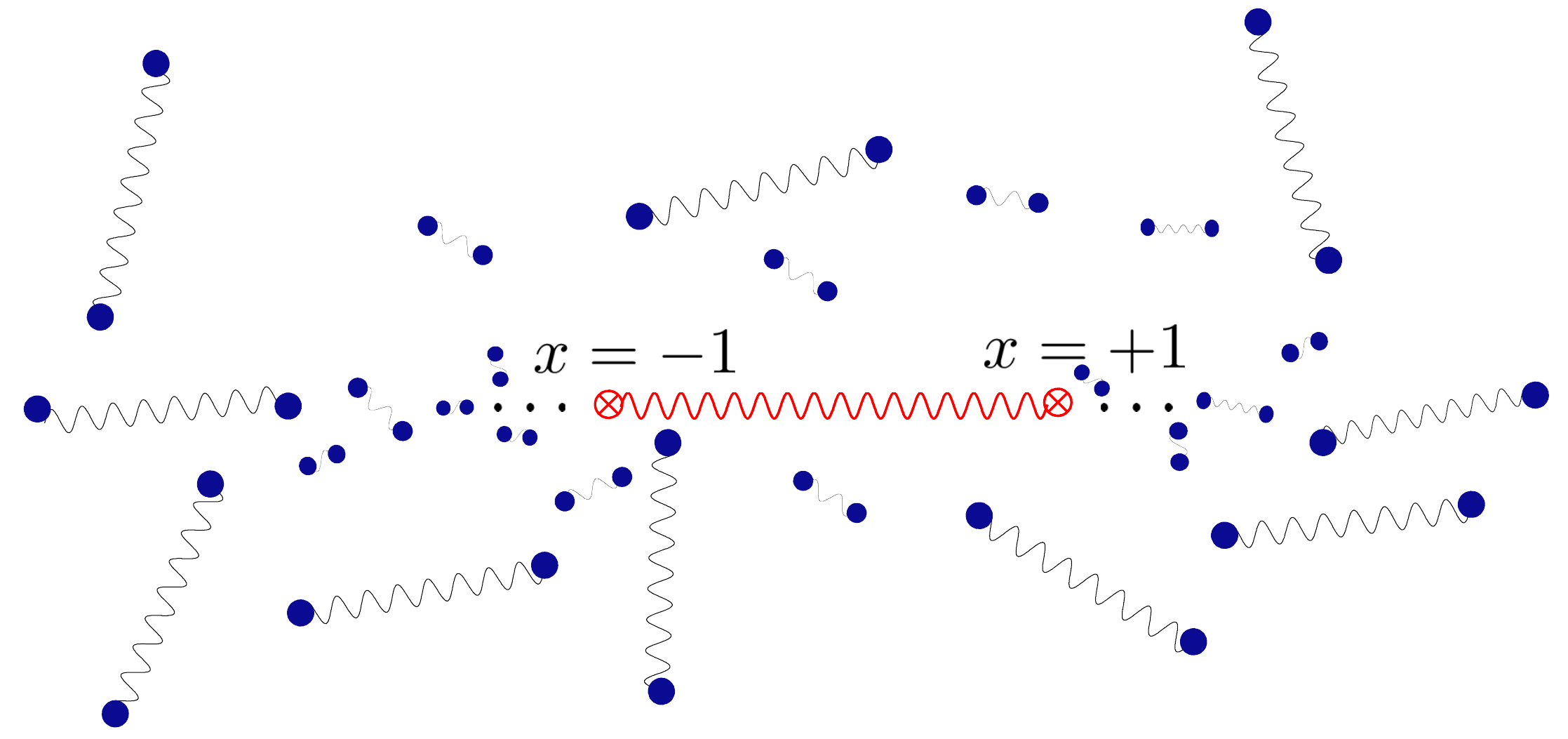}
\end{center}
\caption{ Schematic picture of the spectral curve of an infinite gap solution. } 
\label{infinite-g-pic}
\end{figure}

Before proceeding to  the details, we wish to  emphasize  an important difference between \cite{DV2,Vicedo} and our discussion below. 
While the works \cite{DV2,Vicedo} focused  only on the finite gap solutions, 
namely the solutions  with a finite number of cusp-like points, 
we shall  deal with an enlarged category of solutions which have an infinite 
number of cusp-like points but no node-like points (see \figref{infinite-g-pic}). We will refer to 
 such solutions  as ``infinite gap" solutions.  We exclude the presence of 
node-like points in the above definition because such a point can  be 
universally described by shrinkage of  a branch cut between two 
 cusp-like points.  The  framework  of infinite gap solutions is extremely important 
and useful in controlling  the complete degrees of freedom of the string. 
Of course all the other solutions, including the finite gap solutions, 
can be obtained by certain degeneration limits\fn{Details of  the limiting procedure will be discussed in section 3.2.3.} of such infinite gap solutions.       

Now let us describe the Sklyanin's method, as applied to a string in \ads3.
It is a powerful method for constructing  canonically conjugate variables and is 
  known to be applicable to a wide variety of integrable systems possessing  Lax 
representation.  The main object of concern is the eigenvector $\vec{\psi}$, 
 called the Baker-Akhiezer vector,  of the 
monodromy matrix $\Omega$,  satisfying the eigenvalue equation of the form
\beq{
&\Omega (x;\tau,\sigma )\vec{\psi}(x;\tau ,\sigma )=e^{i\hat{p}(x)}\vec{\psi} (x;\tau ,\sigma )\period \label{eigenvalueeq}
}
Actually, it is of crucial importance to consider the normalized Baker-Akhiezer vector 
  $\vec{h}(x;\tau)$,  defined to be  proportional to $\vec{\psi}(x;\tau, \sigma =0)$
and normalized by the condition 
\beq{
&\vec{n}\cdot\vec{h}=n_1 h_1+n_2 h_2=1\comma\label{nvec}\\
&\vec{h}=\pmatrix{c}{h_1\\h_2}\period 
} 
The constant normalization vector $\vec{n}=(n_1\comma n_2)^{T}$
 will be determined later in section 3.3 from the consideration of global symmetry property. At present, however, it can be chosen arbitrarily. It is known that for a finite
 gap solution associated to a genus $g$ algebraic curve  the normalized Baker-Akhiezer
 vector has $g+1$ poles as a function of $x$. Therefore for an infinite gap solution of our interest it has infinite number of poles.  We will denote the positions 
 of these poles on the spectral curve by $\{\gamma_1,\gamma_2,\cdots \}$. 
Since the monodromy matrix $\Omega$ is constructed out of the string variables, 
 through the relation (\ref{eigenvalueeq}) the positions of the poles $\ga_i$ 
on the spectral curve as well as the quasi-momentum at these poles $\phat(\ga_i)$ 
become dynamical variables. As described in \cite{DV2,Vicedo} and derived 
 slightly more rigorously  in Appendix B,  it turns out that the 
variables $\left(z(\gamma_i)\comma \frac{\sqrt{\lambda}}{4\pi i} \hat{p}(\gamma_i)\right)$, where $z$ is the Zhukovsky variable given in \sref{Zhukovskyvar}, 
 form canonically conjugate pairs   satisfying  the following Poisson bracket relations 
\beq{
&\frac{\sqrt{\lambda}}{4\pi i}\{ z(\gamma_i )\comma \hat{p}(\gamma_j ) \} =\delta_{ij}\comma\label{Poisson-1}\\
&\{ z(\gamma_i )\comma z(\gamma_j )\}=\{ \hat{p}(\gamma_i )\comma \hat{p}(\gamma_j ) \} =0\period\label{Poisson-2}
}
This shows that the filling fractions $S_i$ defined previously
 provide the action variables  of the system. 

To construct the angle variables $\phi_i$ conjugate to $S_i$, we need to find 
 the generating function $F(S_i\comma z(\gamma_i))$ which provides
 the canonical transformation from the pair $\left(z(\gamma_i)\comma \frac{\sqrt{\lambda}}{4\pi i} \hat{p}(\gamma_i)\right)$ to $(\phi_i , S_i)$. 
  Such a function is defined by the 
 following properties:
\beq{
&\frac{\del F}{\del z(\gamma_i )}=\frac{\sqrt{\lambda}}{4\pi i}\hat{p}(\gamma _i)\comma\\
&\frac{\del F}{\del S_i}=\phi_i\period
}
In the present context, the first equation should be viewed as the definition of $F$, while  the second equation should be regarded  as the definition of the angle variables $\phi_i$.  
Therefore, to determine $F$, we need to integrate the first equation with $S_i$'s  fixed. Since the filling fractions are given by the integrals of $\hat{p}dz$ along various cycles on the spectral curve, fixing all the filling fractions is equivalent to  fixing the functional form of $\hat{p}(x)$. Therefore, the integration can be performed as 
\beq{
F(S_i\comma z(\gamma _i))=\frac{\sqrt{\lambda}}{4\pi i}\sum_{i} \int ^{z(\gamma _i)}_{z(x_0)} \hat{p}(x^{\prime})dz^{\prime}\period\label{def-of-F}
}  
The initial point of the integration $x_0$ on the spectral curve can be chosen arbitrarily.  As it will become clear later in section 4, a change of $x_0$ can be absorbed by 
 the change of  overall normalization of the wave function. Similarly, 
a possible integration constant  in $F$, which may depend only on $S_i$, 
can be ignored as it can also be absorbed in the normalization of the wave function. 

Next we compute $\phi_i=\del F/\del S_i$. This requires changing the value 
 of $S_i$ with all the other filling fractions fixed. 
 This is equivalent to adding to $\hat{p}dz$ a one-form whose period integral along $a$-cycles is nonvanishing only for $a_i$. 
Such a one-form should be proportional to a normalized holomorphic differential  $\omega_i$, which satisfies the following properties:
\beq{
\oint _{a_j}\omega_i =\delta_{ij}\comma\hspace{33pt}\oint_{C_s}\omega _i =-1\period 
}
Here  $C_s$ denotes a cycle which surrounds the singular cut. The integral over this
cycle should cancel the contribution from the $a_i$ cycle, since the 
 integral along a large cycle surrounding all the branch cuts should vanish. 
 Using such $\omega_i$, the partial derivative $\del F/\del S_i$  can be expressed as
\beq{
\phi_i =\frac{\del F}{\del S_i} =2\pi \sum _{j}\int^{\gamma_j }_{x _0 } \omega _i \period  
}
Here and hereafter, we regard $\omega_i$ as a differential in $x$.
This is an appropriate generalization of the Abel map,  
which normally  maps an algebraic curve to its Jacobian variety, for non-algebraic 
 curves. When restricted to finite gap solutions, this expression exactly reproduces the definition of the Abel map.

We have now obtained an infinite set of action-angle variables, which satisfy the following canonical form of Poisson bracket relations:
\beq{
\{ \phi_i\comma S_j  \}=\delta_{ij} \comma\hspace{33pt} \{ \phi _i\comma \phi _j\} =\{ S_i\comma S_j\} =0\period
} 
However, there is an  important caveat: Since the above construction is based purely on $J^{r}$, which is invariant under the left global transformation $\mathbb{X}\to V_{L}\mathbb{X}$, the angle variable conjugate to the right global charge $S_0$ cannot be obtained by this method\fn{In other words, the motion of such an angle variable is completely decoupled from the rest and cannot be seen from $J^r$.}. 
To obtain such a variable,  we need to make use of  $J^{l}$.  In an entirely similar
 manner, we can construct  from $J^{l}$  a set of angle variables 
 $\tilde{\phi}_i$, which satisfy 
\beq{
\{ \tilde{\phi}_i\comma S_j \}=\delta_{ij} \comma\hspace{33pt} \{ \tilde{\phi}_i\comma\tilde{\phi}_j\} =\{ S_i\comma S_j\}=0\period
}
The set $\{\phitil_i\}$ contains the desired angle variable $\phitil_0$ conjugate
 to $S_0$. However, it does not contain $\phitil_\infty$, which is conjugate to $S_\infty$. Therefore, to construct a complete set of angle variables, we must 
utilize the two individually incomplete sets,  $\left(  \phi_{i\neq0,\infty} \comma\phi_{\infty}\right)$ and $(\tilde{\phi}_0\comma\tilde{\phi}_{i\neq 0\comma\infty} )$. 
A na\"{i}ve guess would be to use $(\phi_{i\neq 0\comma\infty}\comma\phi_{\infty})$ plus $\tilde{\phi}_{0}$. This, however, is not guaranteed to be correct 
 since $\phi_i$ and $\tilde{\phi}_{0}$ do not  commute in general. 
Nevertheless, we can use  $(\tilde{\phi}_0\comma\phi_{i\neq0,\infty}\comma\phi_{\infty})$ as if they constituted a complete set of angle variables,  for the following 
 reason.  Suppose we find the ``correct" angle variable $\phi_0$ satisfying 
 the following properties:
\beq{
\{ \phi_0\comma S_0\}=1 \comma\hspace{33pt}\{ \phi_{0}\comma S_i\} = \{ \phi_0 \comma \phi_i  \} =0 \hspace{22pt}(i\neq 0)\period
}
Then, from the Poisson bracket relations, we immediately see that the difference 
$\delta\phi_{0}=\phi_0-\tilde{\phi}_0$ commutes with all the 
 action variables, namely $\{ \delta\phi_0 \comma S_i\}=0$ for all $i$. 
This means that it commutes with the worldsheet Hamiltonian, which is made up 
 of the action variables $S_i$, and hence is conserved. 
Therefore $\delta\phi_0$ merely causes a constant shift of the angle variable and it can be absorbed in the normalization of the wave function. Thus, in practice, we can use $(\tilde{\phi}_0\comma \phi_{i\neq 0\comma\infty}\comma\phi_{\infty})$ as a set of angle variables.

\subsubsection{Reduction to finite gap cases}
In this subsection, we explain  how the method of construction of the action-angle
 variables developed above for  infinite gap solutions can be applied to the 
 case of the familiar finite gap solutions\fn{Although it is suggested in \cite{JG} that the finite gap method may be generalized  to describe Wilson loops and correlation functions, here we use the term  ``finite gap solution"  in the conventional  sense. 
Namely, we mean the periodic solution constructed by the usual method\cite{Vicedo} from an  algebraic curve with finite genus, which can serve as the saddle point 
 of a two point function when appropriately complexified.  For a discussion on generalization of the finite gap method, also see section 3.2.4.}  associated with genus $g$ algebraic curve, which are extensively studied in the literature\cite{BKSZ,Vicedo,DV1,DV2} and are believed to correspond to two point functions of various string states. We shall see that  this procedure requires some careful considerations. 

As is well-known, for a finite gap solution of genus $g$,  there are 
$g+2$ non-vanishing filling fractions $\left(S_0\comma S_{\infty}; S_1 \comma \cdots \comma S_g\right)$ and the associated normalized Baker-Akhiezer vector has  $g+1$ dynamical poles. 
To obtain such a solution from an infinite gap solution, we must first 
 set  infinite number of filling fractions to zero, except for $\left(S_0\comma S_{\infty}; S_1 \comma \cdots \comma S_g\right)$, by shrinking the corresponding 
 cuts into node-like points.  Through  this degeneration process, the infinitely many
 poles of the Baker-Akhiezer vectors must somehow ``disappear", leaving $g+1$ dynamical poles of the finite gap solutions. To understand what really  happens, 
it is helpful to study  similar degeneration limit  for  known finite gap solutions\cite{Vicedo}. 
By closely analyzing the motion of the poles in such a degeneration limit, we find that 
actually the unwanted poles do not disappear. 
Instead, they cease to be dynamical\fn{A related discussion is given in  \cite{DV1}.}.
 These nondynamical poles cannot be seen if we use a solution with lower genus from the beginning.  They can  be seen only through the degeneration limit  from a  higher genus solution. This observation strongly suggests that, to obtain a complete set of action-angle variables, we should start from  an infinite gap solution, construct the angle variables from infinitely many poles and then 
consider the limit of those angle variables. Carrying out this procedure, we can trace all the poles including nondynamical ones and obtain the following expression for the angle variables of a finite gap solution with genus $g$: 
  \beq{
  \phi_i =2\pi \sum_{j=1}^{g+1}\int^{\gamma_j}_{x_0}\omega_{i}+2\pi\sum_{J}\int^{\gamma_J}_{x_0}\omega_{i}\period\label{fg-angle}
  }
Here,   $\gamma_j$'s  denote the  dynamical poles,  while $\gamma_J$'s 
 signify the nondynamical ones. 

Let us discuss the nature of the contributions from the nondynamical 
 poles. A detailed argument on the motion of the poles given in Appendix E of \cite{DV1} shows that nondynamical poles are trapped either at node-like points or at cusp-like
 points.  Since such points are discretely placed on the spectral curve, 
the positions of the nondynamical poles $\gamma_J$ do not change under any continuous deformations of the solution which keeps the spectral curve intact.  In particular, they do not change under $SL(2,C)_L\times SL(2,C)_R$  global symmetry transformations.  As we shall discuss later, the only necessary information for the evaluation 
 of the correlation functions is the shift of angle variables under such  global transformations.  Thus,  in practice,  the second term in \sref{fg-angle}  gives the same constant contribution, which can be absorbed into the normalization of the wave function.
 Consequently, the angle variables for the  finite gap solution  can be effectively 
 defined without the second term\fn{The expression  \sref{fg-angle2} coincides with the one derived in \cite{DV2,Vicedo} for finite gap solutions. 
There it was derived within the finite dimensional subspace of the total phase space,
 appropriate  for finite gap solutions with fixed genus.  Our  discussion 
 in this section  corroborates  
the result  of \cite{DV2,Vicedo} from a more general point of view.} as 
\beq{
 \phi_i =2\pi \sum_{j=1}^{g+1} \int^{\gamma_j}_{x_0}\omega_{i}\period\label{fg-angle2}
} 
This  expression  is quite convenient in practice  since we do not 
 have to consider the degeneration limits from the infinite gap solutions. 
Thus we will use \sref{fg-angle2} instead of \sref{fg-angle} as the definition of angle variables for finite gap solutions when we evaluate the correlation functions 
later in section 4.
\subsubsection{Structure of multi-pronged solutions and importance of local asymptotic behavior}
The method of construction of the angle variables given above 
 is for finite gap solutions, which  serve as saddle point configurations 
 for two point functions.  Since we are interested in computing  three and higher
  point functions as well, we must discuss how the method can be 
generalized to such cases. 

Before giving the simple  procedure, which turns out to require only the  knowledge 
 of the local behavior of the saddle point solution in the vicinity of each 
 vertex insertion point, it is instructive to first clarify  the difference of the analytic 
 structures between two-prong and three-prong 
solutions\fn{The discussion to follow 
 is applicable to higher-prong solutions as well.}  in the framework of the
finite gap method. 

Although a solution with three prongs is  much more difficult to 
 construct compared to the corresponding two-pronged solution, 
the behavior around each prong  should be the same if it is generated by 
  the same vertex operator. This implies  that the spectral curve constructed 
 from the local monodromy matrix should be the same as that 
 of the two-point solution.  Therefore the knowledge of the spectral curve alone 
 cannot distinguish between two-point and three-point solutions. 

What can distinguish between the two is the number of dynamical poles 
of the normalized Baker-Akhiezer vector. In the case of a finite gap solution 
 of genus $g$ relevant for a two-point function, there are $g+2$ non-vanishing 
 filling fractions, which are dictated by the spectral curve, 
 and $g+1$ dynamical poles
of the normalized Baker-Akhiezer vector. The reconstruction formula then tells us  that
 these two sets of data determine the (two-point) solution uniquely, up to a 
 global symmetry transformation.  What this implies is that for more general 
 finite gap  solutions,  relevant for three point functions etc., 
the number of dynamical poles can be larger than $g+1$, while 
 the number of branch cuts of finite length on the spectral curve remains to be $g+1$. 
This possibility has been overlooked until quite recently and is first utilized 
 in \cite{JG}  to reconstruct the solution which describes a correlation function of  a circular Wilson loop and a half-BPS operator from the algebraic curve perspective\fn{In \cite{JG}, the authors  reconstructed the solution by requiring the existence of two distinct  poles in the  Baker-Akhiezer vector. Since the spectral curve of this  solution has no branch cuts with finite length, this certainly goes beyond  the ordinary finite gap construction.}.
The easiest way to obtain solutions with more than $g+1$ dynamical poles is to take the degeneration limit of the infinite gap solutions. Although only $g+1$ poles 
remained  dynamical in the special degeneration limit considered in section 3.2.3, 
more general limits can be considered  in which more than $g+1$ dynamical poles survive.  In principle, it is even possible  for the Baker-Akhiezer vector 
 to have  an infinite number of poles  when the spectral curve has only a finite number of branch cuts. This phenomenon is demonstrated for a string in flat spacetime
 explicitly in Appendix H.

Despite the existence of important  structural differences between two- and multi-pronged solutions  as analyzed above,  we now emphasize that as far as  the evaluation of the angle variables needed to compute  the contribution of the wave functions is concerned, 
only the local asymptotic behavior of the solution near the vertex insertion
 point suffices.  This should indeed be the case  because the vertex operator is defined 
locally and it produces the local source term for the equations of motion 
in the form 
\beq{
\del \delbar X^{\mu} -(\del X^{\nu}\delbar X_{\nu}) X^{\mu} = \frac{\pi}{2\sqrt{\lambda}}\sum_i \frac{\delta \log V_i}{\delta X_{\mu}}\period
}
Therefore possible  local behavior around such a point is the same  for two 
and higher point functions. In particular, in the case of LSGKP strings, this follows directly from the boundary condition we impose near the vertex insertion point, 
namely $\al \sim (1/4) \log p\pbar$. More precisely, as we shall describe in detail 
 in the subsequent sections, the crucial  information about the angle variables of the 
 three point solutions needed for the evaluation of the wave functions can be
 extracted  from the behavior of the angle variables for two point functions 
under suitable global symmetry transformations. 
\subsection{Global symmetry and evaluation of the wave function}
Having discussed the method to construct a complete set of action-angle variables,  we now explain how to actually evaluate  the angle variables for a given 
finite gap solution and then compute the wave function in terms of the 
action-angle variables. In both of these processes,  the global symmetry 
of the system will play the key  role.  
\subsubsection{Shift of the angle variables under global symmetry
 transformations}
In this subsection, we shall develop a general method for computing  
the angle variables for a given solution. 
We will describe the method  for a general genus $g$ finite gap solution in \ads3, 
\ie  within the $SL(2,C)$ principal  chiral model we have been dealing with.  
However, as it will be clear, the method can be 
 readily extended to any finite gap solution in more general spacetime. 
Moreover, it is relevant to the multi-point correlation functions, since the behavior of the saddle point solution  in the vicinity of each vertex insertion point is the same as that of two-point solution. 

\begin{figure}[htbp]
\begin{minipage}{0.45\hsize}
  \begin{center}
   \picture{clip,height=3.7cm}{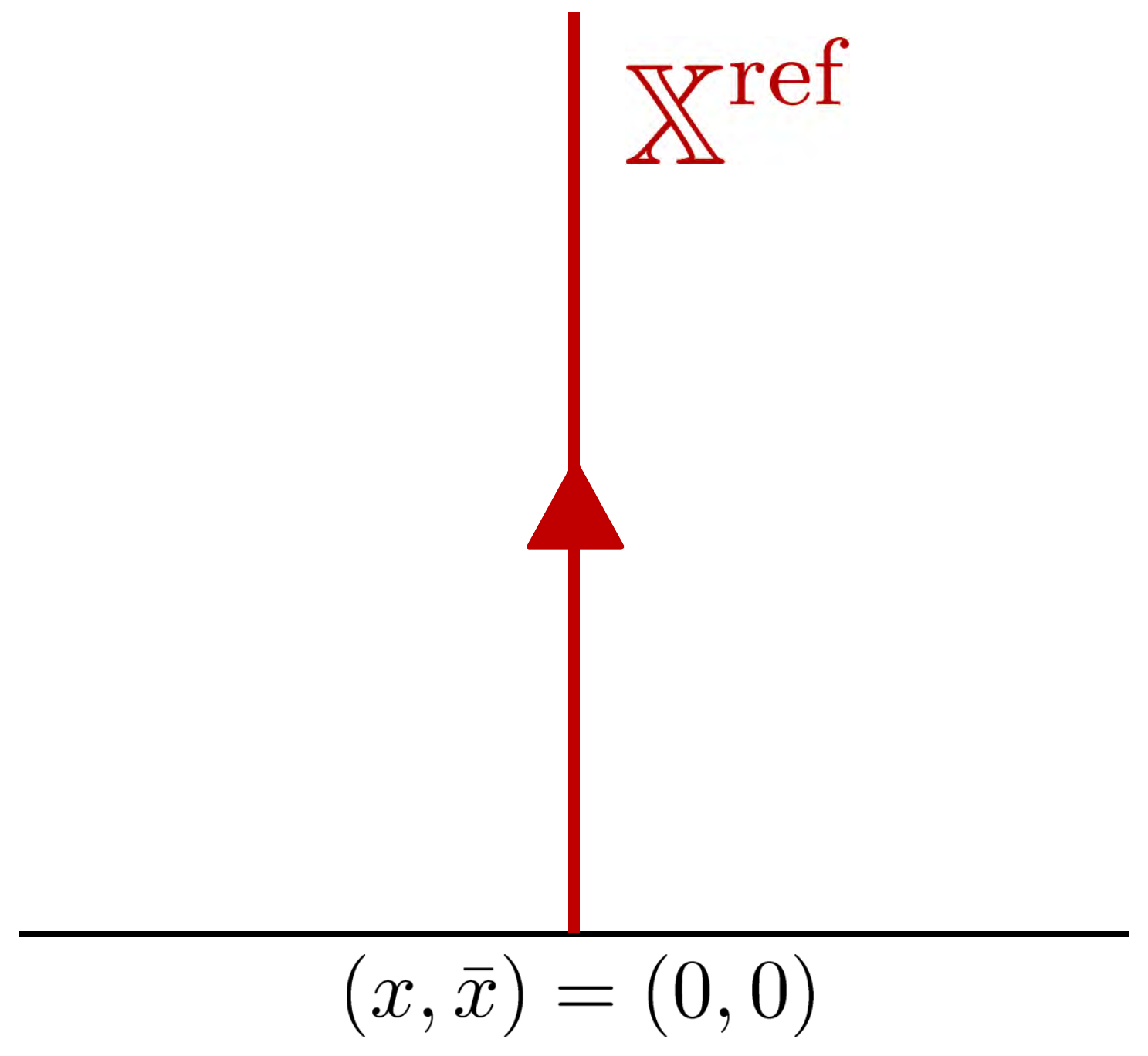}
\end{center}
 \end{minipage}
\hspace{-22pt}
\begin{minipage}{0.1\hsize}
  \begin{flushright}
  $\stackrel{\text{Global Transformation}}{\vspace{11pt}}{\text{\Huge{$\Rightarrow $}}}$
  \end{flushright}
 \end{minipage}
\hspace{22pt} 
\begin{minipage}{0.45\hsize}
  \begin{center} 
  \picture{height=3.7cm,clip}{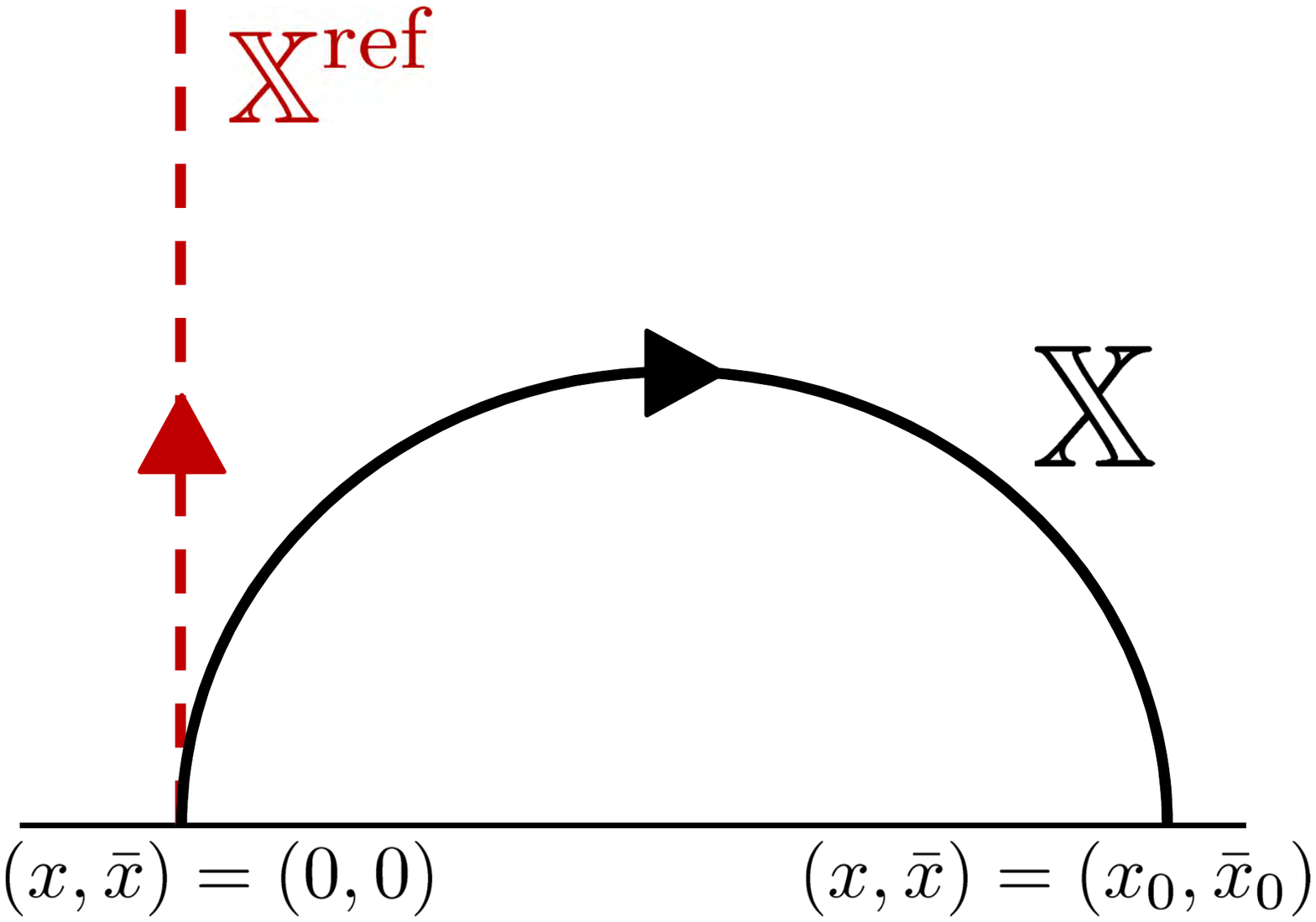}
  \end{center}
 \end{minipage}
\caption{ Schematic picture for producing a general two point solution $\mathbb{X}$ from a reference solution $\refsol$ by a global transformation. } 
\label{bending}
\end{figure}

Our method exploits the simple fact that from a certain ``reference" finite 
gap solution of genus $g$ one can generate solutions of the same type 
 by  the  global $SL(2)_L \times SL(2)_R$ transformations (see \figref{bending}). As we shall show, this procedure is sufficient to generate 
 two point solutions of desired property from which we can compute the 
angle variables of our interest. 
As  the reference solution, it is convenient to take the 
one for which the $SL(2)_L \times SL(2)_R$ charge matrices $Q_L$ and $Q_R$ 
are exactly  diagonal,  namely  $Q_L = L \sig_3 /2, Q_R = -R\sig_3 /2$. 
(For the LSGKP case, the basic solution given in (\ref{gB}) has such a property. )
 We will denote the $SL(2)$ matrix form of such reference 
 solution (as in (\ref{matrixg}))  by  $\refsol$. More general  matrix solution 
$\mathbb{X}'$ can then be generated  as 
\begin{align}
\mathbb{X}' &= V_L \refsol V_R \comma \qquad V_L \in SL(2)_L \comma \quad V_R 
 \in SL(2)_R \period \label{globaltransf}
\end{align}
As already discussed in the previous section, for a genus $g$ finite gap solution, 
 there are $g+2$ non-vanishing action variables, including  the charges 
 $R$ and $L$. 
 We are interested in how the corresponding $g+2$ angle variables  change 
 as we go from $\refsol$ to  $\mathbb{X}'$. 

Such angle variables are extracted from the positions of the poles of the 
suitably normalized eigenvector of the auxiliary linear problem. 
We will first employ the usual {\it left-invariant} Lax connection $J^r(x)$ to formulate 
the auxiliary linear problem. Then, the system is insensitive to the left charge $L$
 and its conjugate angle variable and hence only $g+1$ poles  are visible. 
The information about the left charge and its conjugate will be extracted  later from the similar analysis  using the right-invariant Lax connection $J^l(x)$. 

Let  $\vecpsiref (x; \tau, \sig)$ be the eigenvector of the auxiliary linear problem for the reference solution $\refsol$. The normalized Baker-Akhiezer vector
 $\vechref (x;\tau)$ is proportional to $\vecpsiref (x;\tau,\sig=0)$
 and satisfies the condition
\begin{align}
&\vec{n} \cdot \vechref(x;\tau) = n_1h_1+n_2 h_2 = 1\comma \\
 &\vec{n} = \vecii{n_1}{n_2} \comma \quad \vechref= \vecii{h_1}{h_2} 
\period  \label{normcond}
\end{align}
The choice of the constant normalization vector $\vec{n}$, 
extremely important for the construction of the  proper wave function, 
will be discussed in detail 
 in section 3.3.2.  Until that point, $\vec{n}$ can be arbitrary. 
As the charge matrix $Q_R$
 is diagonal for our reference solution, the corresponding monodromy matrix 
 $\Omegaref (x)$ behaves at $x \rightarrow \infty$ as 
\begin{align}
\Omegaref (x) &= 1 - {2\pi i R \sig_3 \over \sqrt{\lam} x} + O(1/x^2) 
\comma
\end{align}
and hence the normalized eigenvector $\vechref(\infty^\pm;\tau)$ at $x = \pm \infty$ take  the form 
\begin{align}
\vechref(\infty^+;\tau) =\vecii{{1\over n_1}}{0}  \propto \vecii{1}{0} \comma 
\qquad \vechref(\infty^-;\tau)  = \vecii{0}{{1\over n_2}} \propto  \vecii{0}{1} \period  \label{hrefinfty}
\end{align}

Now, under the global transformation (\ref{globaltransf}),  $\vecpsiref$ 
gets transformed into  $V_R^{-1} \vecpsiref $, where we parametrize $V_R$ 
and its inverse  as 
\begin{align}
V_R=\matrixii{v_{11}}{v_{12}}{v_{21}}{v_{22}} \comma \hspace{33pt}V_R^{-1}  = \matrixii{v_{22}}{-v_{12}}{-v_{21}}{v_{11}} 
\period
\end{align}
The vector $\vechref$ gets rotated by the same matrix but in order to retain 
 the normalization condition (\ref{normcond}), we must rescale it 
 appropriately. This gives 
\begin{align}
\vec{h}' (x; \tau)  = {1\over f(x;\tau)} V_R^{-1} \vechref(x; \tau) \comma 
\end{align}
where the rescaling factor $f$ is given by 
\begin{align}
f(x;\tau) &= \vec{n} \cdot \left( V_R^{-1}\vechref(x;\tau) \right)  \nn\\
&= n_1(v_{22} h_1 - v_{12} h_2) + n_2 (-v_{21} h_1 + v_{11}h_2) 
\period  \label{functionf}
\end{align}
Hereafter, we shall suppress the $\tau$-dependence as our  focus will be on the 
 behavior of functions and differentials on the spectral curve parametrized by $x$. 

Let the positions of the poles of $\vechref$ and $\vec{h}'$ on the 
spectral  curve be 
$\{ \ga_1, \ga_2, \ldots, \ga_{g+1}\}$ and $\{ \ga'_1, \ga'_2, \ldots, \ga'_{g+1}\}$ respectively\footnote{ As the number of poles in  the normalized eigenfunction  
 is dictated by the Riemann-Hurwitz theorem, it does not change under the 
 global transformation. See \cite{Vicedo} for details.}.  Then, division by $f$ must remove the poles 
$\{ \ga_1, \ga_2, \ldots, \ga_{g+1}\}$ while  creating   the new poles $\{ \ga'_1, \ga'_2, \ldots, \ga'_{g+1}\}$. In other words, the divisor of $f$ is given by 
\begin{align}
(f) &= \sum_{i=1}^{g+1} (\ga'_i -\ga_i) \period  \label{divisorf}
\end{align}
A natural meromorphic  differential which encodes this information is  
\begin{align}
\varpi &= d (\log f) = {df \over f} \period
\end{align}
From (\ref{divisorf}) $\varpi$ must have poles at $\ga'_i$ and $\ga_i$ 
with residues $1$ and $-1$ respectively. Besides, $\varpi$ may have  
 a  holomorphic part, which can be written as a linear combination of 
the basic holomorphic differentials   $\omega_i$ for $i=1 \sim g$. 
They are assumed to be 
 normalized in the usual way, namely $\int_{a_i} \omega_j = \delta_{ij}$, 
 $\int_{b_i} \omega_j = \Pi_{ij}$, where $\Pi_{ij}$ is the 
 period matrix. 
To express this structure, let us introduce the basic abelian differential of the third kind $\omega_{PQ}$ characterized  by  the properties
\begin{align}
\oint_P \omega_{PQ} &= 1\comma \quad \oint_Q \omega_{PQ} = -1 
\comma \quad \oint_{a_i} \omega_{PQ} =0 \period
\end{align}
Then, $\varpi$ can be written as 
\begin{align}
\varpi &= \sum_{i=1}^{g+1} \omega_{\ga'_i \ga_i} 
 + \sum_{j=1}^g c_j \omega_j \period  \label{expvarpi}
\end{align}
The expansion coefficients  $c_j$ are determined by the integrals of $\varpi$ 
over the  $a_j$-cycles.  As $\varpi$ is a differential of a logarithmic function, 
 the possible contribution must be of the form 
\begin{align}
\int_{a_j} \varpi &= 2\pi i m_j \comma \qquad m_j \in \mathbb{Z} \period
\end{align}
This gives $c_j =  2\pi i m_j$.  Next, consider the integrals of $\varpi$  over 
 the $b_k$-cycles. Again the possible structure is 
\begin{align}
\int_{b_k} \varpi &= 2\pi i n_k \comma \qquad n_k \in \mathbb{Z} \period
\end{align}
From (\ref{expvarpi}) we then get
\begin{align}
\sum_{i=1}^{g+1} \int_{b_k} \omega_{\ga'_i \ga_i} 
&= 2\pi i n_k - 2\pi i \sum_{j=1}^g m_j \Pi_{jk} \period 
\label{intombk}
\end{align}
Now by using a variant of the Riemann bilinear identity\footnote{See 
Corollary 2.42 of \cite{Vicedo}.}, one can rewrite 
\begin{align}
\int_{b_k} \omega_{\ga'_i \ga_i} &= 2\pi i \int_{\ga_i}^{\ga'_i} \omega_k 
\period  \label{idone}
\end{align}
Thus, \sref{intombk} becomes 
\begin{align}
\sum_{i=1}^{g+1} \int_{\ga_i}^{\ga'_i} \omega_k 
= n_k -\sum_{j=1}^g m_j \Pi_{jk}\period
\end{align}
Now note that $n_k$ and $m_j$ are integers which take discrete values. 
On the other hand, the LHS clearly vanishes continuously 
in the limit $\ga_i \rightarrow \ga'_i$. 
Hence, we should set  $n_k = m_j=0$ and conclude 
\begin{align}
\sum_{i=1}^{g+1} \int_{\ga_i}^{\ga'_i} \omega_k  =0\comma 
 \qquad k =1 \sim g \period
\end{align}
What this means is that the angle variables conjugate to the filling fractions 
 $S_k, k=1\sim g$ do not change under the global transformation. 

Therefore, the only angle variable 
 left to be examined is the one associated 
 with the differential $\omega_\infty \equiv {1\over 2\pi i} \omega_{\infty^+\infty^-}$, namely the one conjugate to the charge $R$.  We will denote it by $\phi_\infty$. This can be studied by 
 considering the integral over the contour $b_\infty$ running from 
$\infty^-$ to $\infty^+$. Repeating essentially the same argument made  for $b_k$, 
 except for the evaluation of $\int_{b_\infty} \varpi$, we readily obtain 
\begin{align}
\int_{b_\infty} \varpi = \log \left( {f(\infty^+) \over f(\infty^-)}
\right) = 2\pi i\sum_{i=1}^{g+1} \int_{\ga_i}^{\ga'_i} \omega_\infty
\comma \label{intvarpi}
\end{align}
where we used an identity similar to (\ref{idone}), namely\footnote{See 
 Proposition 2.43 of \cite{Vicedo}. }
 $\int_{b_\infty}\omega_{\ga'_i \ga_i} =2\pi i\int_{\ga_i}^{\ga'_i} \omega_\infty$ 
.  Since the RHS of \sref{intvarpi} expresses  the shift $\iDelta \phi_\infty$ multiplied by $i$ (see \sref{fg-angle2}),  we have  an important  formula 
\begin{align}
\iDelta \phi_\infty &= -i \log \left( {f(\infty^+) \over f(\infty^-)} \right) 
\period   \label{deltaphiinf0}
\end{align}
This can be recognized as  the generalization of the formula given in 
Proposition 8.13 of \cite{Vicedo}, which was derived 
 for the $U(1)_R$ part of the global 
transformation. Our master formula above is valid  for an arbitrary global symmetry 
transformation. 

In the present case, the formula can be made completely explicit by substituting 
 the asymptotic behavior of $\vechref$ at $\infty^\pm$ given in (\ref{hrefinfty}) 
into the form of $f(x)$ in  (\ref{functionf}). This gives
\begin{align}
\iDelta \phi_\infty &= -i \log \left( {v_{22} - {n_2 \over n_1} v_{21}
 \over -{n_1 \over n_2} v_{12} + v_{11}  } \right) \period
\label{iDeltaphiinf}
\end{align}
Likewise,  repeating exactly the same procedure with the {\it right-invariant }
Lax connection, we can compute the shift of the angle variable $\tilde{\phi}_0$,  
conjugate to the $SL(2)_L$ charge $L$, which has been invisible in the 
left-invariant formalism.  We obtain 
\begin{align}
\iDelta \phitil_0 &= -i \log \left( {\vtil_{11} + {\ntil_{2} \over \ntil_1} \vtil_{21} 
\over {\ntil_1 \over \ntil_2} \vtil_{12} + \vtil_{22} } \right) \comma 
\label{iDeltaphitilzero}
\end{align}
where $\vec{\tilde{n}} = (\ntil_1, \ntil_2)^T$ is the normalization vector 
and $\vtil_{ij}$ are the components of the $SL(2)_L$ transformation matrix $V_L$. 
The formulas \sref{iDeltaphiinf} and \sref{iDeltaphitilzero} constitute the main 
result of this subsection. 
\subsubsection{Evaluation of the wave function from global symmetry}
Once we have the information of the action-angle variables, the quantum 
wave function can be constructed in the manner
\beq{
\Psi [\tilde{\phi}_0[\vec{\ntil}],\phi _i[\vec{n}],\phi _{\infty}[\vec{n} ]] \equiv \exp \left( i S_0 \tilde{\phi}_0[\vec{\ntil}]  + iS_{\infty} \phi_{\infty}[\vec{n}]+i\sum_i S_i \phi_i[\vec{n}]\right)\period \label{wave}
}
To determine the wave function completely, however,   we still have to 
specify the following quantities. One is the choice of the normalization 
vectors $\vec{n}$ and $\vec{\ntil}$,  on which the angle variables depend. 
 We shall show below that this is related to the 
  property of the wave function under the global symmetry transformations and there 
 is a definite and proper choice. The other is the choice of the 
origin of the angle variables ( or equivalently, the choice of $x_0$ in \sref{def-of-F}). 
This on the other hand  is intimately related to the overall normalization of the wave function and will be determined when we  evaluate the two point functions in the next section.    

Let us study the  property of the wave function under the global transformation. 
Under the transformation of the solution $\mathbb{X}\, \mapsto \, V_{L}\mathbb{X}V_{R}$, the currents transform as 
  $J^{r}\to V_{R}^{-1}\, J^{r}\, V_{R}$ and $J^{l}\to V_{L}\, J^{l}\, V_{L}^{-1}$. In turn, the left and the right normalized Baker-Akhiezer vectors, which are 
 the eigenvector of the auxiliary linear problem,  transform as 
\beq{
\vec{h}\longmapsto V_{R}^{-1}\cdot  \vec{h} \comma \hspace{33pt}\vec{\tilde{h}}\longmapsto V_{L}\cdot  \vec{\tilde{h}}\period
} 
Now in order to keep the normalization conditions, $\vec{n}\cdot \vec{h}=1$ and $\vec{\tilde{n}}\cdot\vec{\tilde{h}}=1$, $\vec{n}$ and $\vec{\tilde{n}}$ must transform in the opposite way, namely, 
\beq{
\vec{n}\longmapsto V_{R}^{\T}\cdot \vec{n}   \comma \hspace{33pt}\vec{\tilde{n}}\longmapsto  \left( V_{L}^{\T}\right)^{-1}\cdot \vec{\tilde{n}} \comma
}
where the superscript $T$ denotes transposition. 
This means that  the wave function \sref{wave} transforms under the global symmetry transformation in the following way:
\beq{
&\Psi \left[\tilde{\phi} _0[ \vec{\tilde{n}}],\phi _i[\vec{n}],\phi _{\infty}[\vec{n}]\right] \longmapsto \Psi \left[\tilde{\phi} _0[  \text{\small{$\left( V_{L}^{\T}\right)$}}^{\text{\tiny{$-1$}}}\cdot\vec{\tilde{n}}],\phi _i[\text{\small{$ V_{R}^{\T}$}}\cdot \vec{n}],\phi _{\infty}[\text{\small{$ V_{R}^{\T}$}}\cdot \vec{n}]\right]\period  
}

With this formula in mind, we shall  now show that $\vec{n}$ and $\vec{\ntil}$ 
 are dictated by the properties required of the wave function (or the corresponding 
 vertex operator) we wish to construct.  

First we note that  the wave function we want should describe, holographically, a certain single-trace operator in $\mathcal{N}=4$ SYM and hence it should have the same 
transformation property as such an operator. 
 For simplicity,   let us consider the wave function corresponding to a conformal primary operator\fn{Here we restrict our attention to conformal primary operators, namely highest weight operators. This is mainly because finite gap solutions including GKP strings are usually assumed to describe highest weight operators of the gauge theory (see e.g.~\cite{KZ}). It would be an interesting future problem either to find a way to describe non-highest weight operators by classical strings or to rigorously show that classical strings can only describe highest weight operators.} of the gauge theory, 
 carrying the dimension $\Delta$ and the spin  $S$, 
inserted at the origin of the Minkowski space $R^{1,3}$, \ie  the boundary of $AdS_5$:%
\beq{ 
\Psi [\tilde{\phi} _0(\vec{\tilde{n}}),\phi _i(\vec{n}),\phi _{\infty}(\vec{n})]\hspace{22pt}\longleftrightarrow \hspace{22pt}\mathcal{O}^{\Delta\comma S}\left( x^{\mu}=0\right)\period 
}
Now such a conformal primary operator  at the origin of $R^{1,3}$ is completely invariant under the special conformal transformation. Therefore 
  the corresponding wave function should also be invariant under such a transformation given by 
\beq{
&\Psi \left[\tilde{\phi} _0[ \vec{\tilde{n}}],\phi _i[\vec{n}],\phi _{\infty}[\vec{n}]\right] \longmapsto \Psi \left[\tilde{\phi} _0[  \text{\small{$\left( V_{L}^{sc \,\,\T}\right)$}}^{\text{\tiny{$-1$}}}\cdot\vec{\tilde{n}}],\phi _i[\text{\small{$ V_{R}^{sc\,\,\T}$}}\cdot \vec{n}],\phi _{\infty}[\text{\small{$ V_{R}^{sc\,\,\T}$}}\cdot \vec{n}]\right]\comma  
\label{propwave}
}
where $V_{L\comma R}^{sc}$ are the matrices given in Appendix C. 
For $\Psi$ to be invariant, the angle variables  should be invariant and hence the 
 normalization vectors should be unchanged:
\beq{
\vec{n}= V_{R}^{sc\,\, \T}\cdot\vec{n} \comma\hspace{33pt}\vec{\tilde{n}}=\left( V_{L}^{sc\,\, \T}\right)^{-1}\cdot \vec{\tilde{n}}\period\label{propn}
}
From the form of $V_{L\comma R}^{sc}$ we readily find that the 
solutions are  $\vec{n}=(1\comma 0)^{T}$ and $\vec{\tilde{n}}=(0\comma 1)^{T}$.  One subtlety is that precisely for  this special choice  the number of dynamical poles of the Baker-Akhiezer vectors decreases by one.  Thus to avoid this singular point, 
 we should employ the ``regularized" form   
\beq{
\vec{n}=\pmatrix{c}{1\\ \mu } \comma \hspace{33pt}\vec{\tilde{n}}=\pmatrix{c}{ \tilde{\mu} \\1} \comma \label{def-of-n}
} 
where  $\mu$ and $\tilde{\mu}$ are ``regularization parameters", to  be set to zero at the end of the calculation.

It is important to note that the discussion above  shows,  incidentally, 
 that our reference  solution $\refsol$, which starts at the origin of the boundary at $\tau=-\infty$,  precisely corresponds to such an  insertion of 
$\calO^{\Delta, S}(0)$. 
Recall that our reference solution is chosen so that the global charges
 $Q_L$ and $Q_R$ are diagonal.  This  corresponds to the state  annihilated by the lowering operators of $SL(2)_L$ and $SL(2)_R$, which 
are indeed  the special conformal generators from the point of view of the boundary 
theory.  

So far we have only considered the case which corresponds  to the insertion of the 
 operator at the origin.  When we consider the insertions at 
 points other than the origin, the situation changes.  For instance, while 
 a conformal primary operator inserted at $x^{\mu}=0$ is invariant under special conformal transformations, it is no longer so when 
 inserted elsewhere on $R^{1,3}$. This is because the special conformal transformation 
acts also on the coordinates of $R^{1,3}$.  Apparently,   we must reanalyze 
 the choice of the normalization vectors for such cases. 

Fortunately, there is a way to avoid this complication. The key is again the use of 
the global transformation. The same type of operators  inserted at different points are related to each other by a translation on the boundary. Thus the corresponding 
 wave functions should also be likewise related.  Furthermore, for saddle point 
 approximation, evaluating the translated wave function on the original trajectory 
$\mathbb{X}_\ast$ 
is the same as evaluating the original (non-translated) wave function 
 on the inversely translated trajectory  ${\left( V_L^{\T} \right) ^{-1}\mathbb{X}_{\ast}\left( V_R^{\T}\right)^{-1}}$. This is   expressed 
 as (see \figref{translating})
 \beq{
\Psi \left[\tilde{\phi} _0[  \text{\small{$\left( V_{L}^{\T}\right)$}}^{\text{\tiny{$-1$}}}\cdot\vec{\tilde{n}}],\phi _i[\text{\small{$ V_{R}^{\T}$}}\cdot \vec{n}],\phi _{\infty}[\text{\small{$ V_{R}^{\T}$}}\cdot \vec{n}]\right] \Big| _{\mathbb{X}_{\ast}}=\Psi [\tilde{\phi} _0(\vec{\tilde{n}}),\phi _i(\vec{n}),\phi _{\infty}(\vec{n})]\Big| _{\left( V_L^{\T} \right) ^{-1}\mathbb{X}_{\ast}\left( V_R^{\T}\right)^{-1}}\period
}
It turns out to be much easier to consider the transformation of the trajectory  than 
the transformation of the wave function.  In this way, we only need to deal with the 
 case corresponding to the insertion at the origin, for which the previous analysis 
 is valid. This procedure will be carried out explicitly  for the actual computation of 
 the two point and the three point functions in subsequent sections.

Finally, let us return to the general formulas for the shift of the 
 angle variables obtained in \sref{iDeltaphiinf} and \sref{iDeltaphitilzero}. 
Upon  substituting  the appropriate choice of the normalization vectors 
given in \sref{def-of-n}, these formulas become 
\begin{align}
\iDelta \phi_\infty &= - i \log \left( {v_{22} - \mu v_{21}
 \over  -{v_{12} \over \mu} + v_{11}  } \right) \comma 
\label{deltaphiinf} \\
\iDelta \tilde{\phi}_0 &= - i \log \left( {\tilde{v}_{11} + { \tilde{v}_{21} \over \tilde{\mu}}
 \over  \tilde{\mu} \tilde{v}_{12}  + \tilde{v}_{22}  } \right) \period
\label{deltaphizero}
\end{align}
We will demonstrate that these simple and explicit formulas are 
extremely powerful in evaluating the wave functions for various 
correlation functions. 

\section{Computation of two point functions}
In the previous section, we have developed a general method 
of  computing the angle variables for the finite gap solution generated from the 
 reference solution via  a global symmetry transformation. 
The end results  were the simple formulas (\ref{deltaphiinf}) and 
(\ref{deltaphizero}),  which do not depend on the details of the finite gap solution. 

We shall now show that, by making use of these formulas, one can compute 
  semi-classical two point functions for the cases where the saddle point 
configurations  are finite gap solutions.  After describing the essence of the method, which  is quite universal, we will apply it to the general  elliptic GKP strings, 
 \ie without taking the large spin limit.  As it will become clear, the same 
reasoning can in fact be applied in  the  vicinity of each  vertex insertion point 
 of any higher-point function to compute the contribution of the vertex operators. 
This will be demonstrated explicitly  for the three-point function 
 of LSGKP strings in section 5. 
\subsection{General formula for two point functions}
%
Although the logic of our method  is quite general, 
 for  clarity let us consider the case of a two point function of 
 the form $\langle V(x_0,\xbar_0\, ;z_2) V(0,0\, ;z_1) \rangle$  
the semi-classical saddle of which is given by a 
 finite gap solution.  We will compute this correlation function by the use of the state-operator correspondence, without requiring the precise form of the vertex operators. 

In the case of two point functions, the saddle point trajectory $\mathbb{X}$ is 
normally described in terms of the {\it global} cylinder coordinates $(\tau, \sig)$ 
 related to the plane coordinate $z$  by\cite{BuchTseytlin}
\begin{align}
e^{\taug +i\sigg}=\frac{z-z_1}{z-z_2}  \comma \label{gcylinder}
\end{align}
 and it starts from $(0,0)$ at 
$\taug \simeq -\infty$ and ends on $(x_0,\xbar_0)$ at $\taug\simeq + \infty$, both on the boundary of  the \ads3.   On the other hand,  the state-operator correspondence should be described using  the {\it local} cylinder 
 coordinates, namely  $(\tau_1, \sig_1)$ around  $z_1$ and $(\tau_2, \sig_2)$ 
 around $z_2$,  defined by 
\begin{align}
e^{\tauone +i\sigone}=z-z_1 \comma \qquad 
e^{\tautwo+i\sigtwo}=z-z_2 \period
\end{align}
Explicitly, around $z_1$  such a correspondence  is described by 
\beq{
\int \mathcal{D}X\big| _{X = X_0 \text{ at }\tauone = \log \epsilon _1} V(0,0\, ;z_1) e^{-S}=\Psi _1 \left[ X_0\right]\comma 
}
where $X_0$ is the configuration prescribed on a small circle of radius $\ep_1$ 
around $z_1$ and $\Psi_1$ is the wave function corresponding to 
 the vertex operator $V(0,0\, ;z_1)$. 
When the semi-classical approximation is adequate, we can replace the above path integral by the value at the saddle point configuration and the form of the 
 correspondence  simplifies to 
\beq{
V(0,0\, ;z_1)\big| _{X_{\cl}} \exp \left[ -S_{\cl}(\tauone <\log \epsilon _1)\right]=\Psi _1 \left[ X_{\cl}\right]\big| _{\text{at } \tauone = \log \epsilon _1}\period
}
In a similar fashion, the  vertex operator  $V(x_0,\xbar_0\, ;z_2)$ is related to the wave function $\Psi _2$, defined on a small circle of  radius $\epsilon_2$  around 
 $z_2$. 

To evaluate these wave functions, we need to reexpress the trajectory $\mathbb{X}$  parametrized by the global coordinates  in terms of the local coordinates. In the vicinity of the vertex operators, the global coordinates are related to 
 the local coordinates as    
\beq{
\taug +i\sigg \simeq \bracetwo{\log \left( \frac{z-z_1}{z_1-z_2}\right) =\tauone +i\sigone -\log (z_1-z_2) \hspace{22pt} (z\sim z_1)\comma}{\log \left( \frac{z_2-z_1}{z-z_2}\right) =-\tautwo -i\sigtwo +\log (z_2-z_1) \hspace{22pt} (z\sim z_2)\period} \label{globallocal}
}
In the discussion below, it will be quite important to keep track of which type of coordinates are being used in the description of the trajectory $\mathbb{X}$. 
Thus, to make this distinction clear, we shall employ  the square bracket, 
like  $\mathbb{X}[\tau_1,  \sig_1]$, when using the local coordinates, and the 
 round bracket, like $\mathbb{X}(\tau, \sig)$, for the use of the global 
 coordinates.  Using this notation, the trajectory near  $z_1$ and $z_2$ 
can be expressed in two types of coordinates in the following way:
\beq{
&\mathbb{X}[ \tauone \comma \sigone ] \simeq \mathbb{X}(\tauone-\tau_{12}\comma\sigone -\sigma_{12}) &(z\sim z_1)\comma \label{rel1g}\\
&\mathbb{X}[ \tautwo \comma \sigtwo ] \simeq \mathbb{X}(-\tautwo +\tau_{12}\comma -\sigtwo +\sigma_{12}+\pi )  &(z\sim z_2)\period
}
Here  the quantities  $\tau_{12}$ and $\sigma_{12}$ are defined by 
\beq{
& \tau_{12}\equiv\log |z_1-z_2|\comma \hspace{22pt}  \sigma_{12}\equiv\frac{1}{2i}\log\left(\frac{z_1-z_2}{\barz _1-\barz _2}\right)\period
}
Then, including the contribution of the action evaluated in  the global cylinder coordinate,  the semi-classical two point function can be written as 
\begin{align}
&\Psi_2\big|_{\bsc{\mathbb{X}[\tautwo = \log \epsilon _2]}} \,\, \exp \left( -S\big|^{\taug =\tau_f}_{\taug = \tau_i}\right)\,\, \Psi_1\big|_{\bsc{\mathbb{X}[\tauone = \log \epsilon _1]}} \comma
\end{align}
where the initial and the final global times are given by 
\begin{align}
&\tau_i = \log \epsilon_1 -\log |z_1-z_2|  \comma\hspace{33pt}\tau_f = -\log \epsilon_2 +\log |z_1-z_2| \period\label{def-of-tauif}
\end{align}

Now let us  discuss  the actual evaluation of the  wave functions. 
Semi-classical wave functions carrying definite values of the action variables can be constructed and evaluated easily if we employ the  action-angle variables expressed in 
 the local coordinates, as we describe below.

 Consider first the ``initial" wave function $\Psi_1$ corresponding to $V(0,0)$. To make the essence of 
 the argument clear, let us focus on just one generic
 pair of action-angle variables, denoted by $J$ and $\theta$,  without specifying which action-angle variables we are dealing with. 
Then, the wave function is given simply by 
$\Psi _1 [\theta; \tau] = e^{i J \theta -\mathcal{E}\tau}$, 
 where $\mathcal{E}$ is the energy on the worldsheet\fn{Although 
$\mathcal{E}$ vanishes for the relevant saddle point configuration, 
we will keep  $\mathcal{E}$  throughout this subsection to demonstrate explicitly 
 the equivalence of the Virasoro condition and the independence of
 the two-point function on the worldsheet coordinates.}. 
Since  $\calE$ is a function only of $J$ and the momentum 
$\calP$ is of the form $nJ$ with $n$ an integer, as shown in \cite{DV1},  the angle variable $\theta$ depends linearly on the local coordinates  when evaluated on the solution $\mathbb{X}$:
\begin{align}
\theta \big|_{\bsc{\mathbb{X} [\tauone \comma \sigone ]}} &= -i\omega \tauone + n \sigone + \theta\big|_{\bsc{\mathbb{X}[0,0]}} \period 
\label{ang-evolv}
\end{align}
Here,  $\omega$ is the angular frequency given by $\del \mathcal{E}/\del J$ and 
the integer $n$ is the mode number given in  \sref{mode-number}. 
(The presence of the  factor of  $-i$  in the first term on the right hand side is due to 
 the use of the Euclidean worldsheet time $\tau$.) 
By using the method described in subsection 3.2,  it is  possible, in principle, to evaluate \sref{ang-evolv} on a given classical solution. 
In practice, however,  it is much more convenient to make use 
  of the formula developed in subsection 3.3 and compute 
 the angle variables for the solution  relative to those for an appropriate 
reference solution. 
Since the two trajectories can be compared most easily when expressed in terms of the global cylinder coordinates,  we rewrite $\theta\big|_{\bsc{\mathbb{X}[\tauone, \sigone]}}$ using \sref{rel1g} as 
\begin{align}
\theta\big|_{\bsc{\mathbb{X}[\tauone\comma\sigone]}}&\overset{z\sim z_1}{\simeq}\theta\big|_{\bsc{\mathbb{X}(\tauone - \tau _{12}\comma \sigone - \sigma_{12})}}
=\iDelta \theta ^{\mathbb{X}} +\theta \big | _{\bsc{\refsol (\tauone - \tau _{12}\comma \sigone - \sigma_{12})}}\comma\label{deftheta1}
\end{align}
where $\iDelta \theta ^{\mathbb{X}}$ is defined as the difference
\beq{
&\iDelta \theta ^{\mathbb{X}}\equiv \theta \big|_{\bsc{\mathbb{X} (\tauone - \tau _{12}\comma \sigone - \sigma_{12})}}-\theta \big|_{\bsc{\refsol (\tauone - \tau _{12}\comma \sigone - \sigma_{12} )}}\period\label{difang1}
}
Note that $\iDelta\theta^{\mathbb{X}}$ is actually  independent of $\tauone$ and $\sigone$ since both of the angle variables in \sref{difang1} evolve 
 linearly in $\tauone$ and $\sigone$. 
Thus we can define $\iDelta\theta^{\mathbb{X}}$ simply by $\theta \big|_{\bsc{\mathbb{X} (\tau \comma \sigma )}}-\theta \big|_{\bsc{\refsol (\tau \comma \sigma )}}$  at arbitrary values of the global coordinates. 
This observation allows us to compute   $\iDelta \theta ^{\mathbb{X}}$ 
by  the formula developed  in subsection 3.3, \ie  in terms of the parameters of the global transformation 
connecting $\mathbb{X}(\tau\comma \sigma )$ with $\refsol (\tau\comma\sigma)$.  Using \sref{deftheta1} $\Psi_1$ can be evaluated as
\beq{
\Psi_1 \big| _{\bsc{\mathbb{X}[\tau_1 =\log \epsilon_1 ]}}&=\exp \left( i J\theta \big| _{\bsc{\mathbb{X}[\log \epsilon_1\comma 0]}}-\mathcal{E}\log\epsilon_1\right)\nn \\
&=\exp \left( iJ\iDelta\theta^{\mathbb{X}} + iJ\theta \big| _{\bsc{\refsol(\tau_i \comma -\sigma_{12} )}}-\mathcal{E}\log\epsilon_1\right) \nn\\
&=\exp \left( iJ\iDelta\theta^{\mathbb{X}} + iJ\theta \big| _{\bsc{\refsol(\tau_i \comma 0 )}}-iJn\sigma_{12}-\mathcal{E}(\tau_i +\tau_{12})\right) \nn\\
&=\exp\left( iJ\iDelta\theta^{\mathbb{X}}-iJn\sigma_{12}-\mathcal{E}\tau_{12}\right) \Psi_1 \big| _{\bsc{\refsol (\tau_i )}}\label{resulti}\comma
} 
where $\tau_i$ is the initial global time defined in \sref{def-of-tauif}.

\begin{figure}[tbp]
\begin{minipage}{0.43\hsize}
  \begin{center}
   \picture{clip,height=3.7cm}{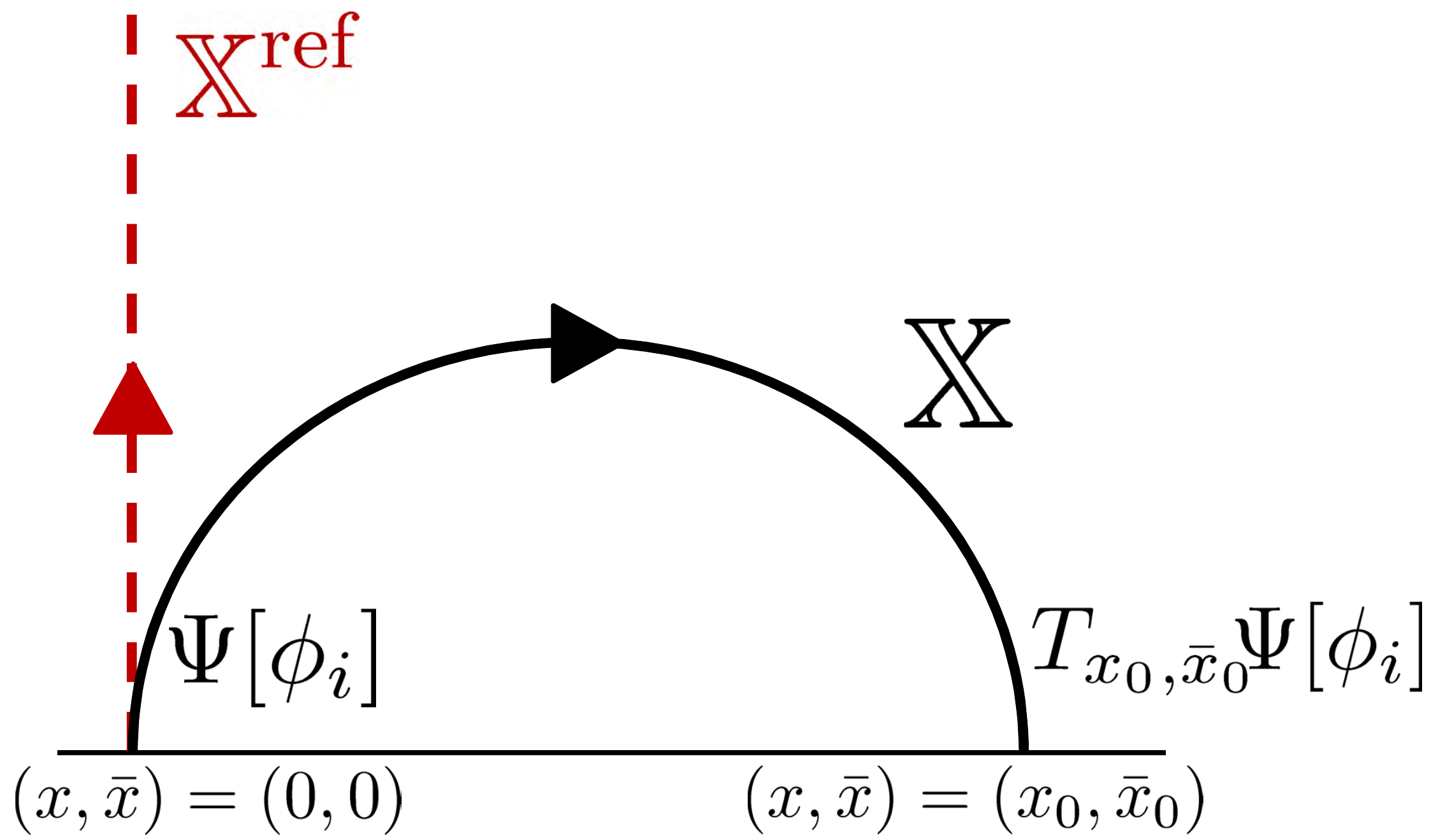}
  \end{center}
 \end{minipage}
\begin{minipage}{0.1\hsize}
  \begin{center}
  $\stackrel{\text{Translation}}{\vspace{11pt}}{\text{\Huge{$\Rightarrow $}}}$
  \end{center}
 \end{minipage}
 \begin{minipage}{0.43\hsize}
  \begin{center}
   \picture{height=3.7cm,clip}{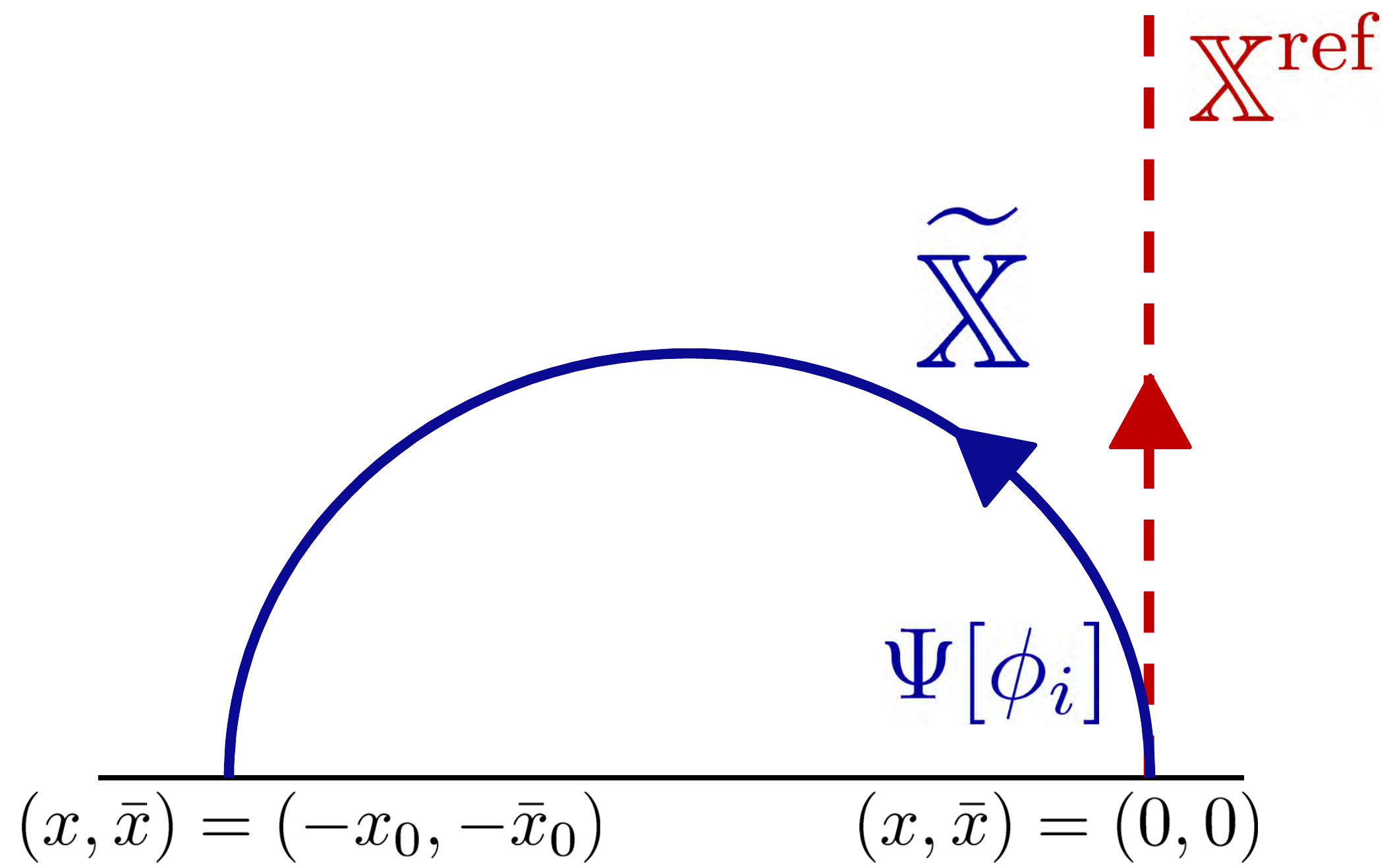}
  \end{center}
 \end{minipage}
\caption{ Schematic picture of how to evaluate the wave functions  at 
two ends of the two point function, relative to their values for the reference 
 solution $\refsol$.  The one at the origin is directly compared to $\refsol$ (left figure).
The one at $(x_0, \xbar_0)$ is evaluated by using the solution $\tilde{\mathbb{X}}$
 obtained by translation and the subsequent reversal of the direction of $\tau$ and $\sig$ (right figure). 
} 
\label{translating} 
\end{figure}
Consider next the ``final" wave function $\Psi _2$, which 
 corresponds to the vertex operator $V(x_0, \xbar_0;z_2)$. 
One subtlety here is that the state-operator correspondence 
 is  made  in  the local cylinder coordinates $(\tautwo\comma\sigtwo)$, 
which, as seen in (\ref{globallocal}), run in directions opposite to the global cylinder coordinates $(\taug\comma \sigg)$. This means that 
we should actually evaluate the final wave function $\Psi_2$  on the {\it reversed solution }$\mathbb{X}(-\tau\comma -\sigma)$ to obtain the correct answer. 
With this in mind,  we can 
 compute the desired contribution in the following way.  
First we note that  $V(x_0,\xbar_0)$ can be obtained from $V(0,0)$ by a translation 
by the  vector $(x_0,\xbar_0)$ and denote it  as $V(x_0,\xbar_0)=T_{x_0,\xbar_0} V(0,0)$.  This means that the corresponding wave functions are also related 
 by the translation as  $\Psi_2 = T_{x_0\comma \bar{x}_{0}}\Psi_1$. 
Then, as can be seen from the \figref{translating}, evaluating $\Psi_2$  on 
the solution $\mathbb{X}$ is the same as evaluating $\Psi_1$ on the inversely translated solution, symbolically denoted as  $T^{-1}_{x_0,\xbar_0} \mathbb{X}$. 
 This is expressed more precisely as 
\beq{
(T_{x_0\comma \bar{x}_{0}}\Psi_1) \big| _{\bsc{\mathbb{X}[\tautwo = \log \epsilon _2]}}=\Psi_1 \big|_{\bsc{T^{-1}_{x_0,\xbar_0} \mathbb{X}[\tautwo = \log \epsilon _2]}}\period\label{defo-psi2}
}
Since $\Psi_1$ is given simply by $e^{iJ\theta-\mathcal{E}\tau}$, we can compute \sref{defo-psi2} by evaluating $\theta$ on $T^{-1}_{x_0,\xbar_0} \mathbb{X}$. To do this,  we first rewrite $\theta\big|_{\bsc{\mathbb{X}[\tautwo\comma\sigtwo]}}$ in terms of the global coordinates  and then reinterpret it as evaluated 
 on the reversed solution.  Explicitly, 
\beq{
\theta\big|_{\bsc{T^{-1}_{x_0,\xbar_0} \mathbb{X}[\tautwo \comma\sigtwo]}}&\overset{z\sim z_2}{\simeq}\theta\big|_{\bsc{T^{-1}_{x_0,\xbar_0} \mathbb{X}(-\tautwo + \tau _{12}\comma -\sigtwo + \sigma_{12}+\pi)}}
=\theta\big|_{\bsc{\tilde{\mathbb{X}}(\tautwo - \tau _{12}\comma \sigtwo - \sigma_{12}-\pi)}}\comma\label{deftheta2beta}}
where $\tilde{\mathbb{X}}$ is the reversed solution defined by 
\beq{\tilde{\mathbb{X}}(\tau\comma\sigma)=T^{-1}_{x_0,\xbar_0} \mathbb{X}(-\tau\comma-\sigma )\period}
Note that, as shown in \figref{translating}, this solution $\tilde{\mathbb{X}}$ starts from the origin just like the reference solution $\refsol$. This allows us to further rewrite \sref{deftheta2beta} as 
\beq{
  \theta\big|_{\bsc{\tilde{\mathbb{X}}(\tautwo - \tau _{12}\comma \sigtwo - \sigma_{12}-\pi)}}&=\iDelta \theta ^{\tilde{\mathbb{X}}} +\theta \big|_{\bsc{\refsol (\tautwo - \tau _{12}\comma \sigtwo - \sigma_{12}-\pi)}}\comma\label{deftheta2}
}
where the difference $\iDelta \theta ^{\tilde{\mathbb{X}}}$ is defined by
\beq{
&\iDelta \theta ^{\tilde{\mathbb{X}}}=\theta \big|_{\bsc{\tilde{\mathbb{X}} (\tau \comma \sigma )}}-\theta \big|_{\bsc{\refsol (\tau \comma \sigma )}}
\period
}
This quantity can be computed, just like   $\iDelta \theta^{\mathbb{X}}$,  by finding the global symmetry transformation which generates  $\tilde{\mathbb{X}}(\tau\comma\sigma)$ from $\refsol(\tau\comma\sigma)$.  
From \sref{defo-psi2}, \sref{deftheta2beta} and \sref{deftheta2}, $\Psi_2$ can be evaluated as
\beq{
\Psi_2 \big| _{\bsc{\mathbb{X}[\tautwo =\log \epsilon_2]}}&=\exp \left( iJ\theta \big| _{\bsc{T_{x_0,\bar{x}_0}\mathbb{X}[\log \epsilon_2\comma 0]}}-\mathcal{E}\log \epsilon _2\right) \nn\\
&=\exp \left( iJ\iDelta\theta^{\tilde{\mathbb{X}}} + iJ\theta \big| _{\bsc{\refsol ( \tau_f \comma -\sigma_{12} -\pi)}}-\mathcal{E}\log \epsilon _2\right) \nn\\
&=\exp \left( iJ\iDelta\theta^{\tilde{\mathbb{X}}} + iJ\theta \big| _{\bsc{\refsol ( \tau_f \comma 0 )}}-iJn(\sigma_{12}+\pi)-\mathcal{E}(-\tau_f +\tau_{12})\right) \nn\\
&=\exp\left( iJ\iDelta\theta^{\tilde{\mathbb{X}}}-iJn(\sigma_{12}+\pi)-\mathcal{E}\tau_{12}\right) \Psi_1 \big| _{\bsc{\refsol (-\tau_f )}}\comma\label{resultf}
}
where $\tau_f$ is as given  in \sref{def-of-tauif}.

Thus, combining the contributions form $\Psi_1$ and $\Psi_2$ given 
 in \sref{resulti} and \sref{resultf} respectively, the net contribution of 
 the wave functions can be written as 
\begin{align}
&\Psi_1\,\Psi_2 \big|_{\mathbb{X}} = (-1)^{Jn} \exp\left( iJ( \iDelta\theta^{\mathbb{X}}+\iDelta\theta^{\tilde{\mathbb{X}}})-2(iJn\sigma_{12}+\mathcal{E}\tau_{12})\right)\Psi_1 \big| _{\scalebox{0.8}{$\refsol (\tau_i )$}}\Psi_1 \big| _{\scalebox{0.8}{$\refsol (-\tau_f )$}}\period\label{finone}
\end{align}
Recall that  the sum of  $Jn$'s  from 
 all the angle variables is identified with the worldsheet momentum $\mathcal{P}$. Therefore, using the fact that the exponent of $  \Psi_1\big|_{\refsol}$,  given by $iJ\theta -\mathcal{E}\tau$, evolves linearly as  $(J\omega-\epsilon) \tau$, \sref{finone} can be rewritten  finally as
\beq{
\Psi_1\,\Psi_2 \big|_\mathbb{X} &= (-1)^{\mathcal{P}}\left(\Psi_1\big| _{\bsc{\refsol (0 )}}\right) ^2  \exp\left( iJ( \iDelta\theta^{\mathbb{X}}+\iDelta\theta^{\tilde{\mathbb{X}}}) -2i \mathcal{P}\sigma_{12}-2\mathcal{E}\tau_{12} -(J\omega-\mathcal{E}) (\tau_f -\tau_i) \right) \nn\\
&=(-1)^{\mathcal{P}}\frac{\left(\Psi_1\big| _{\bsc{\refsol (0)}}\right) ^2 e^{iJ( \siDelta\theta^{\mathbb{X}}+\siDelta\theta^{\tilde{\mathbb{X}}})}}{(z_1-z_2)^{\mathcal{E}+\mathcal{P}}(\bar{z}_1-\bar{z}_2)^{\mathcal{E}-\mathcal{P}}}\exp\left(-(J\omega-\mathcal{E}) (\tau_f -\tau_i)\right)\period \label{finresult2pt}
}

Several important remarks are in order. First, the contribution from the reference solution $\left(\Psi_1\big| _{\bsc{\refsol (0 )}}\right)^2$ controls the  normalization
of  the  two point function.  
Once we properly normalize the  two point function  by determining the value of $\Psi_1 \big| _{\bsc{\refsol (0 )}}$,  the higher-point correlation functions, when computed with the same algorithm, will be properly normalized.  
Second, the dependence on the worldsheet coordinates  $z_1$  and $z_2$ 
 disappears,  together with the sign factor $(-1)^{\mathcal{P}}$,   if the solution satisfies the Virasoro conditions  $\mathcal{E}=\mathcal{P}=0$. 
This confirms the equivalence of the marginality of the vertex operator and the absence of the worldsheet coordinate dependence in the correlation functions discussed  in \cite{BuchTseytlin}. 
Third, note that the form of the dependence on $\tau_i$ and $\tau_f$ is such that 
it  precisely cancels  with the contribution of  the action $S[\theta]\big|_{\tau_i}^{\tau_f}$ constructed in terms of the action-angle variables\fn{In the case of 
 the GKP string, this turns out to hold even when the action is evaluated 
 on the solution expressed in terms of the embedding coordinates. For 
 details, see section 4.2, Appendices D an E.} . 
 Finally,  the expression (\ref{finresult2pt}) tells us that the {\it spacetime} dependence of the two point function, the property of great  interest, comes  from the ``phase shift"  $e^{iJ( \siDelta\theta^{\mathbb{X}}+\siDelta\theta^{\tilde{\mathbb{X}}})}$.

Summarizing, we have established  a simple procedure for computing
 the contribution
 of each  vertex operator in any correlation functions. It consists of the 
 following steps:  Find the behavior of 
 the saddle point solution in the vicinity of the vertex operator in terms of the 
 local cylinder time, translate it to the origin,  and find the global transformation 
 which produces this behavior from that of the  reference solution $\refsol$. 
Then using the master formula we obtain the shift of the angle variables and 
 hence the contribution of the wave function at that point. 
\subsection{Application to the case of elliptic  GKP strings}
To demonstrate the power and ease of our method developed above, 
 let us apply it to the computation of  the two point function for general elliptic GKP strings, \ie  without assuming  the large spin limit. 
 In this case, the reference solution is given by 
\begin{align}
\refsol &=\matrixii{X^{{\rm ref}}_+}{X^{{\rm ref}}}{\Xbar^{{\rm ref}}}{X^{{\rm ref}}_-} 
 = \matrixii{e^{-\theta(\tau)} \cosh \rho(\sig)}{e^{\phi(\tau)} \sinh \rho(\sig)}{e^{-\phi(\tau)} \sinh \rho(\sig)}{e^{\theta(\tau)} \cosh \rho(\sig)}  \comma \label{refsolelGKP}
\end{align}
where $\theta(\tau) = \kappa \tau, \phi(\tau) = \omega \tau$ and 
$\sinh \rho(\sig)$ can be expressed in terms of a Jacobi elliptic function, the 
 detail of which will not be important for us for the essential part of the discussion 
 below\fn{Nevertheless, when we elaborate on some details of the evaluation 
 of the two point function in Appendix E, we will need the explicit form 
 of  the elliptic GKP string solution.} . The only information 
 we need in this subsection is that $\sinh\rho(\sig)$ is a $2\pi$-periodic odd function which vanishes at $\sig=0$.  
As was already explained in section 3.1, we should refer to the $\sig=0$ point of the string when we discuss the emission and the absorption
 of the string by the vertex operators. 
Then the reference solution above starts at $\tau=-\infty$ from the origin 
$(0,0)$ of the boundary of $AdS_3$ and reaches the horizon at $\tau=\infty$. 

We now make a global transformation of the form 
\begin{align}
\mathbb{X} &= \matrixii{X_+}{X}{\Xbar}{X_-} 
 = V_L \refsol V_R 
\end{align}
such that the $\sig=0$ locus of the new solution $\mathbb{X}$ starts from $(0,0)$ and ends at  $(x_0, \xbar_0)$,  both on the boundary (see \figref{bending}). Such a solution plays the role of the saddle point configuration
 for the two point function $\langle V(0,0) V(x_0,\xbar_0) \rangle$. 
The required conditions are
\begin{align}
x\big|_{\sig=0,\tau=-\infty}&= {X \over X_+}\bigg|_{\sig=0,\tau=-\infty}
 = 0 \comma \qquad 
\xbar\big|_{\sig=0,\tau=-\infty} = {\Xbar \over X_+}\bigg|_{\sig=0,\tau=-\infty} = 0
\comma \\
x\big|_{\sig=0,\tau=+\infty} &= {X \over X_+}\bigg|_{\sig=0,\tau=+\infty}
 = x_0 \comma \qquad 
\xbar\big|_{\sig=0,\tau=+\infty} = {\Xbar \over X_+}\bigg|_{\sig=0,\tau=+\infty} = \xbar_0 \period
\end{align}
The  general form of  $V_L$ and $V_R$ which achieve these conditions can be
easily found to be 
\begin{align}
V_L&= V_L^{(1)} \equiv \matrixii{a}{{1\over a \xbar_0}}{0}{{1\over a}} 
\comma \qquad V_R =V_R^{(1)} \equiv  \matrixii{a'}{0}{{1\over a'x_0}}{{1\over a'}} 
\comma \label{LRtwopoint}
\end{align}
where $a$ and $a'$ are arbitrary  parameters. This set of transformations 
 consist of dilatation, rotation and special conformal transformation. 
It does not involve a translation  because   the $\tau=-\infty$ end 
 of the string is tied to the origin $(0,0)$. 
Therefore we may use these forms in 
 the general formulas  given  in (\ref{deltaphiinf})
 and (\ref{deltaphizero})  to obtain the following  finite shifts 
of the angle variables:
\begin{align}
\iDelta\phi_{\infty}^\mathbb{X} &= -i \log {a'}^2 
\comma \qquad \iDelta \phi_{0}^\mathbb{X} = i \log a^2 \period 
\label{shiftphiX}
\end{align}

Next we compute  the angle variables associated with 
 the vertex operator $V(x_0, \xbar_0)$ evaluated on $\mathbb{X}$. According
 to the rule developed previously, all we have to do is find the global symmetry 
 transformation which produces  the configuration, 
\beq{
\tilde{\mathbb{X}}(\tau\comma\sigma) = T^{-1}_{x_0, \xbar_0}\mathbb{X}(-\tau\comma-\sigma)= T^{-1}_{x_0, \xbar_0}V^{(1)}_L \refsol(-\tau\comma-\sigma) V^{(1)}_R\comma
}
from $\refsol(\tau\comma\sigma)$. 
Under the reversal of the coordinates $(\tau\comma\sigma)  \rightarrow (-\tau\comma-\sigma)$, 
the  variables $\theta(\tau)$, $\phi(\tau)$ and $\sinh\rho (\sigma)$ in $\refsol$ flip sign, 
and this leads to the interchange 
$X^{{\rm ref}}_+ \leftrightarrow X^{{\rm ref}}_-, X^{{\rm ref}}\leftrightarrow -\Xbar^{{\rm ref}}$. 
This is effected by the global transformation of the form 
\begin{align}
\refsol (-\tau\comma -\sigma ) &=  V_L^{(2)} \refsol (\tau\comma\sigma )V_R^{(2)}  \comma \qquad 
 V_L^{(2)} = i \sig_2 \comma \quad V_R^{(2)} = -i \sig_2 \period 
\end{align}
As for the translation\footnote{One might wonder at first sight that in the case of 
the spinning string the vertex operators $V(0,0)$ and $V(x_0, \xbar_0)$ 
  carry opposite spin and hence they are not simply related by 
 a translation. This suspicion is  unfounded. In the case where  the spinning string 
 emanates and lands on the same plane, the direction of the spin, as seen 
 from the respective vertex insertion point, is actually the same in the local 
 coordinates and  the vertex operators are related by a simple translation. 
}
 $T^{-1}_{x_0, \xbar_0}$, it is achieved by 
 the matrices $V^{(3)}_L= V_L^{tr}(-\xbar_0)$ and $V^{(3)}_R = V_R^{tr}(-x_0)$, 
 where  $V_L^{tr}(\al)$ and $V_R^{tr}(\albar)$ are given in (\ref{translation}). 
Altogether, we have 
\begin{align}
\tilde{\mathbb{X}}(\tau) &= V_L\refsol V_R \comma 
\end{align}
 where 
\begin{align}
V_L &= V_L^{(3)} V_L^{(1)}V_L^{(2)} =\matrixii{-{1\over a \xbar_0}}{a}{0}{-a \xbar_0} 
  \comma \\
V_R &= V^{(2)}_R V^{(1)}_R V_R^{(3)}
=\matrixii{-{1\over a'x_0}}{0}{a'}{-a'x_0}  \period 
\end{align}
Applying the general formulas for the shifts of the angle variables we obtain 
\begin{align}
\iDelta \phi^{\tilde{\mathbb{X}}}_{0} &= 
i\log \left( {1\over a^2 \xbar_0^2} \right)
= - i  \log (a^2 \xbar_0^2)  \comma \label{shiftphiXtilL}\\
\iDelta \phi^{\tilde{\mathbb{X}}}_{\infty} &= -i \log 
\left( {1\over {a'}^2 x_0^2} \right)
= i \log (a'^2 x_0^2)  \period \label{shiftphiXtilR}
\end{align}

Finally, plugging  the results (\ref{shiftphiX}), (\ref{shiftphiXtilL}) and (\ref{shiftphiXtilR}) into the formula \sref{finone},  we obtain the contribution from the wave functions
 as 
\begin{align}
\Psi_1 \Psi_2 \big|_\mathbb{X}
 &=\exp \left( S_{0}\log (\bar{x}_{0}^2)-S_{\infty}\log (x_0^2)\right)\Psi_1 \big| _{\bsc{\refsol (\tau_i )}}\Psi_1 \big| _{\bsc{\refsol (-\tau_f )}}   \nn\\
&=\frac{\Psi_1 \big| _{\bsc{\refsol (\tau_i )}}\Psi_1 \big| _{\bsc{\refsol (-\tau_f )}}}{x_0^{(\Delta -S)}\bar{x}_0^{(\Delta+S)}} \comma 
\end{align}
where we substituted $S_{\infty}=-R/2= -(\Delta -S)/2\comma S_{0}=L/2=(\Delta + S)/2$. 
As in \sref{finresult2pt}, we can further express the reference wave functions 
at $\tau_i$ and $-\tau_f$ in terms of the value at $\tau=0$. This gives 
\beq{
\Psi_1 \big| _{\bsc{\refsol (\tau_i )}}\Psi_1 \big| _{\bsc{\refsol (-\tau_f )}}
&= \exp \left( \left(\sum _i S_i {\del \calE \over \del S_i}  \right) ( \tau_f-\tau_i)\right) 
 \left(\Psi_1 \big| _{\bsc{\refsol (0 )}}\right) ^2 \period
}
Note that the front factor is precisely of the form 
$\exp\left(S[\phi]\big|^{\tau_f}_{\tau_i}\right)$, where $S[\phi]$ denotes  the action 
constructed in terms of the action-angle variables namely $S[\phi] = 
\int_{\tau_i}^{\tau_f} d\tau (\sum_i S_i \del_\tau \phi_i -\calE) $, where $\calE$ actually vanishes for our solution. 
Therefore this cancels precisely with the contribution of the action 
$\exp\left(-S[\phi]\big|^{\tau_f}_{\tau_i}\right)$ and we obtain 
%
\beq{
\Psi_1 e^{-S[\phi]}\Psi_2 \big|_\mathbb{X}&=\frac{\left(\Psi_1 \big| _{\bsc{\refsol (0 )}}\right) ^2}{x_0^{(\Delta -S)}\bar{x}_0^{(\Delta+S)}}\period\label{finalelgkp}
} 
Upon setting $\Psi_1 \big| _{\bsc{\refsol (0 )}}$ to unity, we get the canonically 
 normalized two point function for a spinning string.  
In Appendix E we will demonstrate  explicitly that 
  this result holds  even when we use the action $S[X]$  constructed in the embedding coordinates.  This  shows  that the possible extra contribution 
discussed in Appendix D can be ignored for the GKP strings. 

\section{Three point functions  for LSGKP strings }
Having developed  a  powerful  method for evaluating  the contribution of the vertex
 operators, we are now ready to complete our  calculation of the three point functions 
 for the LSGKP strings explicitly. 
\subsection{Contribution of the divergent part of the area}
Let us begin with the evaluation of the divergent part of the area, which was  given in (\ref{decompA}) as  $A_{div} = 4 \int d^2z \sqrt{p\pbar}$. 
In order to see clearly how this contribution cancels exactly with 
 the one from the vertex operators, it is useful  to express it as contour integrals using the Riemann bilinear identity. In this regard note  that $A_{div}$ can be written just like $A_{reg}$ in (\ref {Areg}), namely 
\begin{align}
A_{div} &=2i \int \sqrt{p(x)}dz \wedge \sqrt{\pbar(\zbar)} d\zbar 
 = 2i  \int \lam dz \wedge \omega\comma 
\end{align}
where $\lam = \sqrt{p(z)}$ as before and  $\omega =  \sqrt{\pbar(\zbar)} d\zbar$, which is already a closed form. Thus we can make use of 
 the generalized Riemann bilinear identity. Since the derivation, fully described in our previous paper, is lengthy, we do not repeat it here. 
The result  is\fn{In the published version of our previous paper\cite{Kazama:2011cp},  when reexpressing the quantity $\Lambda(z_0)-\Lambda(\widehat{z}_0)\equiv \int ^{z_0}\lambda dz -\int ^{\widehat{z}_0}\lambda dz$ in terms 
 of a contour integral over $\lambda$, we inadvertently made an error and 
 took the contour to be along $d$.  When the analytic 
 domain on which $\Lambda$ is defined is correctly  taken into account, 
 the contour should be along $d+C$ instead. This revision 
 introduces  an additional contribution $\jumpp{C,C}$ and the term $-\sum_{i \ne j} 
\jumpp{C_i, C_j}$ in the formula  (3.38) of \cite{Kazama:2011cp} should be 
 replaced by $+\jumpp{C_i, C_i}$.  Such a correction, shown in the erratum\cite{Kazama:2011cp},  
  is taken into account in 
 deriving the expression shown  in \sref{Adivorig}.} 
\begin{align}
A_{div} &= i \biggl[  2 (\jumpp{C,d} -\jumpp{d,C})
+\sum_{i } \jumpp{C_i,C_i} + 2 \sum_j ( \jumpp{C_j,\ell_j} -\jumpp{\ell_j,C_j} ) \biggr] \nn\\
& - 2i \sum_i \oint_{C_i} \sqrt{\pbar} d\zbar \int_{z_i^\ast}^{z_i}  
\sqrt{p} dz' \comma  \label{Adivorig}
\end{align}
where  
\begin{align}
\jumpp{A,B} &\equiv \int_A \sqrt{p}dz  \int_B \sqrt{\pbar} d\zbar
 \period \label{jumpp}
\end{align}
The contours $C, d, C_i$ and $\ell_i$ are recalled in \figref{contourdj} for 
completeness but except for $C_i$  we will not need them directly as we 
 rewrite this formula into more convenient form below. 
The expression above  is essentially identical to the general formula given in the equation (3.38) of \cite{Kazama:2011cp} for $A_{reg}$ for an $N$ point function.
\nullify{
}
There are, however,  two differences which stem from the replacement $ud\zbar + vdz \rightarrow \sqrt{\pbar} d\zbar$.
(For this reason  we use the notation \sref{jumpp}, which is slightly different 
 from the one used in our previous paper \cite{Kazama:2011cp}.)   First, a constant contribution
$\pi (N-2)/12$ present  in $A_{reg}$, which originates from the singularity of $v$ from around  the zeros of $p(z)$, is absent since here $\sqrt{\pbar}$ is not singular at those zeros.  
Contrarily, the contribution  from a small circle around $z_i$ 
of the form $\oint_{C_i} \omega$  vanished for $\omega = ud\zbar + vdz $ previously  but it  cannot be neglected  here  as   $\omega =\sqrt{\pbar}d\zbar$ has  poles  at $z_i$.  

Now we rewrite (\ref{Adivorig}) into a more convenient form. By expressing
 the contours $C$ and $d_i$  in terms of  basic ``components" such as 
 $C_i, l_i$ and $d$, and using the property  $\jumpp{C_i,C_j}=\jumpp{C_j,C_i}$
 valid  in the present  case, the contours in \sref{Adivorig} can be reassembled into the following form:
\begin{align}
A_{div} &= i \sum_i (\jumpp{C_i,d_i} -\jumpp{d_i, C_i}) 
 + i \sum_i \jumpp{C_i,C_i} - 2i  \sum_i \oint_{C_i} \sqrt{\pbar} d\zbar \int_{z_i^\ast}^{z_i}  
\sqrt{p} dz'  \period
\end{align}
This turned out to be the most suitable  form, since the terms in the first sum 
 precisely cancel against 
 the divergent contributions coming from the vertex operators, 
 as we shall show in section 5.2.  The double integrals in the last sum are easy to evaluate. Since these integrals are performed within the small vicinity of $z_i$, 
 we may use the  approximate form
\begin{align}
\sqrt{p(z)} &= {\delta_i \over z-z_i} + \calO ((z-z_i)^0) \comma 
\end{align}
and its complex conjugate for $\sqrt{\pbar(\zbar)}$. Taking $C_i$ to be 
 a circle of infinitesimal radius $\ep_i$ around $z_i$, the double integral reduces to 
\begin{align}
\oint_{C_i} \sqrt{\pbar} d\zbar \int_{z_i^\ast}^{z_i}  
\sqrt{p} dz'   &= \oint_{C_i} d\zbar 
\left( {\bar{\delta}_i \delta_i \over \zbar-\zbar_i} 
\ln {z-z_i \over z_i^\ast -z_i} \right) + \calO(\ep_i)  \period
\end{align}
This is easily evaluated by introducing the local angle parameter $\theta_i$ around $z_i$ as $i\theta_i = \ln (z-z_i)/(z_i^\ast -z_i)$ and we get 
a finite result 
\begin{align}
\oint_{C_i} \sqrt{\pbar} d\zbar \int_{z_i^\ast}^{z_i}  
\sqrt{p} dz'   &= 2\pi^2 \bar{\delta}_i \delta_i 
\end{align}
This term turns out to exactly cancel  against the term $\jumpp{C_i, C_i}$. 
Thus, we finally obtain the following expression for the divergent part of the area in terms of the contour integrals along $d_i$ and $C_i$:
\beq{
A_{div} &= i \sum_i (\jumpp{C_i,d_i} -\jumpp{d_i, C_i})\period \label{div-final}
}
We shall  show in the next subsection that the same expression emerges  from the calculation of the wave functions,  with opposite sign,  and they will cancel. 
 Thus, we need not perform the complicated integrals  along  $d_j$ appearing in  \sref{div-final}.
\subsection{Contribution of the vertex operators}
The remaining task is to compute the contribution of the vertex operators, or 
 the wave functions, and show that it combines with the divergent part of the 
 area to produce a finite result.  In what follows, we often refer to the results 
 obtained in our previous paper, the essence of which is reviewed in 
section 2. For additional details, see \cite{Kazama:2011cp}. 

Let $\psi^L_a$ and $\psi^R_\adot$ be the normalized solutions of the 
left and the right  auxiliary problems where $p(z)$ is given by (\ref{pz3pt}), 
which carries three double pole singularities of LSGKP type. 
If we can solve for  $\psi^L_a$ and $\psi^R_\adot$ globally, we can 
 obtain the saddle point configuration for the three point function 
 of the LSGKP strings from the reconstruction formula (\ref{reconstformula}), 
which we display again:
\begin{align}
\matrixii{X_+}{X}{\Xbar}{X_-} _{a,\adot} &= 
(\psi^L_a, \psi^R_\adot) \equiv \psi^L_{1,a} \psi^R_{\onedot,\adot} 
 + \psi^L_{2,a} \psi^R_{\twodot,\adot} \period \label{reconstX}
\end{align}
In practice, this is not possible.  However, we can express  the behavior of such 
  saddle point solution near each of the singularities. 
First, near the singularity at $z_i$, we can obtain from (\ref{gaugetrA}) and (\ref{psitwoeq}) the basic ``plus-minus" eigensolutions   $i_\pm(\xi)$ 
of  the auxiliary linear problem with the spectral parameter $\xi$, 
 corresponding to the eigenvalues $e^{\pm i \phat_i(\xi)}$ of the  monodromy matrix $M_i$. These plus-minus eigenfunctions were defined only up to $\xi$-independent overall factor  in our previous paper. However, to analyze the asymptotic behavior of the solution in detail  we need to adopt a definite normalization. Here we choose the normalization  such that the $i_\pm$ solutions asymptote to the following forms near the vertex 
insertion points:
\begin{align}
1_{\pm}&\sim \varphi^{1} _{\pm} \comma \quad 2_{\pm}\sim \mp \varphi^{2} _{\mp} \comma \quad 3_{\pm}\sim  \varphi^{3} _{\pm} 
\comma \label{def-o-pm}\\
\varphi ^{i}_+(\xi) &= {1\over \sqrt{2}} e^{i\kappa _i (\xi^{-1}w -\xi \wbar)}
 \vecii{ -i e^{(w-\wbar)/8}}{ e^{-(w-\wbar)/8} } \comma \\
 \varphi ^{i} _- (\xi) &= {1\over \sqrt{2}} 
e^{-i\kappa _i (\xi^{-1}w -\xi \wbar)}
 \vecii{- e^{(w-\wbar)/8} }{ i e^{-(w-\wbar)/8}}
\period  \label{ipim}
\end{align}
Here $w$ is the local cylinder coordinate given by  $w =\tau+i\sig = \ln (z-z_i) $. Note that  the definition of $2_{\pm}$ is slightly different from that of 
$1_{\pm}$ and $3_\pm$, reflecting the different  behavior of $\sqrt{p(z)}$ near $z_i$  on the first sheet, namely $\sqrt{p(z)} \simeq 
i\kappa_i/2(z-z_i)$ for $i=1,3$ and $\sqrt{p(z)} \simeq 
-i\kappa_i/2(z-z_i)$ for $i=2$.  
$i_\pm(\xi)$ above are normalized under the $SL(2)$ invariant 
 product as $\slprod{i_+}{i_-} =-\slprod{i_-}{i_+}=1$. 
From these $\xi$-dependent solutions,  the  solutions to the original left and right  auxiliary linear problems are obtained as 
\begin{align}
i_\pm^L &= i_\pm (\xi=1) \comma \qquad i_\pm^R = e^{-i\pi/4} \matrixii{0}{-i}{1}{0} i_\pm(\xi=i) \period \label{ipmLR}
\end{align}
Since they form a  complete basis, we can expand the unknown global 
 solutions $\psi^L_a, \psi^R_\adot$  in the following way:
\begin{align}
\psi^L_a &= \slprod{\psi_a^L}{i_-^L} i_+^L - \slprod{\psi_a^L}{i_+^L}i_-^L
\comma \qquad 
 \psi^R_\adot = \slprod{\psi^R_\adot}{i_-^R} i_+^R
 - \slprod{\psi^R_\adot}{i_+^R}i_-^R \period \label{exppsiLR}
\end{align}
Although the coefficients,  such as $\slprod{\psi_a^L}{i_-^L}$, 
cannot be computed directly, these are  useful representations  of $\psi^{L,R}$ 
 in the vicinity of the insertion points. Plugging (\ref{exppsiLR}) in the 
 formula (\ref{reconstX}) and using (\ref{ipim}) and (\ref{ipmLR}), 
 we obtain the following solutions in the vicinity of $z_i$:
\begin{align}
X_+ & \simeq  e^{\kaphat_i \tau} \urm (
 \ulp \sh_i -\ulm \ch_i) 
+ e^{-\kaphat_i \tau} \urp ( \ulm \sh_i -\ulp  \ch_i ) \comma \label{solXplus} \\
X_- & \simeq  e^{\kaphat_i \tau} \drm  ( 
\dlp \sh_i-\dlm  \ch_i ) 
+  e^{-\kaphat_i \tau} \drp  ( \dlm \sh_i -\dlp  \ch_i )  \comma   \\
X&\simeq  e^{\kaphat_i \tau} \drm  ( \ulp \sh_i -\ulm  \ch_i 
) 
+ e^{-\kaphat_i \tau} \drp  ( \ulm \sh_i-\ulp  \ch_i )  \comma   \\
\Xbar &\simeq  e^{\kaphat_i \tau} \urm (
 \dlp \sh_i -\dlm  \ch_i ) 
+  e^{-\kaphat_i \tau} \urp ( \dlm \sh_i -\dlp  \ch_i )   \period   \label{solXbar}
\end{align}
In the above formulas, to make the expressions compact,  we have  introduced the 
following abbreviations:
\beq{
\ulpm &\equiv \slprod{\psi_1^L}{\hatL{i}_\pm}
\comma \qquad \dlpm \equiv \slprod{\psi_2^L}{\hatL{i}_\pm} \comma 
\quad \hatL{i}_\pm \equiv {1\over \sqrt{2}}(\pm i_+^L + i^L_-) \comma \label{hatipm}\\
\urpm &\equiv \slprod{\psi_\onedot^R}{i^R_\pm} 
\comma \qquad \drpm \equiv \slprod{\psi_\twodot^R}{i^R_\pm} 
\comma \\
\ch_i &\equiv \cosh \kaphat_i \sig \comma \qquad \sh_i \equiv \sinh \kaphat_i \sig
\comma\\
\hat{\kappa}_{1,3}&=\kappa_{1,3}\comma\qquad\qquad \hat{\kappa}_2=-\kappa_2\period
}
Note that $\hat{i}_{\pm}^{L}$ defined above satisfy  the same normalization condition as $i_{\pm}^{L}$, that is,  $\slprod{\hat{i}_{+}^{L}}{\hat{i}_{-}^{L}}=1$.
The position $(\xsup{i}, \xbarsup{i})$ on the boundary 
 from which the $i$-th string state emanates  at the local cylinder time $\tau=-\infty$ 
 can be expressed easily in terms of the coefficients $\prone$'s and $\prtwo$'s. 
Taking into account that $\kaphat_i =\kappa_i >0$ for $i=1,3$ while 
$ \kaphat_2 = -\kappa_2 <0$, we obtain 
\begin{align}
x^{(i)} &= {X \over X_+} \bigg|_{\tau=-\infty, \sig=0} 
 =\bracetwo{ \drp / \urp\quad \mbox{for $i=1,3$} }{ \drm / \urm
\quad \mbox{for $i=2$}} \comma \label{xsupi}\\
\xbar^{(i)} &= {\Xbar \over X_+} \bigg|_{\tau=-\infty, \sig=0}
=\bracetwo{ \dlp / \ulp \quad \mbox{for $i=1,3$} }{ \dlm / \ulm 
\quad \mbox{for $i=2$}} \period \label{xbarsupi} 
\end{align}

We now show, by applying the method 
 developed in the previous section,  that the contribution of the wave functions
 to the three point function can be expressed in terms of the 
 coefficients $\ulpm$ and $\urpm$. 
The first step is to translate the solution 
near $z_i$,  given in (\ref{solXplus})-(\ref{solXbar}),  by 
 the vector $(-\xsup{i}, -\xbarsup{i})$. This is effected  by a transformation 
 with the matrices $V_L^{tr}(-\xsup{i})$ and $V_R^{tr}(-\xbarsup{i})$
 given in Appendix C. 
For $i=1,3$,  after the translation we get 
\begin{align}
\tilde{\mathbb{X}} =& V_L^{tr}(-\xbarsup{i})\mathbb{X}(\tau) V_R^{tr}(-\xsup{i})
\nn\\
=&   e^{-\kappa_i \tau}\!\!
\matrixii{\!\!\! \urp (  \ulm \sh_i -\ulp  \ch_i) }{\!\! 0\!\! }{\!\! {\urp\over \ulp} \sh_i }{\!\! 0\!\!} \!\! +e^{\kappa_i \tau}\!\!\matrixii{\!\! \! \urm (\ulp \sh _i-\ulm \ch _i)} {\!\! \frac{\ulp \sh _i-\ulm \ch _i}{\urp}\!\!}{\!\!\!-\frac{\urm}{\ulp}\ch _i}{\!\!-\frac{\ch _i}{\urp \ulp}\!\!} \comma 
\end{align}
where the Schouten identity
\beq{\slprod{\chi_1}{\chi_2}\slprod{\chi_3}{\chi_4} 
 + \slprod{\chi_1}{\chi_4}\slprod{\chi_2}{\chi_3} + \slprod{\chi_1}{\chi_3}\slprod{\chi_4}{\chi_2} =0}
has been  utilized. This should be 
 compared with the reference solution around $z_i$ 
\begin{align}
\refsol =  \matrixii{e^{-\kappa_i\tau } \cosh (\kappa_i \sigma)}{e^{\kappa_i\tau }\sinh (\kappa_i \sigma)}{e^{-\kappa_i\tau } \sinh (\kappa_i \sigma)}{e^{\kappa_i\tau }\cosh (\kappa_i \sigma)}  \comma 
\end{align}
at  $\tau \simeq -\infty$. 
Then we can see that $\tilde{\mathbb{X}}$ and $\refsol$ are related by the $SL(2)_L
  \times SL(2)_R$  transformation of the form 
\begin{align}
\tilde{\mathbb{X}} &= V_L \refsol V_R   \label{calXtil13} \comma \\
V_L &= \matrixii{a}{b}{0}{{1\over a}} \comma \qquad 
V_R = \matrixii{a'}{0}{c'}{{1\over a'}} \comma  \label{VLVR13}\\
a&=i\ulp \comma \quad 
b=  - i\ulm \comma \quad 
a' =  i \urp \comma \quad c' =-i \urm \period
\end{align}
Applying the general formulas \sref{deltaphiinf} and \sref{deltaphizero}  for the shift of the angle variables, 
we immediately obtain 
\begin{align}
\iDelta \phi_{0}^{(i)} &= -i \log \left( {1\over a^2} \right) =
i \log\left( - (\ulp )^2  \right)\comma  \\
\iDelta \phi_\infty ^{(i)}  &= -i\log \left( { a'^2} \right) =-i \log 
\left( -(\urp)^2 \right) \period
\end{align}
We now repeat the similar analysis for $i=2$ case. Looking at the form of the 
solutions (\ref{solXplus})-(\ref{solXbar}), we see immediately that 
compared to the previous case we need to change the signs of the 
 superscripts, such as $\prone^\pm \rightarrow \prone^\mp$, 
 for all the coefficients. This change of the signs gives rise to additional $(-1)$ factors when applying the Schouten identity.   
In this way  we  obtain 
\begin{align}
&\tilde{\mathbb{X}} \!= e^{-\kappa _2 \tau}\!\!
\matrixii{\!\!\! \urm ( \ulp \sh_i -\ulm  \ch_i ) }{\!\! 0\!\! }{\!\! -{\urm\over \ulm} \sh_i }{\!\! 0\!\!} \!\! +\! e^{\kappa _2 \tau}\!\!\matrixii{\!\! \! \urp (\ulm \sh _i-\ulp \ch _i)} {\!\! \frac{\ulp \ch _i-\ulm \sh _i}{\urm}\!\!}{\!\!\!\frac{\urp}{\ulm}\ch _i}{\!\!-\frac{\ch _i}{\urm \ulm}\!\!} 
\period
\end{align}
The transformation from  the reference solution is again 
given by (\ref{calXtil13}) and  (\ref{VLVR13}), this time with the parameters 
\begin{align}
a&=i \ulmtwo \comma \quad 
b= i \ulptwo \comma \quad 
a' = - i \urmtwo \comma \quad c' =-i \urp \period
\end{align}
This gives  the shift of the angle variables of the form 
\begin{align}
\iDelta \phi_0^{(2)} &= -i \log \left( {1\over a^2} \right) =
i \log\left( - (\ulmtwo )^2  \right)\comma  \\
\iDelta \phi_\infty^{(2)}  &= -i \log \left( a'^2 \right)
 =-i \log \left( - (\urmtwo )^2 \right) \period
\end{align}
Altogether,  the contribution of the wave functions relative to that of the reference trajectory is given by 
\begin{align}
\exp \left( i\sum_{i=1}^3  S_0^{(i)} \iDelta \phi _0^{(i)} + S_{\infty}^{(i)} \iDelta \phi^{(i)}_\infty
\right) \comma 
\end{align}
 and is expressed solely in terms of the invariant products $\ulpone, \ulmtwo, 
\ulptri, \urpone, \urmtwo, \urptri$. 

Thus our main task will be to evaluate  these coefficients. 
As shown in  (\ref{xsupi}) and (\ref{xbarsupi}), they are related to 
 the positions of the vertex operators on the boundary. 
The quantities of essential significance  for the correlation functions are rather the 
 differences of these positions  $\xsup{i} -\xsup{j}$ and $\xbarsup{i} -\xbarsup{j}$. Let us express them  in terms of the above coefficients. 
For instance, $\xsup{1} -\xsup{2}$ is given by $(\drpone \urmtwo -\drmtwo \urpone)/\urpone \urmtwo$. By using the Schouten identity,  we can identify  the numerator as $-\slprod{1_+^R}{2_-^R}$ and 
 hence the expression is simplified to  $\xsup{1} -\xsup{2}=-\slprod{1_+^R}{2_-^R}/ \urpone \urmtwo$. In a similar fashion, one can express all  the  relative positions in terms of the $SL(2)$ invariant products 
 $\ulpone, \ulmtwo, 
\ulptri, \urpone, \urmtwo, \urptri$ and $\slprod{1_+^R}{2^R_-}, \slprod{1_+^R}{3^R_+}, 
\slprod{2_-^R}{3^R_+},\slprod{\hatL{1}_+}{\hatL{2}_-},\slprod{\hatL{1}_+}{\hatL{3}_+}, 
\slprod{\hatL{2}_-}{\hatL{3}_+}$.  These six relations in turn allow us 
 to express\footnote{ This is characteristic of the three point functions. 
 For  $N$-point functions, the number of relative positions is $N(N-1)$, while 
 the number of overlaps of the type $\slprod{\psi^R_\onedot}{i_\pm^R}$
 and $\slprod{\psi^L_1}{\hatL{i}_\pm}$  (where the choice of $+$ or $-$ is 
 unique  for each $i$) 
 is $2N$. These numbers match precisely 
 for $N=3$.}
 the six coefficients 
$\ulpone, \ulmtwo, 
\ulptri, \urpone, \urmtwo, \urptri$ in terms of the relative positions 
 and the overlaps of the local eigenfunctions $\slprod{i^{R,L}_\pm}{j^{R,L}_\pm}$. The result is given by 
\begin{align}
(\ulpone)^2 &= { {-\slprod{\hatL{1}_+}{\hatL{2}_-} \slprod{\hatL{3}_+}{\hatL{1}_+}  \over \slprod{\hatL{2}_-}{\hatL{3}_+} 
 }{(\xbarsup{2} -\xbarsup{3}) \over (\xbarsup{1}-\xbarsup{2})(\xbarsup{3}-\xbarsup{1})}  } \comma \\
(\ulmtwo)^2 &=  { -\slprod{\hatL{1}_+}{\hatL{2}_-} \slprod{\hatL{2}_-}{\hatL{3}_+}  \over \slprod{\hatL{3}_+}{\hatL{1}_+} } 
 {(\xbarsup{3} -\xbarsup{1}) \over (\xbarsup{1}-\xbarsup{2})(\xbarsup{2}-\xbarsup{3})}  \comma \\
(\ulptri )^2&= {- \slprod{\hatL{3}_+}{\hatL{1}_+} \slprod{\hatL{2}_-}{\hatL{3}_+}  \over \slprod{\hatL{1}_+}{\hatL{2}_-} } 
 {(\xbarsup{1} -\xbarsup{2}) \over (\xbarsup{3}-\xbarsup{1})(\xbarsup{2}-\xbarsup{3})}  \comma \\
(\urpone )^2&=  { -\slprod{1^R_+}{2_-^R} \slprod{3^R_+}{1_+^R}  \over \slprod{2^R_-}{3_+^R} } 
 {(\xsup{2} -\xsup{3}) \over (\xsup{1}-\xsup{2})(\xsup{3}-\xsup{1})}  \comma \\
(\urmtwo) ^2&=  { -\slprod{1^R_+}{2_-^R} \slprod{2^R_-}{3_+^R}  \over \slprod{3^R_+}{1_+^R} } 
 {(\xsup{3} -\xsup{1}) \over (\xsup{1}-\xsup{2})(\xsup{2}-\xsup{3})}  \comma \\
(\urptwo )^2 &=  { -\slprod{3^R_+}{1_+^R} \slprod{2^R_-}{3_+^R}  \over \slprod{1^R_+}{2_-^R} } 
 {(\xsup{1} -\xsup{2}) \over (\xsup{3}-\xsup{1})(\xsup{2}-\xsup{3})} 
\period
\end{align}
This is an important set of formulas. The left hand sides carry  the information 
 about how  the global solutions  $\{\psi^L_a, \psi^R_\adot\}$ 
overlap with the local  solutions near each insertion point.  These quantities 
 cannot be computed directly since we do not know the global solutions. 
This, however,  is converted 
 on the right hand side to the information about the relative positions 
 and the overlaps of the eigenfuntions at different insertion points. 
Furthermore, when one makes  proper identification of  the states $i_\pm$ with
the small solutions $s_i$ around $z_i$, one recognizes  that the ratios 
of their  invariant  products are computable using the results 
obtained in our previous paper.

First consider such a ratio present in $\urpone$.  Since $i^R_\pm$ type 
 solutions are constructed from $i_\pm(\xi=i)$ by an $SL(2)$ transformation as in (\ref{ipmLR}),  the 
 imaginary part of $\xi$ is positive and the proper
 identification is  $1_+\propto s_1 \comma 2_- \propto s_2 \comma 3_+\propto s_3 $. To derive the precise relation between $i_{\pm}$ and $s_i$, we recall the 
form of the small solutions $s_i(\xi)$ given in \cite{Kazama:2011cp}:
\beq{
s_i(\xi )&= \frac{e^{-i\pi /4}}{\sqrt{2}} \pmatrix{c}{e^{\alpha /2}\left(\pm\xi\del  S_{(i),odd}^{+}\right)^{-1/2}\\  e^{-\alpha /2}\left(\pm\xi\del  S_{(i),odd}^{+}\right)^{1/2}\left( 1\pm \frac{\delbar  S_{(i),even}}{\delbar  S^{+}_{(i),odd}}\right)}\exp \left[ \pm S_{(i),odd}^{+}\right] \period\label{siasymp}
}
Here $S_{(i),odd}^{+}$ is given by the power expansion in $\xi$ as 
\beq{
S^{+}_{(i),odd}&=\frac{1}{\xi} \int _{z_i^{\ast}} dz\sqrt{p} + \xi \left(\frac{1}{2}\int _{z_i^{\ast}} dz\{ \frac{1}{\sqrt{p}}(\del \hat{\alpha} )^2 -\del(\frac{1}{\sqrt{p}}\del \hat{\alpha}) \} +\int _{z_i^{\ast}} d\barz \sqrt{\bar{p}}e^{-2\hat{\alpha}}\right) +\calO(\xi^2) \comma 
}
where  $z_i^{\ast}$ is a point close to $z_i$ separated by $\epsilon_i$. By analyzing the behavior of \sref{siasymp} around $z_i$ and comparing it with \sref{def-o-pm}, we find that for ${\rm Im}\,\xi>0$, 
\beq{
1_+ = e^{i(\xi^{-1}-\xi )\kappa_1 \log \epsilon_1 }s_1 \comma\quad 2_-= e^{i(\xi^{-1}-\xi ) \log \epsilon_2 }s_2\comma\quad 3_+= e^{i(\xi^{-1}-\xi ) \log \epsilon_3 }s_3 \period \label{ident}
}
Therefore we can reexpress  the ratio in $\urpone$ in terms of the small solutions as 
\begin{align}
{ \slprod{1^R_+}{2_-^R} \slprod{3^R_+}{1_+^R}  \over \slprod{2^R_-}{3_+^R} }  &=e^{4\kappa_1 \log \epsilon _1}{\slprod{s_1}{s_2} \slprod{s_3}{s_1} \over \slprod{s_2}{s_3}}(\xi=i)\label{ratiomod}
\period
\end{align}
We recognize that the ratio on the right hand side of \sref{ratiomod} is nothing but the expression already encountered  in (\ref{ratioinvprod}) 
 in the computation of the regularized area and hence can be evaluated: 
In \cite{Kazama:2011cp}, the part regular at $\xi=0,\infty$ is determined  by the use of the Wiener-Hopf decomposition, while the singular part is given simply by the first two terms of \sref{ratioinvprod}, \ie 
  integrals of $\sqrt{p}$ and $\sqrt{\bar{p}}$ along the contour $d_1$. 
The situation is similar for $\urmtwo$ and $\urptri$. 

Next we consider the ratios appearing in $\ulpm$. We will evaluate the relevant invariant products by again relating them to those between the small solutions and then 
utilizing the results in our previous paper. 
However, there are two significant differences from the previous case 
for $\be^\pm_i$'s.  The first difference stems from the fact 
 that the eigenfunctions $i^L_\pm$ in the ratios are obtained at $\xi=1$. 
At this parameter point, there is no clear distinction between the small  and the big solutions since the eigenfunctions asymptotically behave as $\exp \left[ \pm\kappa_i(w -\bar{w})\right]$.  A simple way to solve this problem 
 is to take the limit from the region where  ${\rm Im}\, \xi >0$. Then we can identify the small solutions unambiguously as \sref{ident}.  Having specified the relation between $i_{\pm}$ and $s_i$, our next task is to evaluate the invariant products of $i_{\pm}$ by relating them to those of the small solutions.  At this point, the second 
 difference arises because the relevant invariant products 
 involve states other than $s_i$ as well. 
 For instance,  $\slprod{\hatL{1}_+}{\hatL{2}_-}
= \slprod{1^L_+ + 1^L_-}{-2^L_+ +2^L_-}/2$ contains $1_-$ and $2_+$, which 
 are not small solutions. Thus we cannot simply apply the results of our previous paper, where only the products between the small solutions were calculated. 
 To determine the relevant  invariant products, we take the following steps. 
First, the invariant products involving only the small solutions, $\slprod{1_{+}^{L}}{2_{-}^{L}},\slprod{2_{-}^{L}}{3_{+}^{L}}$ and $\slprod{3_{+}^{L}}{1_{+}^{L}}$, 
can be computed just like the corresponding quantities in the R sector. 
The only difference here is that  for $\xi=1$ the prefactors $\exp\left(i(\xi^{-1}-\xi )\kappa_i \log \epsilon_i \right)$ in \sref{ident} are simply unity. 
Next recall that in \cite{Kazama:2011cp} all the invariant products of $i_{\pm}$ are determined up to three unfixed constants, $c, A_2$ and $A_3$, which are connected to the normalization of $i_{\pm}$. 
 By using the results for the three invariant products obtained above, we can now completely determine these constants. Then, this information in turn can be used to determine the rest of the invariant products. For details of the calculation, we refer the reader to Appendix F.

We are now ready to present the complete results for the contribution of 
 the wave functions.  Before showing the explicit details, however, it is useful 
to recall  its general structure and discuss some of its features.  It is of the form 
\beq{\Psi _1\Psi_2\Psi_3 \big| _{\mathbb{X}}=\exp \left( i\sum_{i=1}^3  S_0^{(i)} \iDelta \phi _0^{(i)} + S_{\infty}^{(i)} \iDelta \phi^{(i)}_\infty\right) \prod_{i=1}^{3} \Psi \big|_{\bsc{\refsol (\log \epsilon_i)}}\period\label{wavetotal}}
As in the case of the two point function,  the contribution from the reference wave functions $\Psi \big|_{\bsc{\refsol (\log \epsilon_i)}}$  contains  divergences, 
 which however will cancel against the ones from the action. 
On the other hand, the first term, which is of the form of  the ``phase shift" 
expressed in terms of the angle variables, is completely finite. 
Let us elaborate on this point. 
Recall that this term was evaluated by rewriting it in terms of  the invariant products between the small solutions plus the prefactors shown in \sref{ratiomod}. 
As discussed in our previous work,  the invariant products between the small solutions at general $\xi$ are composed of the part regular at $\xi=0, \infty$ 
and the singular part. 
As for the L sector, the singular part of the products between the small solutions vanishes by setting $\xi=1$ and the prefactors are simply unity. Hence  the phase shift 
coming from the L sector is manifestly finite. On the other hand, for the R sector, the singular part does not vanish. It is given essentially by the contour integral connecting two singularities and produces the divergences proportional to $\log \epsilon_i$. These divergences  precisely cancel against the contribution from the prefactors given by $\exp (4\kappa_i\log \epsilon_i)$ in \sref{ratiomod}. Therefore, in total, the phase shift for the R sector is also finite.  To present the results in a convenient form, however, 
 we shall make  the following slight rearrangement of the divergent contributions. 
From the result of Appendix E, we can calculate 
the ratio $\Psi\big| _{\refsol(\log \epsilon_i)} /\Psi\big| _{\refsol(0)}$. This turns out to be exactly the inverse of the contribution from the prefactors in \sref{ratiomod}. 
Therefore it is convenient to cancel these $\log \epsilon_i$ divergences explicitly. 

Combining altogether,  the total contribution of the wave functions can be
 presented  in the following form:
 \beq{
\Psi _1\Psi_2\Psi_3\big| _{\mathbb{X}} =& \frac{C_{{\rm w.f.}}}{(x^{1}-x^{2})^{\ell _1^{-}+\ell_2 ^{-}-\ell _3^{-}}(x^{2}-x^{3})^{\ell _2^{-}+\ell_3 ^{-}-\ell _1^{-}}(x^{3}-x^{1})^{\ell _3^{-}+\ell_1 ^{-}-\ell _2^{-}}}\nn\\
&\times  \frac{\left(\Psi \big|_{\refsol (0)}\right)^3}{(\xbar^{1}-\xbar^{2})^{\ell _1^{+}+\ell_2 ^{+}-\ell _3^{+}}(\xbar^{2}-\xbar^{3})^{\ell _2^{+}+\ell_3 ^{+}-\ell _1^{+}}(\xbar^{3}-\xbar^{1})^{\ell _3^{+}+\ell_1 ^{+}-\ell _2^{+}}}\period\label{wave-final}
}
Here, the quantities $\ell_{i}^{\pm}$ are essentially the charges $L$ and $R$ carried by each of the vertex operator  defined by 
\beq{ \ell _{i}^{-} =  (\Delta^{(i)} - S^{(i)})/2 \comma 
\qquad \ell _{i}^{+} \equiv (\Delta^{(i)} + S^{(i)})/2 \period
}
The log of the coefficient $C_{{\rm w.f.}}$ is given by
\beq{
\log C_{{\rm w.f.}} = &H_{-}\left[h(x,\xi =i ) \right] +H_{+}\left[h(x,\xi =1 ) \right] \nn\\ &+ \frac{i\sqrt{\lambda}}{2}\sum_{j=1}^{3} \hat{\kappa}_j \left(\int _{d_j} \sqrt{p}dz -\int _{d_j} \sqrt{\bar{p}}d\barz \right) +\sum _j \ell _j^{+} \log \tilde{c} \comma\label{wave-final2}
 }
where the constant $\tilde{c}$ is of the form
\begin{align}
\tilde{c}= 1-\sqrt{\frac{\sinh(\pi (-\kappa_1+\kappa_2+\kappa_3))\sinh(\pi (\kappa_1-\kappa_2+\kappa_3))\sinh(\pi (\kappa_1+\kappa_2-\kappa_3))}{\sinh(\pi (\kappa_1+\kappa_2+\kappa_3))}} \period
\end{align}
Finally the functions  $H_{\pm}[\ast]$, which take a function as the argument, 
 are defined as 
\beq{
& H_{\pm}\left[f(x)\right] \! \equiv 2\sum_{j=1}^{3} \!  \ell^{\pm}_j f(\kappa_j )  -\left( \sum_{j=1}^{3}\ell^{\pm} _j\right)  f(\sum_{j=1}^{3}\frac{\kappa_j}{2})-( -\ell^{\pm} _1 +\ell^{\pm} _2 +\ell^{\pm} _3 ) f(\frac{-\kappa_1+\kappa_2+\kappa_3}{2})\nn\\
&\qquad\qquad -(\ell^{\pm} _1 -\ell^{\pm} _2 +\ell^{\pm} _3)f(\frac{\kappa_1-\kappa_2+\kappa_3}{2})-(\ell^{\pm} _1 +\ell^{\pm} _2 -\ell^{\pm} _3)f(\frac{\kappa_1+\kappa_2-\kappa_3}{2})\comma 
}
and the actual argument functions  $h(x,\xi=i)$ and $h(x,\xi=1)$ 
 are defined  in Appendix F.  Concerning the contributions for $\log C_{{\rm w.f.}}$, 
 the first and the third terms are from the R sector whereas the second and the fourth terms are from the L sector. 

Let us make  a comment  on some  important features of \sref{wave-final} and \sref{wave-final2}.  First, one should note that the integrals along $d_j$ present in \sref{wave-final2}, such as $\hat{\kappa}_j\int _{d_j}\sqrt{p}dz$, exactly cancel the whole contribution from the divergent part of the area \sref{div-final}. 
Second,  the form of the result \sref{wave-final} manifestly shows that the spacetime dependence of the three point function comes purely from the wave functions. 
This is as expected  because the remaining 
 contribution from the area of the worldsheet is invariant under the global symmetry
 transformations. 
\subsection{Final result  for the three point functions and their properties}
Now we combine all the contributions and obtain the final result for  the three point function.  
Let us recall the various contributions we have calculated. The finite part of the action is calculated in our previous paper. Using the function  $H_{-}$, it can be  expressed as 
\beq{
\log C_{{\rm area}} =-\frac{\sqrt{\lambda}}{2\pi} A_{{\rm fin}} =-\frac{7\sqrt{\lambda}}{12}+H_{-}\left[ K(x) \right]\period
}
The divergent part of the action is studied in section 5.1 and the result for this contribution is given by \sref{div-final}. Lastly the contributions from the wave functions are calculated in the previous subsection and the results are given by \sref{wave-final} and \sref{wave-final2}. 

When these contributions are put together, with the normalization condition $\Psi\big|_{\refsol}=1$  determined by the calculation of two point functions in section 4.2, 
 the structure constant of the three point function is obtained as 
\beq{
\log C^{{\rm LSGKP}}= -\frac{7\sqrt{\lambda}}{12} + H_{-}[\widetilde{K}(x)]+H_{+}[h(x,\xi =1) ] +\sum _j \ell _j^{+} \log \tilde{c}\comma \label{FINAL}
}
where $\tilde{K}(x)$ and $h(x,\xi =1) $ are given by 
\beq{
&\tilde{K}(x)\equiv K(x)+h(x,\xi = i)=\frac{1}{2\pi}\int _{-\infty}^{\infty}\! d\theta\, \frac{\cosh 2\theta}{\cosh \theta }\log \left( 1- e^{-4\pi x\cosh \theta }\right) \comma \\
&h(x,\xi=1 )=-\frac{1}{2}\log \left( 1-e^{-4\pi x}\right)  \period
}

It is intriguing to observe that a part of our final result  \sref{FINAL} 
 above has the same structure as the universal contribution from the 
$AdS_2$ part calculated in \cite{JW} for operators without Lorentz spins. 
To see this, we must actually subtract from the result of \cite{JW}  ``the regularized divergent contribution" of $S^{5}$  in order to extract the purely  $AdS_2$ part. 
 After such a subtraction, 
 the result of \cite{JW} becomes essentially the same as $H_{-}[\widetilde{K}(x)]$, which in our case comes solely from the right-charge,  \ie $\Delta-S$,  sector. 
From the technical point of view, this coincidence may not be so surprising, 
because both works are based on the same equation \sref{mSGeq} with essentially the same boundary condition. A notable difference is, however, that 
 while we explicitly take into account the contribution of the wave functions, 
 the work of \cite{JW} does not deal directly with such contribution. 
In their approach, such effects appear to be indirectly taken into account by the 
way the worldsheet divergence is regularized using a cut-off in the target space.
 It would be  important  to understand the precise relation between the contribution from the wave functions and the effect of such a cut-off in the target space, 
preferably from a physical point of view.  On the other hand,  there is 
 no counterpart of our left-charge sector in \cite{JW}, which comes from 
 the  presence of the spin in addition to the dilatation charge.  In general, 
 to get the complete dependence on the boundary coordinates 
 for  systems carrying quantum numbers other than the conformal dimension, 
one must consider the effect of the wave functions explicitly, as we have done. 

Another property of our final result worth mentioning is its behavior 
 under the limit where it is expected to reduce to a two point function. 
Specifically, consider the limit  in which $\kappa_3$ becomes zero with 
  $\kappa_2 \rightarrow \kappa_1$ at the same time. 
 One might think that such a limit is beyond the domain of validity of our large charge approximation. However, if we set $\kappa_3$ to zero exactly, the saddle point trajectory should become  the one for the two point  function and hence $C^{LSGKP}$ should  reduce precisely to unity 
if all our normalization procedures have been correct. 
What actually happens is that if we take the limit $\kappa_3 \rightarrow 0$ naively
 in \sref{FINAL}, we get $\log C^{{\rm LSGKP}}=-\sqrt{\lambda}/2$, which 
 is different from the expected value $0$. Let us trace the origin of this 
contribution.  As discussed in our previous paper, 
 this contribution comes from the following 
 integral on the worldsheet 
\beq{
\int d^2 z \del\delbar \alpha \period\label{boundaryintegralalpha}
} 
As the integrand is a total derivative, it can be reexpressed as the boundary contour integral  $\int(\del\alpha dz-\delbar \alpha d\barz) $. The boundary in this case consists of the three small circles with radius $\epsilon_{i}$ around the vertex insertion points $z_i$  and the large circle  at infinity. Using the behavior $\exp(2\alpha) \sim\sqrt{p\bar{p}}$ in the vicinity of the insertion points, one finds that the integral around 
each insertion point yields  $\pi$, which contributes 
 to the structure constant the value $-\sqrt{\lam}/2$. Since it is a constant 
 independent of $\kappa_i$,  the contribution  from around $z_3$ remained  in the
naive limit $\kappa_3\rightarrow 0$. Now the resolution of the puzzle is clear. 
The calculation sketched above is valid only when $\kappa_3 \gg \ep_3$. 
When we take $\kappa_3$ to zero, we must actually fix $\ep_3$ in order for 
 the singularity at $z_3$ to properly disappear.  If we do this, then there is no 
 contribution from around such a regular point and we obtain  $\log C^{{\rm LSGKP}} =0$ as desired. 
\section{Discussions}
In this paper, in order to complete the calculation of the three point function 
 for the LSGKP strings initiated in our previous work, 
we have developed a general method for evaluating the contribution 
 of the vertex operators for semi-classical heavy string states, which correspond   to 
operators with large quantum numbers  in $\mathcal{N}=4$ SYM.  
In this method,  we first construct the action-angle variables via the 
  Sklyanin's method and then use them to evaluate the wave functions,  which correspond to the vertex operators  through the state-operator correspondence. We have tested this method for the two point functions and then applied it to the three point function for the LSGKP
 strings. Combined with the result  of our 
 previous work, we obtained the complete answer for 
 the three point functions with the expected spacetime dependence. 

Let us now mention several directions for further research.  

Our present work has  shown that the action-angle variables constructed 
{\it \`a la} Sklyanin can be extremely useful. This makes it important 
 to construct such variables for a string in the full $AdS_5\times S^{5}$
 spacetime.  For the bosonic sector, this should not be so difficult\fn{Generalization of the Sklyanin's method to the case of $GL(N)$  is already discussed in \cite{Sklyanin}.}. 
 But  the extension to the fermionic sector,  which should 
be necessary  for quantum consideration,  may be nontrivial. 

As concerns the use of the action angle variables, 
it should  also be of great interest to reanalyze the calculation on 
 the gauge theory side in terms of such variables, as already  suggested in \cite{Gromov:2011jh}. 
This will help us  explore  the connection between our results on the string theory 
side and the new integrable structures recently discovered on the SYM side\cite{Escobedo, Escobedo2, Caetano, Gromov:2011jh, Gromov:2012vu, Serban:2012dr, Kostov:2012jr, Foda, Foda2, Kostov2, Morphism}. 

In order to facilitate the comparison between the calculations of the three  point 
 functions on the string side with those on the gauge theory side,  it is desirable to 
 generalize the result of the gauge theory side from  the $SU(2)$ sector to more general sectors, especially to the $SL(2)$ sector.  An attempt in  this direction is briefly discussed in the appendix of \cite{Gromov:2011jh}. However,  to obtain a general expression for three point functions of three long non-BPS operators, further detailed study will have to be performed.  
An alternative way to compare the results on two sides is to generalize our result for 
the string in \ads3 to the case in other spacetimes,  in particular to a string in
 $AdS_2\times S^{3}$, for which the corresponding analysis in the SYM theory 
 is easier.  Our general method of computing the contribution of the vertex 
 operators can be applied to such cases as well.   Since the action-angle variables for a string on $R\times S^{3}$ are  already constructed, it  should  actually be straightforward for a string in $AdS_2\times S^{3}$. 
However, the evaluation of the action may be technically  involved,  since the equations  we obtain after the Pohlmeyer reduction are more  complicated than for a string in \ads3.  The task  may be simplified  if we could calculate the contribution from the action directly from the coset connection without resorting to the Pohlmeyer reduction.  The direct evaluation of the action from the coset connection will also enable us to unify the calculation of the action and the vertex operators and it will  help us 
 uncover  the structure  underlying the entire  calculation. 

Another important project would be to evaluate the four point functions and 
understand their characteristic  properties, in particular  how the crossing 
 symmetry is realized.  Although qualitatively new features will come in, 
 our method of calculation based on  the global symmetry 
 transformation  should prove useful in such an investigation. 

We are currently pursuing some of these directions and we hope to report our 
 progress in the near future. 
\par\bigskip\noindent
{\large\bf Acknowledgment}\par\smallskip\noindent
The research of  Y.K. is supported in part by the 
 Grant-in-Aid for Scientific Research (B) 
No.~20340048, while that of S.K. is supported in part 
 by JSPS Research Fellowship for Young Scientists,   from the Japan 
 Ministry of Education, Culture, Sports,  Science and Technology. 
\appendix
\setcounter{equation}{0}
\def\apsection#1{\subsection*{{\bfall #1}}\addcontentsline{toc}{subsection}{#1}}
\section*{Appendices}
\addcontentsline{toc}{section}{Appendices}
\apsection{A:\quad Derivation of the  singularity structure 
 of  $\hat{p}(x)$}
\renewcommand{\theequation}{A.\arabic{equation}}
\renewcommand{\thefigure}{A.\arabic{figure}}
In this appendix, we give a derivation of the singularity structure of the quasi-momentum $\phat(x)$ quoted in (\ref{sing-euc}).  

Consider first the singularity at $x=1$ associated with that of the left-invariant Lax connection $J^r_z(x) =j_z/(1-x)$, where $j_z$ is a $2\times 2$ matrix  given by  $j_z = \mathbb{X}^{-1} \del \mathbb{X}$. 
 An important property of $j_z$ for the string in \ads3 is that 
it must satisfy the Virasoro condition $\Tr [ j_z j_z]=0$. This does not of course mean 
 that $j_z$ vanishes. It means that $j_z$ must be similar to a special Jordan block 
$P$, namely 
\begin{align}
j_z &= u P u^{-1} \comma \qquad P \equiv \matrixii{0}{1}{0}{0} 
\period \label{jzP}
\end{align}
Here $u $ is a $2\times 2$ matrix which depends on the worldsheet
 coordinate $z$. Note that $P$ is nilpotent and so is $j_z$.  To diagonalize 
 the monodromy matrix  to obtain $\phat(x)$, it is convenient to go to a 
 gauge  where the connection is not degenerate. For this purpose
 we consider the gauge  transformed Lax connection $\Jtil^r_z$ given by 
\begin{align}
\Jtil^r_z &= u^{-1} J^r_z u + u^{-1} \del u = {P \over 1-x} + A\comma
\end{align}
where $A \equiv u^{-1} \del u$. Near $x=1$,  
 the eigenvalues $\lam_\pm$ of $\Jtil^r_z$ are easily computed  as 
\begin{align}
\lam_\pm &= \pm \sqrt{{A_{21} \over 1-x} } + \calO((x-1)^0) \period
\label{lampm}
\end{align}
A square root type singularity appeared\footnote{This type of phenomenon 
commonly occurs when the dominant term is a Jordan block and the subdominant term  is non-degenerate. }.  The matrix element $A_{21}$ can actually 
be expressed in terms of $j_z$. To see this, consider $\del j_z$. Differentiating 
 (\ref{jzP}) 
 we easily get $\del j_z = (\del u u^{-1}) j_z - j_z (\del u u^{-1})$. Then, 
taking the nilpotency of $j_z$ into account, we get 
\begin{align}
\half \trace (\del j_z \del j_z) &=  \trace (  (\del u u^{-1}) j_z (\del u u^{-1}) j_z) 
\nn\\
&=  \trace  ( \del u P u^{-1} \del u P u^{-1}) = \trace (APAP) =  (A_{21})^2
\period \label{A21} 
\end{align}
From (\ref{lampm}) and (\ref{A21}), the quasi-momentum $\phat(x)$ 
near $x=1$ can be expressed as
\beq{
\hat{p}(x)&= \frac{2\pi \kappa _{+}}{\sqrt{1-x}}+O\left( (x-1)^0\right)\comma
}
with 
\beq{
\kappa_{+}=\frac{1}{2\pi i} \oint dz \left( \half \Tr \left( \del j_z \del j_z\right) \right)^{1/4} \period
}
For the behavior near $x=-1$, a similar  analysis of  $j_{\barz}$ leads to
\beq{
\hat{p}(x)&= \frac{2\pi \kappa _{-}}{\sqrt{1+x}}+O\left( (x+1)^0\right)\comma\\
\kappa_{-}&=\frac{1}{2\pi i} \oint dz \left(\half  \Tr \left( \delbar j_{\barz} \delbar j_{\barz}\right) \right)^{1/4}\period
}
\apsection{B:\quad Remarks on the derivation of the 
canonical variables $(z(\gamma_i), \hat{p}(\gamma_i))$}
\setcounter{equation}{0}
\renewcommand{\theequation}{B.\arabic{equation}}
\renewcommand{\thefigure}{B.\arabic{figure}}
The starting point of the construction of the action-angle variables described
  in section 3.2 was the existence of the canonical pair of variables $(z(\gamma_i), \hat{p}(\gamma_i))$.  Although the derivation of  the canonical Poisson brackets between these variables was  given sometime ago in \cite{DV2} for a string in 
 $R \times S^3$, we would like to make some clarifying remarks on this derivation as applied to our system, namely a string in \ads3. 

The first remark is that as we need not make the gauge-fixing of the target space 
 time coordinate as was done in \cite{DV2},  we may use the Poisson bracket 
 instead of  the Dirac bracket.  

The second remark is more important.  In the derivation given in \cite{DV2}, 
 the fact that the position of the poles of the Baker-Akhiezer function on 
 the spectral curve,  denoted here\footnote{In \cite{DV2} it was 
 denoted by $x_{\hat{\ga}_i}$. We use a simplified notation here.} 
  by $\ga_i$,  
are dynamical variables made out 
 of the string coordinates was not duly taken into account in some of the steps. 
 Consequently, although the conclusions were correct, the derivation was somewhat misleading. Below we show that by treating  the dynamical nature of $\ga_i$'s properly, 
 with a  change of  the order of arguments,   one can give a 
  satisfactory derivation. 

As  in \cite{DV2}, it is convenient to perform a similarity transformation 
 on the monodromy matrix $\Omega$ and its eigenvector $\vec{h}$ so  that the 
normalization condition for the components of $\vec{h}$ becomes simple. 
As discussed in section 3.3.2, the appropriate normalization vector for our problem 
 is  $\vec{n}=(1,\mu)^{\T}$ and the corresponding similarity transformation 
 is given by 
\beq{
\tilde{\Omega}(x)\equiv \pmatrix{cc}{1&\mu\\ 0&1}\Omega(x)\pmatrix{cc}{1&-\mu\\ 0&1} \equiv  \pmatrix{cc}{\tilde{\mathcal{A}}(x)&\tilde{\mathcal{B}}(x)\\ \tilde{\mathcal{C}}(x)&\tilde{\mathcal{D}}(x)} \period
}
The components of  $\tilde{\Omega}$ satisfy the 
same algebraic relations as  the components of $\Omega$, as  in  \cite{DV2}, 
 which read 
\beq{
\{ \tilde{\mathcal{B}}(x)\comma \tilde{\mathcal{B}}(x^{\prime})\}=&0\comma \label{apb-1}\\
\{ \tilde{\mathcal{A}}(x)\comma \tilde{\mathcal{B}}(x^{\prime})\}=&\left( \tilde{\mathcal{A}}(x)\tilde{\mathcal{B}}(x^{\prime})+\tilde{\mathcal{A}}(x^{\prime})\tilde{\mathcal{B}}(x)\right)\hat{r}(x\comma x^{\prime})\nn\\
&+\left( \tilde{\mathcal{A}}(x)\tilde{\mathcal{B}}(x^{\prime})+\tilde{\mathcal{D}}(x^{\prime})\tilde{\mathcal{B}}(x)\right)\hat{s}(x\comma x^{\prime})\comma\label{apb-2}\\
\{ \tilde{\mathcal{A}}(x)\comma \tilde{\mathcal{A}}(x^{\prime})\}=&\left( \tilde{\mathcal{B}}(x)\tilde{\mathcal{C}}(x^{\prime})-\tilde{\mathcal{B}}(x^{\prime})\tilde{\mathcal{C}}(x)\right)\hat{s}(x\comma x^{\prime})\period\label{apb-3}
}
Here  $\hat{r}(x,x^{\prime})$ and $\hat{s}(x,x^{\prime})$ are the coefficients of the Maillet's $r$-$s$ matrices\cite{Maillet}  given by
\beq{
\hat{r}(x,x^{\prime})&\equiv -\frac{2\pi}{\sqrt{\lambda}}\frac{x^2 +x^{\prime ^2}-2x^2 x^{\prime ^2}}{(x-x^{\prime})(1-x^2)(1-x^{\prime ^2})}\comma\\
\hat{s}(x,x^{\prime})&\equiv -\frac{2\pi}{\sqrt{\lambda}}\frac{x +x^{\prime }}{(1-x^2)(1-x^{\prime ^2})}\period
}
At the poles of the normalized Baker-Akhiezer vector, 
the components of $\Omtil$ satisfy the following relations\footnote{They follow 
 simply from the eigenvalue equation, the normalization condition and 
the value of $\trace \tilde{\Omega}$. }
\beq{
\tilde{\mathcal{B}}(\gamma _i)= 0\comma \quad \tilde{\mathcal{D}}(\gamma_i)=\tilde{\mathcal{A}}(\gamma_i )^{-1}=e^{ i\hat{p}(\gamma _i)}\period\label{apb-4}
}
 Now by studying these relations 
 at the positions of the poles $x=\ga_i, x'=\ga_j$  one can obtain  formulas necessary  to derive the canonical brackets for $(z(\gamma_i), \hat{p}(\gamma_i))$. 
However, the limit  such as $x \rightarrow \ga_i$ must be taken with care. 
Such substitutions must be done  {\it after } the computation of the Poisson brackets, since in obtaining  the relations \sref{apb-1}-\sref{apb-3} 
the quantities $x$ and $x'$ have been treated as non-dynamical  numbers. 
This was not duly taken into account in some of the procedures in \cite{DV2}. 
 For instance, the analysis in \cite{DV2} starts with the naive substitutions 
 of $x\rightarrow \ga_i$ an $x' \rightarrow \ga_j$ in \sref{apb-3}, including 
 those in the Poisson bracket on the left hand side. In this way the authors 
first derives the relation $\{ \tilde{\mathcal{A}}(x)\comma \tilde{\mathcal{A}}(x^{\prime})\}=0$. This is not justified. 

To derive the needed formulas properly, it is convenient to start from the 
analysis of \sref{apb-1} instead. 
Since $\tilde{\mathcal{B}}(x)$ has zeros at $\gamma_i$ and $\gamma_j$ $(i\neq j)$, it can be expressed  as $\tilde{\mathcal{B}}=(x-\gamma_i) \mathcal{B}^{\prime}$ or $\tilde{\mathcal{B}}=(x-\gamma_j) \mathcal{B}^{\prime\prime}$. 
 The functions  $\mathcal{B}^{\prime}$ and $ \mathcal{B}^{\prime\prime}$ 
are not known but what is important is that they have the properties  $\mathcal{B}^{\prime}(\gamma_i)\neq 0$ and $\mathcal{B}^{\prime\prime}(\gamma_j)\neq 0$. Then, \sref{apb-1} can be rewritten as
\beq{
&(x-\gamma_i )(x^{\prime}-\gamma_j )\{ \mathcal{B}^{\prime}(x)\comma \mathcal{B}^{\prime\prime}(x^{\prime})\} -(x^{\prime}-\gamma_j )\mathcal{B}^{\prime}(x)\{ \gamma_i \comma \mathcal{B}^{\prime\prime}(x^{\prime})\}\nn\\
&-(x-\gamma_i )\mathcal{B}^{\prime\prime}(x^{\prime})\{\mathcal{B}^{\prime}(x) \comma \gamma_j\} +\mathcal{B}^{\prime}(x)\mathcal{B}^{\prime\prime}(x^{\prime}) \{ \gamma_i \comma \gamma_j\}=0\period
}
Now at this stage  we can take  the limit $x\to \gamma_i$ and $x^{\prime}\to \gamma_j$. Then the first three terms vanish manifestly and from the last term 
 we obtain the relation 
\beq{
\{\gamma_i \comma \gamma_j\}=0\period\label{apb-6}
}
Next step is to  consider \sref{apb-2}. Here again, we should  substitute  
the expansion $\tilde{\mathcal{A}}(x) =\tilde{\mathcal{A}}(\gamma_i) + (x-\gamma_i)\mathcal{A}^{\prime}(x)$ as well as the ones for $\mathcal{B}^{\prime}$ and $ \mathcal{B}^{\prime\prime}$. Then similarly to the case of \sref{apb-1}, 
the limit $x\to \gamma_i$ and $x^{\prime}\to \gamma_j$ can be taken 
easily and, making use of the relation \sref{apb-6}, we can deduce 
the important relation 
\beq{
\{ \tilde{\mathcal{A}}(\gamma _i)\comma \gamma _j\} = \frac{4\pi }{\sqrt{\lambda}}\tilde{\mathcal{A}} (\gamma_i ) \frac{\gamma_i^2}{\gamma_i^2 -1} \delta _{ij}\period\label{apbimp-2}
} 
Finally, similar calculation for \sref{apb-3} leads to 
\beq{
\{\tilde{\mathcal{A}}(\gamma_i )\comma \tilde{\mathcal{A}}(\gamma_j )\}=0\period\label{apbimp-3}
}
 Both \sref{apb-6} and \sref{apbimp-2} are needed to obtain this result. 
The rest of the analysis is the same as in \cite{DV2} and one proves 
that $(z(\gamma_i), \hat{p}(\gamma_i))$ constitute a canonical pair  of variables. 

\apsection{C:\quad $SL(2,C)_L \times SL(2,C)_R$ transformations  as conformal transformations}
\setcounter{equation}{0}
\renewcommand{\theequation}{C.\arabic{equation}}
\renewcommand{\thefigure}{C.\arabic{figure}}
In this appendix, we make a summary of the actions  of the global 
symmetry transformations  which play crucial roles  in the main text. 
As we are interested in the saddle point configurations, which are in general 
 complex, we will not impose any reality conditions on the coordinates. 
Hence we will be concerned with $SO(4,C)$  which is isomorphic to 
 $G =SL(2,C)_L \times SL(2,C)_R$.  Below we  consider  these transformations from
 the point of view of conformal transformations. 

Define the $2\times 2$ matrix $\mathbb{X}$ and the action of $G$ as in 
(\ref{matrixg})-(\ref{sltransfg}), namely, 
\begin{align}
\mathbb{X} &\equiv \matrixii{X_+}{X}{\Xbar}{X_-} \comma \qquad 
\det \mathbb{X}=1 \comma \\
X_+ &= X_{-1} + X_4 \comma \qquad X_- = X_{-1} - X_4 
\comma \\
X &= X_1 + iX_2 \comma \qquad \Xbar = X_1 -iX_2 \comma  \\
\mathbb{X}' &= V_L \mathbb{X} V_R \comma \qquad V_L \in SL(2)_L \comma \quad 
 V_R \in SL(2)_R \period 
\end{align}
Then the basic transformations  are identified  as follows:
\nxt
(1)\ Dilatation
\begin{align}
X_+ &\rightarrow \lam X_+ \comma \quad 
X_- \rightarrow {1\over \lam} X_- \comma \quad 
X, \Xbar:\ \mbox{invariant}  \\
V_L^d(\lam) &= \matrixii{\sqrt{\lam}}{0}{0}{ {1\over \sqrt{\lam}}} 
\comma \quad V_R^d(\lam) =  \matrixii{\sqrt{\lam}}{0}{0}{ {1\over \sqrt{\lam}}} 
\period
\end{align}
(2)\ Rotation
\begin{align}
X &\rightarrow \xi  X \comma \quad 
\Xbar \rightarrow {1\over \xi}  \Xbar \comma \quad X_\pm:\ \mbox{invariant}
\\
V_L^r(\xi) &= \matrixii{\sqrt{\xi}}{0}{0}{{1\over \sqrt{\xi}}}
\comma \quad V_R^r(\xi) =  \matrixii{{1\over \sqrt{\xi}}}{0}{0}{\sqrt{\xi}}
\period
\end{align}
(3)\ Translation
\begin{align}
X &\rightarrow  X+ \al X_+ \comma \quad 
\Xbar \rightarrow \Xbar + \albar X_+ \comma 
\quad X_+:\ \mbox{invariant}  \\
X_- &\rightarrow X_- + \al \Xbar + \albar X + \albar \al X_+ \comma \\
V^{tr}_L(\albar) &= \matrixii{1}{0}{\albar }{1}
\comma \quad V_R^{tr}(\al) = \matrixii{1}{\al }{0}{1} \period
\label{translation}
\end{align}
(4)\ Special conformal transformation
\begin{align}
X & \rightarrow X + \be X_- \comma \quad \Xbar \rightarrow 
\Xbar + \bebar X_-  \comma \quad X_-:\ \mbox{invariant} \\
X_+ &\rightarrow X_+ + \bebar X + \be \Xbar + \bebar\be X_- \comma \\
V_L^{sc}(\be) &= \matrixii{1}{\be}{0}{1} \comma \quad 
V_R^{sc}(\bebar) = \matrixii{1}{0}{\bebar}{1}  \period
\label{specialconformal}
\end{align}
\apsection{D:\quad Effect of the canonical change of variables 
 on the correlation function}
\setcounter{equation}{0}
\renewcommand{\theequation}{D.\arabic{equation}}
\renewcommand{\thefigure}{D.\arabic{figure}}
Here we elaborate on  the effect of the canonical change of variables 
 on the calculation of the correlation functions mentioned in section 3.1. 
 As is well-known,  in classical mechanics  the change of  canonical variables from 
a set $(q,p)$ to a new set $(q^{\prime},p^{\prime})$ is effected by 
the transformation of the action of the form 
\beq{
S(q,p) =S^{\prime}(q^{\prime},p^{\prime})+\int d\tau \frac{dF}{d\tau}\comma
}
where $F= F(q,p^{\prime})$ is the generating function of the transformation 
 giving $p$ and $q'$ in terms of $(q,p')$ in the manner
\beq{
\frac{\del F}{\del q} = p\comma\quad\frac{\del F}{\del p^{\prime}}&=q^{\prime}\period
}
Although the  added term $\int d\tau (dF/d\tau)$ does not affect the classical equation of 
 motion, it can  produce a contribution when evaluated on a given trajectory 
 if $F(\tau_f) \ne F(\tau_i)$:
\beq{
S(q,p)\big| _{\tau=\tau_i}^{\tau=\tau_f} =S^{\prime}(q^{\prime},p^{\prime})\big| _{\tau=\tau_i}^{\tau=\tau_f} + F(\tau_f )-F(\tau_i )\period\label{apd-0}
}

The existence of such additional contribution can also be seen  from a quantum-mechanical point of view. 
Consider the  amplitude $\calA$ for the transition from a state $\Psi _1$ to 
a state $\Psi_2$ in the path integral formulation. In terms of the variables $(q,p)$, it is expressed as
\beq{
\calA=\int dq_i dq_f\int \mathcal{D}q\big|_{q(\tau_i) =q_i\comma \, q(\tau_f) = q_f}  e^{-S[q]\big|_{\tau=\tau_i}^{\tau=\tau_f}}\bar{\Psi} _2 (q_f)\Psi _1 (q_i)\comma 
}
where  the wave functions are defined by $
\Psi_1 (q_i) \equiv \braket{q_i | \Psi _1}$,  $\Psi_2 (q_f) \equiv \braket{q_f | \Psi _2}$ and $\bar{\Psi}$ denotes  the complex conjugate of $\Psi$. 
Suppose that the semiclassical approximation is valid and the path integral is dominated by a saddle point configuration $q_{\ast}$. Then, we have 
\beq{
\calA=e^{-S[q_{\ast}]\big|_{\tau=\tau_i}^{\tau=\tau_f}}\bar{\Psi} _2 (q_{f\ast})\Psi _1 (q_{i\ast})\period\label{apd-1}
}
Now if we express the same semi-classical amplitude in terms of a different 
  set of canonical variables $(q^{\prime},p^{\prime})$, we likewise get 
\beq{
\calA=e^{-S^{\prime}[q^{\prime}_{\ast}]\big|_{\tau=\tau_i}^{\tau=\tau_f}}\bar{\Psi}^{\prime} _2 (q^{\prime}_{f\ast})\Psi^{\prime} _1 (q^{\prime}_{i\ast})\comma\label{apd-2}
}
where the wave functions $\Psi^{\prime}$ are given by
$\Psi^{\prime}_1 (q^{\prime}_i) \equiv \braket{q^{\prime}_i | \Psi _1}$, 
$\Psi^{\prime}_2 (q^{\prime}_f) \equiv \braket{q'_f | \Psi _2}$. 
Comparing the two expressions \sref{apd-1} and \sref{apd-2}, we find  the following relation between the two actions evaluated on $q_\ast$:
\beq{
S[q_{\ast}] \big|_{\tau=\tau_i}^{\tau=\tau_f}=S^{\prime}[q^{\prime}_{\ast}]\big|_{\tau=\tau_i}^{\tau=\tau_f}-\log \left( \Psi_1^{\prime}(q^{\prime}_{i\ast})/ \Psi_1(q_{i\ast})\right) -\log \left( \bar{\Psi}_2^{\prime}(q^{\prime}_{f\ast})/ \bar{\Psi}_1(q_{f\ast})\right)\period\label{apd-3}
}
This can be simplified  by using the relation between the two wave functions $\Psi$ and $\Psi^{\prime}$, namely
\beq{
\Psi _1 ^{\prime}(q^{\prime}_{i})=\braket{q^{\prime}_i | \Psi _1} = \int dq_i \braket{q^{\prime}_i|q_i}  \braket{q_i | \Psi _1} =\int dq_i \braket{q^{\prime}_i|q_i}  \Psi_1 (q_i)\period  
}
In the semiclassical approximation, this relation is also dominated by the saddle point value and we may write
\beq{
\Psi _1 ^{\prime}(q^{\prime}_{i\ast}) = \braket{q^{\prime}_{i\ast}|q_{i\ast}} \Psi_1 (q_{i\ast})\period\label{apd-4}
}
Using the formula \sref{apd-4} and the similar one for $q_f$, the relation \sref{apd-3} can be reexpressed as
\beq{
S[q_{\ast}] \big|_{\tau=\tau_i}^{\tau=\tau_f}=S^{\prime}[q^{\prime}_{\ast}]\big|_{\tau=\tau_i}^{\tau=\tau_f}-\log \braket{q^{\prime}_{i\ast}|q_{i\ast}}+\log \braket{q^{\prime}_{f\ast}|q_{f\ast}}\period\label{apd-5}
}
Comparing \sref{apd-0} and \sref{apd-5}, we find the useful identification 
\beq{
F(\tau_f ) =\log \braket{q^{\prime}_{f\ast}|q_{f\ast}}\comma\quad F(\tau_i ) =\log \braket{q^{\prime}_{i\ast}|q_{i\ast}}\period
}

The discussion above clearly shows that, when we evaluate the action and the wave functions in terms of  different canonical variables, we must add appropriate 
 contribution  of the form $\log \braket{q^{\prime}|q}$ 
 to $\log \calA$ in order to  obtain the correct amplitude. Explicitly, 
\beq{
\log \calA =& -S[q_{\ast}] \big|_{\tau=\tau_i}^{\tau=\tau_f} +\log \Psi^{\prime}_1 (q^{\prime}_{i\ast})+\log \bar{\Psi}^{\prime}_2 (q^{\prime}_{f\ast})\nn\\
&+\log \braket{q^{\prime}_{f\ast}|q_{f\ast}} -\log \braket{q^{\prime}_{i\ast}|q_{i\ast}}\period
} 
This remark also applies to our  calculation of the correlation functions in this paper, 
 where we make the change of variables from the original embedding 
coordinates $X_\mu(\sig)$  to the angle variables $\phi_i $ when computing 
 the wave functions. 
Therefore,  in principle, we need to add   additional contributions of the form 
$ \log \braket{X(\sigma )| \phi _i}$.  In the case of the correlation functions 
 for the GKP strings, however,  we will show in Appendix E that these terms 
 actually do not contribute to the final result.  Arguments for more general cases 
 are yet to be developed. 
\apsection{E:\quad Some details of  the two point function of the elliptic GKP strings}
\setcounter{equation}{0}
\renewcommand{\theequation}{E.\arabic{equation}}
\renewcommand{\thefigure}{E.\arabic{figure}}
In this appendix,  we present some details of  the evaluation of the two point function of the elliptic GKP strings  left out in the discussion of section 4.2.  In particular, we will explicitly compute the 
contributions of the action $S [X]\big| _{\tau=\tau_i}^{\tau=\tau_f}$ and of the reference 
 wave functions $\Psi _{\refsol (\tau_i)}$ and $\Psi _{\refsol (-\tau_f)}$. 
 We will show that the dependence on $\tau_i$ and 
 $\tau_f$ exactly cancel among these contributions  and argue that the extra contribution 
of the type $\log \langle X(\sig)| \phi_i \rangle$ discussed in Appendix D 
does not affect the two point function in the case of the GKP strings. 

Let us recall the reference solution for  the elliptic GKP strings, shown   
previously in \sref{refsolelGKP}. It is of the form  
\beq{
\refsol(\tau ) = \pmatrix{cc}{e^{-\kappa \tau}\cosh \rho (\sigma )&e^{\omega \tau}\sinh \rho (\sigma )\\e^{-\omega \tau}\sinh \rho (\sigma )&e^{\kappa \tau}\sinh \rho (\sigma )}\comma\label{ape-0}
}
where $\kappa$, $\omega$, $\cosh \rho (\sigma)$ and $\sinh \rho (\sigma)$ are 
given by 
\beq{
\kappa \equiv \frac{2k}{\pi} \mathcal{K}(k^2) \comma & \quad \omega \equiv \frac{2}{\pi} \mathcal{K}(k^2)\comma\\
\cosh \rho (\sigma) \equiv \frac{1}{\sqrt{1-k^2}}\dn \left( \omega (\sigma +\pi /2)\right)\comma &\quad \sinh \rho (\sigma) \equiv \frac{k}{\sqrt{1-k^2}}\cn \left( \omega (\sigma +\pi /2)\right)\period
}
Here $k (\le 1)$ is the elliptic modulus, $\mathcal{K}(k^2)$ is the 
the complete elliptic integral of the first kind and $\dn(u)$ and $\cn(u)$ are the Jacobi 
 elliptic functions. 

 For this solution the action can be easily 
calculated as 
\beq{
S[X] \big| _{\tau=\tau_i}^{\tau=\tau_f}=\frac{\sqrt{\lambda}}{\pi}\int d^2 z \del \vec{X}\delbar \vec{X}=\frac{\sqrt{\lambda}}{2\pi} \kappa ^2 (\tau_f-\tau_i )\int d\sigma \,\sn^2 \left (\omega(\sigma +\pi /2)\right)\period\label{ape-1}
}
Explicit evaluation of the integral, though not difficult, is unnecessary as 
 it will be seen to be canceled by the contribution of the wave functions. 

As for the evaluation of the wave functions $\Psi \big| _{\refsol (\tau_i)}$ and $\Psi \big|_{\refsol (-\tau_f)}$, it suffices to express them in terms of the value 
at $\tau=0$, namely $ \Psi\big| _{\refsol(0)}$.  This can be achieved 
by the use of the universal formula derived in section 3.3  since 
 the time evolution of the reference solution can be effected   by a global 
symmetry transformation.  It  consists of a dilatation 
and a rotation in the form 
\beq{
\refsol(\tau )= \pmatrix{cc}{e^{(\omega -\kappa)\tau/2}&0\\0&e^{-(\omega -\kappa)\tau/2}}\refsol (0) \pmatrix{cc}{e^{-(\omega +\kappa)\tau/2}&0\\0&e^{(\omega +\kappa)\tau/2}}\period
}
This allows us to compute the difference of the angle variables 
 as 
\beq{
 &\iDelta \phi _{0} = \phi _{\infty}\big| _{\refsol (\tau )}-\phi _{\infty}\big| _{\refsol (0 )} = -i (\omega -\kappa) \tau\comma\label{ape-2}\\
&\iDelta \phi _{\infty} = \phi _{\infty}\big| _{\refsol (\tau )}-\phi _{\infty}\big| _{\refsol (0 )}=-i (\omega +\kappa)\tau\period\label{ape-3}
}
On the other hand,  the action variables $S_0$ and $S_{\infty}$ for the elliptic GKP string solution are given  by $S_0 = (\Delta +S)/2$ and $S_{\infty} = -(\Delta-S)/2$,  
where the global charges $\Delta$ and $S$ are expressed as 
\beq{
&\Delta = \frac{\sqrt{\lambda}}{2\pi}\kappa\int _0^{2\pi} d\sigma  \cosh ^2 \rho (\sigma) \comma \qquad 
S=\frac{\sqrt{\lambda}}{2\pi}\omega \int _{0}^{2\pi }d\sigma   \sinh ^2 \rho (\sigma) \period \label{DeltaS}
}
From the expressions \sref{ape-2}-\sref{DeltaS} and with the use of  some 
basic  identities for the elliptic functions, we can 
 express the wave functions at $\tau_i$ and $-\tau_f$ in terms of the 
 one at $\tau=0$ as 
\beq{
\Psi \big|_{\refsol (\tau_i)} &= \exp \left( iS_0 \iDelta \phi_0 +iS_{\infty} \iDelta \phi_{\infty} \right) \Psi \big|_{\refsol(0)}\nn\\
&=\exp \left( -\frac{\sqrt{\lambda}\kappa^2\tau _i}{2\pi} \int _{0}^{2\pi} d\sigma \,\sn ^2 (\omega (\sigma +\pi/2) ) \right)\Psi \big|_{\refsol(0)}\comma\label{ape-4}\\
\Psi \big|_{\refsol (-\tau_f)} &= \exp \left( iS_0 \iDelta \phi_0 +iS_{\infty} \iDelta \phi_{\infty} \right) \Psi _{\refsol(0)}\nn\\
&=\exp \left( \frac{\sqrt{\lambda}\kappa^2\tau _f}{2\pi} \int _{0}^{2\pi} d\sigma \,\sn ^2 (\omega (\sigma +\pi/2) ) \right)\Psi \big|_{\refsol(0)}\period\label{ape-5}
}
Combining \sref{ape-4} and \sref{ape-5} with \sref{ape-1}, we see that 
 the dependence on $\tau_i$ and $\tau_f$ precisely cancel and we get 
\beq{
\Psi\big| _{\refsol (\tau_i)}\Psi\big| _{\refsol (-\tau_f)} \exp\left( -S[X] \big| _{\refsol}\right) =\left( \Psi\big| _{\refsol(0)}\right)^2\comma 
}
as announced in \sref{finalelgkp}. 

We now discuss what this result implies concerning 
 the nature of the additional contribution  of the form $\log \braket{X(\sigma) | \phi_i}$ at $\tau_i$ and $\tau_f$ discussed in Appendix D. 
Such a contribution must be considered since 
the action is evaluated with the string coordinates $X_\mu(\sig)$ 
while  the wave functions are computed in terms of the action-angle variables. 
First note that since the saddle point configuration is characterized 
by the values of the action-angle variables, the expression
 $\log \braket{X(\sigma) | \phi_i}$ can be viewed as a function of $S_i$ 
and $\phi_i$. In fact since $S_i$ are constant for our solution we need to 
 consider only the dependence on the angle variables $\phi_0$ and $\phi_\infty$.
  We then  recall  the  fact that 
the values of such angle variables can be shifted by arbitrary parameters 
 $a$ and $a'$ as  in \sref{shiftphiX}, \sref{shiftphiXtilL} and \sref{shiftphiXtilR}. 
Nevertheless the final result for the two point function given in \sref{finalelgkp} 
is independent of such parameters. This implies that the contribution 
 of the term  $\log \braket{X(\sigma) | \phi_i}$ for the GKP solution does
not   depend on  $\phi_0$ and $\phi_\infty$. Such a constant 
 contribution can then be absorbed in  the normalization of the wave function 
and hence does not affect the result for the two point function. 
\apsection{F:\quad Evaluation of $\braket{\hat{i}_{\pm}^{L}\comma \hat{j}_{\pm}^{L}}$}
\setcounter{equation}{0}
\renewcommand{\theequation}{F.\arabic{equation}}
\renewcommand{\thefigure}{F.\arabic{figure}}
In this appendix, we give some  details of the evaluation of $\braket{\hat{i}_{\pm}^{L}\comma \hat{j}_{\pm}^{L}}$. 
As outlined in subsection 5.2,  we first calculate the invariant products 
involving only the small solutions, namely $\slprod{1_{+}^{L}}{2_{-}^{L}},\slprod{2_{-}^{L}}{3_{+}^{L}}$ and $\slprod{3_{+}^{L}}{1_{+}^{L}}$, in the same manner as the corresponding quantities in the R sector. 
Since the  prefactors $\exp\left(i(\xi^{-1}-\xi )\kappa_i \log \epsilon_i \right)$ in \sref{ident} are simply unity for $\xi=1$, these products are obtained as 
\beq{
&\log \slprod{1_+^{L}}{2_-^{L}}=\hat{h}(\kappa _1) +\hat{h}(\kappa _2)-\hat{h}(\frac{\kappa _1 +\kappa _2+\kappa _3}{2})-\hat{h}(\frac{\kappa _1 +\kappa _2-\kappa _3}{2})\comma\\
&\log \slprod{2_-^{L}}{3_+^{L}}=\hat{h}(\kappa _2) +\hat{h}(\kappa _3)-\hat{h}(\frac{\kappa _1 +\kappa _2+\kappa _3}{2})-\hat{h}(\frac{-\kappa _1 +\kappa _2 +\kappa _3}{2})\comma\\
&\log \slprod{3_+^{L}}{1_+^{L}}=\hat{h}(\kappa _3) +\hat{h}(\kappa _1)-\hat{h}(\frac{\kappa _1 +\kappa _2+\kappa _3}{2})-\hat{h}(\frac{\kappa _1 +\kappa _2 -\kappa _3}{2})\period
}
Here the function  $\hat{h}(x)$ is obtained  from the function $h(x,\xi)$ 
defined  by 
\beq{
h(x,\xi )\equiv &-\frac{1}{\pi i} \int _{0}^{\infty} d\xi^{\prime} \frac{1}{\xi^{\prime^{2}}-\xi^2}\log \left( 1- \exp \left( -2\pi x(\xi^{\prime^{-1}}+\xi^{\prime})\right) \right)\comma \label{apf-1}
}
by the limiting procedure $\lim_{\epsilon\to +0} h(x,1+i\epsilon)$. 
Although $h(x,\xi)$  is a complicated function for general values of $x$ and $\xi$,  
it simplifies enormously in the limit above as 
\beq{
\hat{h}(x)=&-\frac{1}{2\pi i} \left[ \int _{0}^{\infty} d\xi^{\prime} \frac{1}{\xi^{\prime^{2}}-(1+i\epsilon )}\log \left( 1- \exp \left( -2\pi x(\xi^{\prime^{-1}}+\xi^{\prime})\right) \right) \right.\nn\\
&\left. + \int _{0}^{\infty} d\xi^{\prime^{-1}} \frac{1}{\xi^{\prime^{-2}}-(1+i\epsilon )}\log \left( 1- \exp \left( -2\pi x(\xi^{\prime^{-1}}+\xi^{\prime})\right) \right) \right]\nn\\
=&-\frac{1}{2\pi i} \int _{0}^{\infty} d\xi^{\prime} \left(\frac{1}{\xi^{\prime^{2}}-(1+i\epsilon )}-\frac{1}{\xi^{\prime^{2}}-(1-i\epsilon )}\right)\log \left( 1- e^{  -2\pi x(\xi^{\prime^{-1}}+\xi^{\prime}) }\right) \period\label{apf-2}
}
Since the first factor in the integrand is recognized as $\delta (\xi^{\prime^2}-1)$, 
$\hat{h}(x)$ finally reduces to a simple function  given by 
\beq{
\hat{h}(x)=-\frac{1}{2}\log \left( 1- e^{-4\pi x} \right)\period 
} 
 
Next we recall that in \cite{Kazama:2011cp} the invariant products of $i_{\pm}$ are determined up to three  constants, $c, A_2$ and $A_3$, which are connected to the normalization of $i_{\pm}$. Specifically, they are related to the 
 invariant products obtained above as (see equations (3.69), (3.72) and (3.65) 
 of \cite{Kazama:2011cp}) 
\beq{
&\slprod{1_+^{L}}{2_-^{L}}=\frac{1}{2A_2 \sinh 2\pi \kappa _2}\comma \\
&\slprod{2_-^{L}}{3_+^{L}}=\frac{A_3 \left( e^{2\pi(\kappa_1-\kappa_3)}-e^{2\pi\kappa _2}\right)}{2A_2\sinh 2\pi \kappa _2} = -\frac{A_3\sinh \left( \pi (-\kappa _1+\kappa_2+\kappa_3 ) \right)}{A_2\sinh 2\pi \kappa_2}e^{\pi(\kappa_1+\kappa_2-\kappa_3)}\comma\\
&\slprod{3_+^{L}}{1_+^{L}}=-cA_3\period
} 
From these relations, we get
\beq{
&A_2=\frac{1}{2\sinh 2\pi \kappa_2 \braket{1_+\comma 2_-}}\comma\\
&A_3=-\frac{\braket{2_-\comma 3_+}}{\braket{1_+\comma 2_-}}\frac{e^{-\pi (\kappa_1+\kappa_2 -\kappa_3)}}{2\sinh \left( \pi (-\kappa_1 +\kappa_2 +\kappa_3 )\right)}\comma \\
&c=2\frac{\braket{3_+\comma 1_+}\braket{1_+ \comma 2_+}}{\braket{2_- \comma 3_+}}e^{\pi (\kappa_1+\kappa_2-\kappa_3 )}\sinh \left( \pi (-\kappa_1 +\kappa_2 +\kappa_3)\right) \period 
}
With these expressions for $c\comma A_2$ and $A_3$,  we can 
 compute  the remaining invariant products given in \cite{Kazama:2011cp}.
  For instance, the products involving big solutions around $z_1$ and $z_2$, 
namely  $\slprod{1_-^{L}}{2_+^{L}}$, $\slprod{1_+^{L}}{2_+^{L}}$ and $\slprod{1_- ^{L}}{2_- ^{L}}$, are obtained as 
\beq{
&\slprod{1_-^{L}}{2_+^{L}}=-\frac{1}{\braket{1_+\comma 2_-}}\frac{\sinh \left( \pi (\kappa_1+\kappa _2+\kappa _3)\right)\sinh \left( \pi (\kappa _1 +\kappa _2 -\kappa _3)\right)}{\sinh 2\pi \kappa _1\sinh 2\pi \kappa _2}\comma\\
&\slprod{1_+^{L}}{2_+^{L}}= \frac{\braket{3_+\comma 1_+}}{\braket{2_- \comma 3_+}}\frac{\sinh \left( \pi (-\kappa_1+\kappa_2 +\kappa_3 )\right)}{\sinh 2\pi \kappa _2}e^{\pi (\kappa _1+\kappa _2-\kappa_3)}\comma\\
&\slprod{1_- ^{L}}{2_- ^{L}}=\frac{\braket{2_- \comma 3_+}}{\braket{3_+\comma 1_+}}\frac{\sinh \left( \pi (\kappa_1 -\kappa_2 +\kappa_3 )\right)}{\sinh 2\pi \kappa _1}e^{-\pi (\kappa _1+\kappa _2-\kappa_3)}\period
} 
Combining these results, the invariant products between $\hat{i}^{L}_{\pm} \equiv (\pm i^{L}_{+}+i^{L}_{-})/\sqrt{2}$ are given by
\beq{
&\slprod{\hat{1}_{+}^{L}}{\hat{2}_{-}^{L}}=\tilde{c}\sqrt{\frac{\sinh \left(\pi (\kappa_1+\kappa_2-\kappa_3)\right)\sinh \left(\pi (\kappa_1+\kappa_2+\kappa_3)\right)}{\sinh (2\pi \kappa_1)\sinh (2\pi \kappa_2)}}\comma\\
&\slprod{\hat{2}_{-}^{L}}{\hat{3}_{+}^{L}}=\tilde{c}\sqrt{\frac{\sinh \left(\pi (-\kappa_1+\kappa_2+\kappa_3)\right)\sinh \left(\pi (\kappa_1+\kappa_2+\kappa_3)\right)}{\sinh (2\pi \kappa_2)\sinh (2\pi \kappa_3)}}\comma\\
&\slprod{\hat{3}_{+}^{L}}{\hat{1}_{+}^{L}}=\tilde{c}\sqrt{\frac{\sinh \left(\pi (\kappa_1-\kappa_2+\kappa_3)\right)\sinh \left(\pi (\kappa_1+\kappa_2+\kappa_3)\right)}{\sinh (2\pi \kappa_3)\sinh (2\pi \kappa_1)}}\comma
}
where $\tilde{c}$ is the expression given by 
\beq{
\tilde{c}= 1-\sqrt{\frac{\sinh(\pi (-\kappa_1+\kappa_2+\kappa_3))\sinh(\pi (\kappa_1-\kappa_2+\kappa_3))\sinh(\pi (\kappa_1+\kappa_2-\kappa_3))}{\sinh(\pi (\kappa_1+\kappa_2+\kappa_3))}}\period
}
\apsection{G:\quad Two point function  for a  particle in Euclidean $AdS_3$}
\setcounter{equation}{0}
\renewcommand{\theequation}{G.\arabic{equation}}
\renewcommand{\thefigure}{G.\arabic{figure}}
In section 4, we developed, through  somewhat abstract reasoning  based on integrability,  a powerful 
 method of computing the two point 
 functions, which are applicable even to the cases where the action-angle variables 
are difficult to construct  explicitly.  It is  instructive, however, to check the method 
 by applying it to a non-trivial system for which the action-angle variables 
{\it  can}  be obtained  analytically. 

One such system is a particle in \ads3. If we employ the global coordinates 
$(\theta, \phi, \rho)$,  related to the embedding coordinates $X_\mu$ by 
\beq{
&X_{-1}=\cosh \rho \cosh \phi \comma &X_{4}=\cosh \rho \sinh \phi \comma\\
&X_1=\sinh \rho \cos \theta \comma &X_2=\sinh \rho \sin \theta \comma 
}
the action takes the form 
\beq{
S=\int d\tau \,\frac{1}{2}\left( -\dot{\rho}^2 - \cosh ^2 \rho\, \dot{\phi}^2 +\sinh ^2 \rho \,\dot{\theta}^2 \right)\period
}
The action variables corresponding to the motions  of $\theta$, $\phi$ and $\rho$ 
are defined as 
\beq{
J_{\theta}=\frac{1}{2\pi}\oint p_{\theta }d\theta \comma \quad  J_{\phi} = \frac{1}{2\pi}\oint p_{\phi} d\phi \comma \quad J_{\rho} = \frac{1}{2\pi}\oint p_{\rho} d\rho \period
}
Since the motions of $\phi$ and $\rho$ are actually not periodic,  
we  need to perform the analytic continuations   $\tilde{\rho}=i\rho$ and $\tilde{\phi}=i\phi$ to define the periodic integrals above. The angle variables 
 $(\varphi_\theta, \varphi_\phi, \varphi_\rho)$  conjugate 
 to $(J_\theta, J_\phi, J_\rho)$  can be constructed as 
\beq{
\frac{\del F}{\del J_{\theta}} =\varphi_{\theta} \comma\quad\frac{\del F}{\del J_{\phi}} =\varphi_{\phi}\comma\quad\frac{\del F}{\del J_{\rho}} =\varphi_{\rho}\comma 
}
where $F \equiv\int d\theta \, p_{\theta}+\int d\phi \, p_{\phi}+ \int d\rho \, p_{\rho}$ is the generating function. To compute these integrals,  one must 
express  the momenta $(p_\theta, p_\phi, p_\rho)$ 
in terms of the action variables and the coordinates $(\theta, \phi, \rho)$. 
The general form of the angle variables obtained through this procedure are rather 
complicated,  but for  $J_{\theta}=J_{\rho}=0$  they take the 
  following simple form:
\beq{
&\varphi_{\phi}=-\phi +\ln \cosh \rho \comma\quad\varphi_{\theta}= \theta +\ln \sinh \rho \comma\quad\varphi_{\rho}=-2i \ln \sinh \rho \period
}
We  can now evaluate these angle variables explicitly on the reference solution $\refsol$, 
given by 
\beq{
\refsol = \pmatrix{cc}{e^{-\kappa \tau}&0\\0&e^{\kappa \tau}}\comma
}
and on the  transformed solution $\mathbb{X} \equiv V_{L} \refsol V_{R}$, 
with  $V_{L,R}$ given in  \sref{LRtwopoint}, just as in section 4. The results  for the shifts are 
\beq{
&\iDelta\varphi^{\mathbb{X}}_{\phi}=-\log (a a^{\prime} )\comma \quad \iDelta\varphi^{\tilde{\mathbb{X}}}_{\phi} =\log (a a^{\prime} x_0 \bar{x}_0 )\period
}
These shifts contribute to the two point function as 
\beq{
\exp \left( -\Delta (\iDelta\varphi^{\mathbb{X}}_{\phi}+\iDelta\varphi^{\tilde{\mathbb{X}}}_{\phi})\right) = \frac{1}{x_0^{\Delta}\bar{x}_0^{\Delta} }\comma
}
which  reproduces the correct spacetime behavior. The remaining contributions, \ie those from the action and the reference wave functions, cancel as in section 4.2 and we obtain the properly normalized two point function.
\apsection{H:\quad Exact solution describing a  scattering of three spinning strings in flat space and its action-angle variables}
\setcounter{equation}{0}
\renewcommand{\theequation}{H.\arabic{equation}}
\renewcommand{\thefigure}{H.\arabic{figure}}
Construction of  three-pronged solutions  in (the subspace of) $AdS_5 \times S^5$ 
is an important challenging   problem.  As discussed in section 3.2.4, their analytic 
 structure is expected to be qualitatively quite different from that of the  two point solutions.  
To give support to this observation, we present below an exact solution describing a scattering of three spinning strings in flat space and analyze its local behavior.  This 
 confirms  some important structures concerning  the 
action-angle variables. 

A solution describing  three interacting strings spinning in the $x_1$-$x_2$ plane 
is given by 
\beq{
X^{\mu} = -i\left( k^{\mu}_{1} \ln |z| +k^{\mu}_{2} \ln |z-1|+k^{\mu}_{3} \ln |z-\infty | \right)\comma \hspace{11pt} \mu \neq 1,2 \comma\\
X= \frac{w_3}{2i}(z-\bar{z} )\comma \quad \bar{X} = \frac{w_1}{2i}\left(\frac{1}{z}-\frac{1}{\bar{z}} \right)+\frac{w_2}{2i}\left(\frac{1}{z-1}-\frac{1}{\bar{z}-1} \right)\period 
}
Here, as indicated, $X^\mu$ denotes the directions other than the plane 
of rotation, $X$ and $\Xbar$ stand for  $X_1+iX_2$ and its complex conjugate 
 respectively and the momentum vectors $\vec{k}_i$ and the parameters $w_i$, 
which are related to the spins of the prongs, 
must satisfy the following conservation laws and the on-shell conditions demanded by the Virasoro conditions:
\begin{align}
&\vec{k}_1+\vec{k}_2+\vec{k}_3 =0 \comma \quad w_1+w_2=w_3  
\comma 
\nn\\
&(\vec{k}_{1})^2 +w_1 w_3 = (\vec{k}_{2})^2 +w_2 w_3 =(\vec{k}_{3})^2 +(w_3)^2 =0 \period 
\end{align}
Let us study its local behavior by focusing on the vicinity of the singularity 
$z=0$. The expansion around this point reads 
\beq{
&X=\frac{w_3}{2i}(z-\bar{z} ) \comma\quad\bar{X}=\frac{w_1}{2i}\left(\frac{1}{z}-\frac{1}{\bar{z}} \right) -\frac{w_2}{2i}(z-\bar{z} ) + O(|z|^2)\period\label{aph-1}
}
This should be compared with the well-known two-point spinning string solution 
of   \cite{Tseytlin} given by 
\beq{
&X=\frac{w}{2i}(z-\bar{z} ) \comma\quad \bar{X}=\frac{w}{2i}\left(\frac{1}{z}-\frac{1}{\bar{z}} \right)\period\label{aph-2}
}
There are two important differences  to be noted. First, the Fourier coefficient in front of  the structure $(z-\zbar)$ and the one in front of $(1/z-1/\zbar)$ are 
different in \sref{aph-1}, while they are the same in \sref{aph-2}. 
Since the log of the ratio of such Fourier coefficients describes the shift of the 
 angle variable, this means that the presence of the other  vertex 
 operator generated  a shift of the angle variable in the case of the three-pronged 
solution. This type of phenomenon was 
observed also in the calculation of the correlation functions performed in 
sections 4 and 5.  The second feature is that for  \sref{aph-1} 
there are an infinite number of additional Fourier modes excited 
 in $\Xbar$ besides  $(1/z-1/\zbar)$. However, since there are 
 no corresponding modes in $X$, these additional excitations do not
 contribute to the action variable, namely the spin,  given by 
\beq{
S=\frac{i}{4\pi \alpha^{\prime}}\int _{0}^{2\pi}d\sigma (X\dot{\bar{X}}-\dot{X}\bar{X})\period
}
This means that the infinite number of action variables corresponding 
 to such additional  Fourier modes must vanish. 
Therefore, the  solution above embodies the general feature expected of 
 the solution for the higher-point functions.  Namely, such a solution 
 has (possibly infinitely many)  dynamical angle variables for which 
 the conjugate action variables are zero,  in addition to those associated with 
 the  action variables  which are finite.  This suggests that solutions for higher-point functions in \ads3 may be constructed also by introducing infinitely many additional degenerate cuts
 on the spectral curve. 

\end{document}